\documentclass[12pt]{article}
\usepackage[utf8]{inputenc}
\usepackage{graphicx, caption, subcaption}
\usepackage{lipsum}
\usepackage[english]{babel}
\usepackage{enumerate}
\usepackage{enumitem}
\usepackage{moreenum}
\usepackage{amssymb}
\usepackage{amsmath}
\usepackage{mathtools}
\usepackage{braket}
\usepackage{tensor}
\usepackage{slashed}
\usepackage{float}
\usepackage{feynmf}
\usepackage{mathrsfs}
\usepackage{comment}
\usepackage{authblk}
\usepackage{cite}
\usepackage[a4paper, total={6.75in, 7.5in}]{geometry}
\usepackage{relsize}
\usepackage{scalerel}
\usepackage{stackengine,wasysym}
\usepackage[unicode,linktoc=all,hidelinks,colorlinks=true,linkcolor=blue,citecolor=blue,allcolors=blue]{hyperref}
\usepackage{tikz}
\usetikzlibrary{decorations.pathmorphing} 
\usepackage{placeins} 

\renewcommand{\eqref}[1]{\hyperref[#1]{Eq. \textup{(\ref{#1})}}}
\newcommand{\secref}[1]{\hyperref[#1]{Section}~\textup{\ref{#1}}}
\newcommand{\figref}[1]{\hyperref[#1]{Fig. \textup{\ref{#1}}}}

\newcommand\reallywidetilde[1]{\ThisStyle{%
  \setbox0=\hbox{$\SavedStyle#1$}%
  \stackengine{-.1\LMpt}{$\SavedStyle#1$}{%
    \stretchto{\scaleto{\SavedStyle\mkern.2mu\AC}{.5150\wd0}}{.6\ht0}%
  }{O}{c}{F}{T}{S}%
}}

\newcommand{\pvec}[1]{\vec{#1}\mkern2mu\vphantom{#1}}
\providecommand{\keywords}[1]{\textbf{Keywords: } \textbf{#1}}
\newcommand{\highvec}[1]{\overset{\smash{\raisebox{-3.75pt}{$\vec{\phantom{#1}}$}}}{#1}}

\title{\textbf{Subleading soft radiation during scattering of dressed states in QED}}
\author[1]{\textbf{Stavros Christodoulou}
    \footnote{schris14@ucy.ac.cy}}
\author[1]{\textbf{Nicolaos Toumbas}
    \footnote{nick@ucy.ac.cy}}
\affil[1]{Department of Physics, University of Cyprus, Nicosia, 1678, Cyprus}
\date{\today}

\begin{document}
\defineshorthand{;}{?}
\maketitle
\pagestyle{plain}

\begin{abstract}
    We study soft photon emission during scattering of Faddeev-Kulish charged states in QED at leading order in perturbation theory. The charged asymptotic particles are accompanied by clouds of an infinite number of soft photons of energy less than a characteristic infrared scale $E_d$. When the corresponding ``dressing'' functions are suitably corrected to subleading order in the soft momentum expansion, as advocated in recent work by Choi and Akhoury, we show explicitly that the emission of additional radiative soft photons with energy less than $E_d$ is completely suppressed. Moreover, the dressing renders the elastic amplitudes infrared-finite, order by order in perturbation theory, regulating the infrared divergences due to virtual soft photons at the energy scale $E_d$. Therefore, the characteristic energy scale of the soft photons in the clouds provides an effective infrared cutoff, allowing for the formulation of an infrared finite S-matrix.    

    
\end{abstract}
\vspace{1em}
\keywords{Infrared divergences, Soft theorems, Dressed state formalism}

\section{Introduction}
    \label{sec:intro}


Infrared divergences in QED \cite{Bloch:1937pw,Yennie:1961ad,Weinberg:1965nx,Gell-Mann:1954wra,Low:1954kd,Low:1958sn,Burnett:1967km,Dollard:1964jmp,Chung:1965zza,Greco:1967zza,Kibble:1968sfb,Kibble:1968lka,Kibble:1968npb,Kibble:1968oug,Kulish:1970ut,Strominger:2017zoo} 
obstruct the formulation of an S-matrix in terms of the conventional Fock basis of asymptotic states. 
Indeed, at any finite order in perturbation theory, scattering amplitudes between Fock states, which include only a finite number of soft photons, suffer from infrared divergences due to virtual photons in the loops, as well as soft real photon emission.
When summed to all perturbative orders, the virtual divergences exponentiate, leading to the vanishing of the Fock-basis amplitudes \cite{Weinberg:1965nx}. As a result, an infinite number of soft photons must be present in the initial and final states, in order to have non-vanishing S-matrix amplitudes. The vanishing of the conventional Fock-basis amplitudes in QED can be understood as a consequence of symmetry \cite{Lysov:2014csa,He:2014cra,Kapec:2015ena,Campiglia:2015qka,Kapec:2017tkm,Strominger:2017zoo,Gabai:2016kuf}. Large gauge transformations, where the gauge parameters asymptote to angle dependent constants at infinity, lead to an infinite number of conservation laws and Ward identities, which fail to be satisfied by the conventional initial and final charged Fock space configurations. Since the conservation laws cannot be satisfied, the corresponding scattering amplitudes vanish to all orders. These results do not pertain only to QED, but generalize to generic gauge theories with long-range interactions and gravity \cite{He:2014laa,Strominger:2014pwa,Banks:2014iha,Strominger:2013jfa}. The construction of a well-defined S-matrix and the understanding of the correlation between its hard and soft parts continue to pose a challenging problem in the formulation of these theories.  

One route to follow is to focus on inclusive processes where during the scattering of charged particles, an arbitrary number of soft photons with total energy below a certain value can be emitted \cite{Bloch:1937pw,Weinberg:1965nx,Yennie:1961ad}. Experimentally, the study of such inclusive processes is well-motivated, since soft photons with energies below a certain cutoff set by the resolution of the detectors cannot be observed. The corresponding inclusive cross-sections are infrared finite order by order in perturbation theory, with the infrared divergences due to virtual soft photons in the loops canceling against the infrared divergences due to the emission of real soft photons \cite{Bloch:1937pw,Weinberg:1965nx,Yennie:1961ad}. However, this route does not resolve the issue of obtaining a well defined S-matrix and does not provide a proper framework for describing the asymptotic particle states in the presence of long-range gauge interactions.

Faddeev and Kulish (FK) proposed another route to 
obtain infrared-finite observables in QED. Their approach involves dressing the asymptotic charged particle states with coherent clouds of soft photons \cite{Kulish:1970ut,Tomaras_2020,Akhoury:2011kq,Bachas:1995ik}. The clouds contain an infinite number of soft photons with energies that do not exceed a characteristic energy scale $E_d$, which we take to be very small. The FK states are, in fact, the proper eigenstates of the asymptotic Hamiltonian, which includes the slowly decaying parts of the interaction Hamiltonian (when written in the interaction picture) due to the long-range nature of the Coulombic force. 
Similar constructions were previously proposed by Dollard, Chung, and Kibble \cite{Dollard:1964jmp,Chung:1965zza,Greco:1967zza,Kibble:1968sfb,Kibble:1968oug,Kibble:1968npb,Kibble:1968lka,Carney:2018ygh}. 
The FK amplitudes associated with the scattering of dressed charged particles are finite, order by order in perturbation theory. Essentially, soft photons in the clouds regulate infrared divergences due to soft virtual photons in the loops, with the characteristic energy scale $E_d$ providing an infrared cutoff. Moreover, the clouds of soft photons render the charges associated with large gauge transformations of the initial and final dressed states independent of the momenta of the charged particles, allowing for the associated conservation laws to be satisfied \cite{Gabai:2016kuf} (see also \cite{Choi:2017ylo} for the analogous discussion in the gravitational case). Therefore, these dressed amplitudes remain non-vanishing to all orders in perturbation theory \cite{He:2014cra,Campiglia:2015qka,Kapec:2015ena,Mirbabayi:2016axw,Kapec:2017tkm,Gomez:2018war,Hirai:2019gio,Tomaras_2020,Hirai:2020kzx,Furugori:2020vdl,Hirai:2022yqw,Nguyen:2023ibj, Gabai:2016kuf}. 

Since the dressing provides an infrared cutoff of order $E_d$ for the elastic scattering amplitudes, thus regulating the infrared divergences due to virtual soft photons in the loops \cite{Kulish:1970ut,Tomaras_2020}, one wonders if a similar picture pertains for the radiative amplitudes, describing the emission of additional soft photons. These are extra soft photons, distinct from those already contained in the clouds.
In particular, if the characteristic energy scale $E_d$ behaves as a universal infrared cutoff, we should expect the emission of soft photons with energies below $E_d$ to be completely suppressed. In other words, any additional photons produced by radiation during scattering processes can be distinguished from the soft photons in the clouds, which carry energy below $E_d$, and will not lead to extra infrared divergences in the corresponding amplitudes at any finite order in perturbation theory. Moreover, only the soft photons in the clouds will exhibit strong entanglement with the hard particles, as shown in \cite{Gomez:2017rau,Tomaras_2020,Irakleous:2021ggq,Toumbas:2023qbo}. Finally, if this is indeed the case, the dressed state formalism can reproduce the inclusive Bloch–Nordsieck rates, computed in the conventional Fock basis, which remain finite, free of infrared divergences, order by order in perturbation theory.



In recent work, Choi and Akhoury \cite{Choi:2019rlz} constructed dressed states, where the FK dressing functions are corrected to subleading order in the soft momentum expansion and to first order in the QED coupling, using conservation laws associated with large gauge transformations and soft theorems at tree level and subleading order in the momentum expansion. Their results indicate that such dressed amplitudes associated with the emission and absorption of additional soft photons (or gravitons) with energies below $E_d$ are suppressed. 
Based on these results, they showed
that the dressed-state formalism yields the inclusive Fock-basis Bloch–Nordsieck rates and cross-sections. 

Our goal in this work is to calculate explicitly dressed amplitudes in QED, corresponding to the emission of soft radiative photons at tree level. Once the dressing factors associated with the charged particles are determined to subleading order in the soft momentum expansion, we demonstrate that the emission of soft photons with energies below $E_d$ is completely suppressed as advocated in \cite{Choi:2019rlz}.
We analyze the scattering of a dressed electron by a heavier dressed muon, of a dressed electron by a photon, and of a dressed electron by a dressed positron. For each case, we show that the elastic dressed amplitude is given by the infrared-finite part of the corresponding Fock-basis amplitude, thereby confirming that the dressing regulates the infrared divergences due to virtual soft photons. We proceed to show that, during such processes, the emission of soft photons with energies below $E_d$ is indeed suppressed at tree level. 
These results confirm the equivalence between the dressed-state formalism and the Bloch–Nordsieck method. 
For this result, it is crucial to implement the correction to the FK dressings to subleading order in the soft momentum expansion \cite{Choi:2019rlz}. It will be interesting to carry out the analysis at the one-loop order, in order to confirm that the suppression of such radiative amplitudes continues to hold. If this is true, the entanglement between these extra radiative soft photons and the dressed hard particles is small, as argued in \cite{Tomaras_2020,Irakleous:2021ggq,Toumbas:2023qbo}.

The paper is organized as follows. In \secref{sec:2}, we construct FK states to leading order in the soft momentum expansion and define the FK S-matrix elements. We consider both generic elastic and radiative photon-emitting processes. In \secref{sec:3}, we study soft radiation during electron--muon, electron--positron, and electron--photon scattering at tree level, and we apply the arguments in \cite{Choi:2019rlz} to construct dressings to subleading order in the soft momentum expansion. We demonstrate explicitly that the elastic dressed amplitudes are equivalent to the infrared-finite part of the corresponding Fock-basis amplitudes, and that the amplitudes for additional soft radiation are completely suppressed. We summarize our results and discuss their implications and open problems in \secref{sec:conc}. Details about notation and conventions can be found in \hyperref[sec:A]{Appendix A}. In \hyperref[sec:A]{Appendix B}, we review the subleading soft photon theorem. \hyperref[sec:A]{Appendix C} contains details concerning the calculation of single soft photon emitting amplitudes during scattering of dressed states. 

\section{Infrared structure of QED and the dressed state formalism}
    \label{sec:2}
    As shown by Faddeev and Kulish, dressed states are crucial in addressing the issue of infrared divergences and in providing an infrared-safe framework for describing particle interactions in QED \cite{Kulish:1970ut}. These states, formed by dressing the Fock space charged particles with clouds of soft photons, are eigenstates of the asymptotic Hamiltonian. The asymptotic Hamiltonian differs from the usual free Hamiltonian because it incorporates the slowly decaying parts of the interaction Hamiltonian (when written in the interaction picture). As a result, the S-matrix elements between dressed states are non-vanishing and free of infrared divergences to all orders and order by order in perturbation theory \cite{Chung:1965zza, Kibble:1968lka, Kibble:1968npb, Kibble:1968oug, Kibble:1968sfb,Kulish:1970ut}.

In this section, we construct the FK states associated with charged Fock states to leading order in the soft momentum expansion and review various properties of the corresponding amplitudes, following \cite{Tomaras_2020}. We first demonstrate that the elastic dressed amplitudes do not contain any infrared divergences. These amplitudes are equal to the infrared-finite part of the corresponding undressed elastic amplitudes. Then, we study the emission of an additional single soft photon during scattering of dressed charged states. The leading soft factor, which contains a pole as the energy of the emitted photon vanishes, is not present in the FK dressed amplitudes. However, these amplitudes are not suppressed in the limit of arbitrarily soft momentum, as expected in order for the dressed state formalism to yield the Bloch - Nordsieck inclusive cross-sections. This motivates us to consider subleading corrections to the dressing functions, as advocated in \cite{Choi:2019rlz}.

\subsection{Construction of Faddeev-Kulish states}
    The FK states are obtained by acting with the unitary operator $e^{R_f}$ on Fock basis states \cite{Kulish:1970ut,Gabai:2016kuf,Tomaras_2020,Gomez:2017rau}, where
\begin{equation}\label{eq.2.1.1}
    R_f=
        \int\frac{d^3\vec{p}}
            {(2\pi)^3(2\omega_p)}\
        \hat{\rho}(\Vec{p})
        \int
            \reallywidetilde{d^3k}\
            \bigg(
                f(\highvec{p},\Vec{k})
                    \cdot
                a^{\dagger}(\Vec{k})-
                h.c.
            \bigg)
\end{equation}
and 
\begin{equation}\label{eq.2.1.2}
    \hat{\rho}(\pvec{p})=
        \sum_s
            b_s^{\dagger}(\pvec{p})b_s(\pvec{p})-
            d_s^{\dagger}(\pvec{p})d_s(\pvec{p})
\end{equation}
is the charge density operator; $b_s^{\dagger}(\pvec{p})$ and $d_s^{\dagger}(\pvec{p})$ create a charged fermionic particle and an antiparticle with momentum $\pvec{p}$, spin polarization $s$, and energy $\omega_p=\sqrt{|\vec{p}|^2+m^2}$, respectively. 
Moreover, $a^{\dagger}_r(\pvec{k})$ creates a photon with momentum $\pvec{k}$ and polarization vector $\epsilon_r^{\mu}(\pvec{k}),\ r=0,1,2,3$, and
\begin{equation}\label{eq.2.1.4}
    f(\highvec{p},\Vec{k})
        \cdot
    a^{\dagger}(\Vec{k})=
    \sum_r
        f^{\mu}(\highvec{p},\Vec{k})\
        \epsilon_{r\mu}^*(\Vec{k})\
        a^{\dagger}_r(\Vec{k})
\end{equation}
where 
\begin{equation}\label{eq.2.1.5}
    f^{\mu}(\highvec{p},\Vec{k})=
        e\bigg(
            \frac{p^{\mu}}{p\cdot k}-
            c^{\mu}
        \bigg)
        e^{-ip\cdot k t_0/p^0}
        ,\hspace{5em}
        c^{\mu}=\frac{1}{2k^0}
            (-1,\hat{k})
\end{equation}
The dressing function $f^{\mu}(\Vec{k},\highvec{p})$ is singular as $|\Vec{k}|\rightarrow0$, since it has a pole. Furthermore, $t_0$ is a time reference scale, which can be set to zero\footnote{
    Since the integral in \eqref{eq.2.1.1} is governed by low momentum modes with $|\Vec{k}|\leq E_d$, we can choose $E_dt_0<<1$ and approximate $e^{-ip\cdot k t_0/p^0}$ with unity.
}, and $c^{\mu}$ is a null vector satisfying $c\cdot k = 1$. As a result, the dressing function is transverse.  

The integration measure for the photon momenta is given by
\begin{equation}\label{eq.2.1.3}
    \int\reallywidetilde{d^3k}=
        \frac{1}{2}
        \int
            \frac{d\Omega_k}
                {(2\pi)^3}
        \int_{\lambda}^{E_d}
            \omega_kd\omega_k
\end{equation}
and includes integration over the direction of the unit vector $\hat{k}$ and over the magnitude of $\vec{k}$, which is equal to the energy of the photon: $\omega_k=|\vec{k}|$. The lower limit of the $\omega_k$-integral is given by the infrared cutoff of the theory, $\lambda$, which will be taken to zero at the end of the computations\footnote{
    Logarithmic infrared divergences in physical quantities will be displayed as powers of $\log\lambda$.
}. The upper limit of the integral restricts the photon energies in the clouds not to exceed the characteristic scale $E_d$, which we take to be sufficiently small for the soft photon theorems to apply \cite{Gomez:2018war}. In particular, we take $\lambda < E_d \leq \Lambda$, where $\Lambda$ is the infrared reference energy scale that characterizes virtual soft photons \cite{Weinberg:1965nx}. In other words, virtual soft photons that run in the loops are those that carry energy less than or equal to $\Lambda$. Both scales $E_d$ and $\Lambda$ are taken to be sufficiently smaller than the mass $m_e$ of the electron. In the limit $\lambda \to 0$, we keep the ratios $E_d/\Lambda$, $E_d/m_e$ and $\Lambda/m_e$ fixed, with $E_d/m_e,\,\Lambda/m_e <<1$. 


The dressed state associated with a multi-particle/antiparticle state $\alpha=\{e_i,\vec{p}_i,s_i\}$ can be written in a product form \cite{Tomaras_2020}
\begin{equation}\label{eq.2.1.6}
    \ket{\alpha}_d=
        \ket{\alpha}
            \times
        \ket{f_{\alpha}}
\end{equation}
with
\begin{equation}\label{eq.2.1.7}
    \ket{f_{\alpha}}=
        \mathcal{N}_{\alpha}\
        e^{
            \int
            \reallywidetilde{d^3k}\
                f_{\alpha}(\Vec{k})
                    \cdot
                a^{\dagger}(\Vec{k})
        }
        \ket{0}
    ,\hspace{2em}
    \mathcal{N}_{\alpha}=
        e^{
            -\frac{1}{2}
            \int
            \reallywidetilde{d^3k}\
                f^*_{\alpha}(\Vec{k})
                    \cdot
                f_{\alpha}(\Vec{k})
        }
\end{equation}
Here, $\ket{\alpha}$ denotes the undressed multi-particle/antiparticle state, and 
$\ket{f_{\alpha}}$ is a coherent photon state describing the clouds surrounding the charged particles. The normalization constant $\mathcal{N}_{\alpha}$ ensures that 
$\bra{f_\alpha}f_{\alpha}\rangle =1$. 
Note that the single-particle/antiparticle dressing function $f^{\mu}(\highvec{p},\Vec{k})$ has been replaced by its multi-particle/antiparticle counterpart
\begin{equation}\label{eq.2.1.8}
    f_{\alpha}(\Vec{k})=
        \sum_{i\in\alpha}
        e_i
        \bigg(
            \frac{p_i^{\mu}}
                {p_i\cdot k}-
            c^{\mu}
        \bigg)
        e^{-ip\cdot k t_0/p_i^0}
\end{equation}
When $t_0=0$, the second term yields $Q_{\alpha}c^{\mu}$, where $Q_{\alpha}$ is the total charge of the state $\alpha$. As a result, the terms proportional to $c^{\mu}$ vanish for states with zero net charge. As we mentioned earlier, we choose to set $t_0=0$ for the computations throughout this work. 

The overlap between two coherent-photon states dressing an incoming charged state $\alpha=\{e_i,\Vec{p}_i,s_i\}$ and an outgoing charged state $\beta=\{e_i',\pvec{p}_i',s_i'\}$, respectively, with $Q_{\alpha}=Q_{\beta}$,
is given by \cite{Tomaras_2020}
\begin{equation}\label{eq.2.1.9}
    \braket{f_{\beta}|f_{\alpha}}=
        \bigg(
            \frac{\lambda}{E_d}
        \bigg)^
            {\mathcal{B}_{\beta\alpha}}
\end{equation}
where 
\begin{equation}\label{eq.2.1.10}
    \mathcal{B}_{\beta\alpha}=
        -\frac{1}{16\pi^2}
        \sum_{i,j}
        \eta_i\eta_j
        e_ie_j
        \upsilon_{ij}^{-1}
        \ln\bigg(
            \frac{1+\upsilon_{ij}}
                {1-\upsilon_{ij}}
        \bigg)
\end{equation}
is a positive kinematical factor, $\eta_i=+1(-1)$ for outgoing (incoming) charged particles, and
\begin{equation}\label{eq.2.1.11}
    \upsilon_{ij}=
        \bigg[
            1-\frac{m_i^2m_j^2}{(p_ip_j)^2}
        \bigg]^{1/2}
\end{equation}
is the magnitude of the relative velocity of the $j$-th particle with respect to the $i$-th. See also \cite{Weinberg:1965nx}. Since $\mathcal{B}_{\beta\alpha}>0$, the overlap vanishes in the limit $\lambda\rightarrow0$ (where $E_d$ is kept fixed or approches zero more slowly than $\lambda$). 

\subsection{Formulation of the Faddeev-Kulish S-matrix}
    Here, we analyze elastic and radiation-emitting FK amplitudes \cite{Tomaras_2020}. We present evidence that the elastic amplitudes are infrared-finite, order by order and to all orders in perturbation theory. The radiation-emitting amplitudes are free from the leading singularities that arise when the energies of the emitted real photons vanish.

\subsubsection{Elastic FK amplitudes}
    Consider a scattering process $\alpha\rightarrow\beta$, with no soft photons of energy less than $E_d$ in the initial and final states. In this section, we review the leading order results concerning the corresponding dressed amplitude presented in \cite{Tomaras_2020}. Drawing from insights outlined in \cite{Chung:1965zza}, we compute the S-matrix element between the incoming and outgoing dressed states:
\begin{equation}\label{eq.2.2.1.1}
    \tilde{S}_{\beta\alpha}=
        \tensor[_d]{\braket{\beta
            |S|
        \alpha}}{_d}
\end{equation}
The S-matrix element between the corresponding undressed states is denoted by
\begin{equation}\label{eq.2.2.1.2}
    S_{\beta\alpha}=
        \bra{\beta}S\ket{\alpha}
\end{equation}
Expanding the exponential operators of the coherent-photon states yields the following expression for the elastic FK amplitude
\begin{equation}\label{eq.2.2.1.3}
    \tilde{S}_{\beta\alpha}=
        \mathcal{N}_{\beta}
        \mathcal{N}_{\alpha}
        \sum_{n,m=0}^{\infty}
        \frac{1}{m!n!}\
        \bra{\beta}
        \prod_{l=1}^m
            \int
            \reallywidetilde{d^3q_l}\
            \big[
                f^*_{\beta}(\vec{q}_l)
                    \cdot
                a({\vec{q}_l})
            \big]\
        S\
        \prod_{s=1}^n
            \int
            \reallywidetilde{d^3k_s}\
            \big[
                f_{\alpha}(\vec{k}_s)
                    \cdot
                a^{\dagger}({\vec{k}_s})
            \big]
        \ket{\alpha}
\end{equation}
\eqref{eq.2.2.1.3} is given as a double sum of scattering amplitudes involving $n$ incoming soft photons and $m$ outgoing ones. These amplitudes are weighted by a factor $1/(m! n!)$. 

There is always a possibility that a certain number of soft photons from the initial and final FK states do not interact with the charged particles/antiparticles. Let the number of these soft photons be $l,\ 0\le l\le \rm{min}(m, n)$. Then, the contribution of these non-interacting soft photons to the FK amplitude is given by
\begin{equation}\label{eq.2.2.1.4}
    l!\ \bigg(
        \int\reallywidetilde{d^3q}\
        \big[
            f^*_{\beta}(\vec{q}
                \cdot
            f_{\alpha}(\vec{q})
        \big]
    \bigg)^l
\end{equation}
Next, the remaining $n'=n-l$ soft photons from the initial state must be absorbed by an external fermion line. Likewise, the remaining $m'=m-l$ soft photons from the final state need to be emitted by an external fermion line. Since these $n'$ incoming and $m'$ outgoing photons are soft -- $E_d$ is taken to be sufficiently small -- we are justified in approximating their contribution to the dressed amplitude using the leading soft photon theorem, which can be recast into the following convenient form
\begin{equation}\label{eq.2.2.1.5}
\begin{split}
    \lim_{\omega_k\rightarrow0}
        \bra{\beta}
            a_r(\vec{k})S
        \ket{\alpha}=
        \big[
            f_{\beta}(\vec{k})-
            f_{\alpha}(\vec{k})
        \big]
            \cdot
        \epsilon_r^*(\vec{k})\
        \bra{\beta}
            S
        \ket{\alpha}+...
        \hspace{0.25em}
    \\
    \lim_{\omega_k\rightarrow0}
        \bra{\beta}
            Sa^{\dagger}_r(\vec{k})
        \ket{\alpha}=-
        \big[
            f^*_{\beta}(\vec{k})-
            f^*_{\alpha}(\vec{k})
        \big]
            \cdot
        \epsilon_r(\vec{k})\
        \bra{\beta}
            S
        \ket{\alpha}+...
\end{split}
\end{equation}
where the charge conservation law $Q_{\alpha}=Q_{\beta}$ was used. The ellipses stand for smooth, non singular terms in the limits $\lambda,|\vec{k}_{\gamma}|\rightarrow0$, at least at tree-level\footnote{Subleading contributions to the soft photon theorem will be considered in \secref{sec:3}.}. By taking into account that there are $(n'+l)!/(n'! l!)$ ways to choose $l$ photons from the initial state of $n$ soft photons, and that there are $(m'+l)!/(m'! l!)$ ways to choose $l$ photons from the final state of $m$ soft photons, we deduce that the interacting soft photons contribute the following factors\footnote{
    Additional logarithmic divergences in the soft theorem may arise at the loop order in perturbation theory and at subleading order in the soft momentum expansion. The contributions of these divergent terms to the factors in \eqref{eq.2.2.1.6} are integrable and suppressed (since they are of order $E_d$). Hence, we disregard them in our leading-order analysis.
}
\begin{equation}\label{eq.2.2.1.6}
\begin{split}
    \frac{(m'+l)!}{m'!l!}
        \bigg(
        \int\reallywidetilde{d^3q}\
        f^*_{\beta}(\vec{q})
            \cdot
        \big[
            f_{\beta}(\vec{q})-
            f_{\alpha}(\vec{q})
        \big]
        \bigg)^{m'}
        \hspace{0.25em}
    \\
    \frac{(n'+l)!}{n'!l!}
        \bigg(-
        \int\reallywidetilde{d^3q}\
        \big[
            f^*_{\beta}(\vec{q})-
            f^*_{\alpha}(\vec{q})
        \big]
            \cdot
        f_{\alpha}(\vec{q})
        \bigg)^{n'}
    \hspace{-0.125em}
\end{split}
\end{equation}
to the dressed amplitude. The double sum over the $(m,n)$ configurations in \eqref{eq.2.2.1.3} can be rearranged as a sum over all possible configurations of $(l,m', n')$. That is,
\begin{equation}\label{eq.2.2.1.7}
\begin{split}
    \tilde{S}_{\beta\alpha}=
        \mathcal{N}_{\alpha}
        \mathcal{N}_{\beta}
        \sum_{l,m',n'=0}^{\infty}
        \frac{1}{(m'+l)!(n'+l)!}&
        \frac{(m'+l)!}{m'!l!}
        \frac{(n'+l)!}{n'!l!}\
        l!
            \bigg(
            \int\reallywidetilde{d^3q}\
            \big[
                f^*_{\beta}(\vec{q})
                    \cdot
                f_{\alpha}(\vec{q})
            \big]
            \bigg)^l
        \\
        \times\bigg(
            &\int\reallywidetilde{d^3q}\
            f^*_{\beta}(\vec{q})
                \cdot
            \big[
                f_{\beta}(\vec{q})-
                f_{\alpha}(\vec{q})
            \big]
            \bigg)^{m'}
        \\
        \times\bigg(-
            &\int\reallywidetilde{d^3q}\
            \big[
                f^*_{\beta}(\vec{q})-
                f^*_{\alpha}(\vec{q})
            \big]
                \cdot
            f_{\alpha}(\vec{q})
            \bigg)^{n'}
        S_{\beta\alpha}
\end{split}
\end{equation}
Certain combinatorial factors in \eqref{eq.2.2.1.7} cancel. As a result, the dressed amplitude factorizes into a product of three single-sums, with each sum exponentiating to yield \cite{Tomaras_2020}
\begin{equation}\label{eq.2.2.1.8}
    \tilde{S}_{\beta\alpha}=
        \braket{f_{\beta}|f_{\alpha}}\
        e^{
            \int
            \reallywidetilde{d^3k}\
            \big[f^*_{\beta}(\vec{q})-f^*_{\alpha}(\vec{q})\big]
                \cdot
            \big[f_{\beta}(\vec{q})-f_{\alpha}(\vec{q})\big]
        }\
        S_{\beta\alpha}
\end{equation}
where we have used the fact that the dressing functions are real for $t_0=0$. The exponent of the exponential factor in \eqref{eq.2.2.1.8} is given by
\begin{equation}\label{eq.2.2.1.9}
    \int
        \reallywidetilde{d^3q}\
        \big[f^*_{\beta}(\vec{q})-f^*_{\alpha}(\vec{q})\big]
            \cdot
        \big[f_{\beta}(\vec{q})-f_{\alpha}(\vec{q})\big]=
    \sum_{ij}
        \eta_i\eta_je_ie_j
        \int\reallywidetilde{d^3q}\
        \frac{p_i\cdot p_j}                        
            {(p_i\cdot q)(p_j\cdot q)}=
        \ln\bigg(
            \frac{E_d}{\lambda}
        \bigg)^
            {2\mathcal{B}_{\beta\alpha}}
\end{equation}
Overall, the FK elastic amplitude reduces to \cite{Tomaras_2020}
\begin{equation}\label{eq.2.2.1.10}
    \tilde{S}_{\beta\alpha}=
        \braket{f_{\beta}|f_{\alpha}}\
        \bigg(
            \frac{E_d}{\lambda}
        \bigg)^
            {2\mathcal{B}_{\beta\alpha}}\
        S_{\beta\alpha}=
        \bigg(
            \frac{E_d}{\lambda}
        \bigg)^
            {\mathcal{B}_{\beta\alpha}}\
        S_{\beta\alpha}
\end{equation}
Now, as shown in \cite{Weinberg:1965nx}, the exponentiation of virtual infrared divergences gives
\begin{equation}\label{eq.2.2.1.11}
    S_{\beta\alpha}=
        \bigg(
            \frac{\lambda}{\Lambda}
        \bigg)^{\mathcal{B}_{\beta\alpha}}
        e^{i\phi_{\beta\alpha}}
        S_{\beta\alpha}^{(\Lambda)}
\end{equation}
with $S_{\beta\alpha}^{(\Lambda)}$ being the Fock basis scattering amplitude without the contributions from the virtual soft photons. The phase $\phi_{\beta\alpha}$, being real \cite{Weinberg:1965nx}, does not contribute to the square of the absolute value of the amplitudes or the corresponding rates. Finally,
\begin{equation}\label{eq.2.2.1.12}
    \tilde{S}_{\beta\alpha}=
        \bigg(
            \frac{E_d}{\Lambda}
        \bigg)^{\mathcal{B}_{\beta\alpha}}
        e^{i\phi_{\beta\alpha}}
        S_{\beta\alpha}^{(\Lambda)}
\end{equation}
is non-zero and free of infrared divergences in the limit $\lambda\rightarrow0$ with the ratio $\Lambda/E_d$
kept fixed \cite{Tomaras_2020}. It is always possible to choose $\Lambda =E_d$, in which case the absolute value of the dressed amplitude is equal to the infrared-finite part of the corresponding undressed Fock-basis amplitude.

\subsubsection{Soft radiation and FK amplitudes} 
    The FK amplitude describing soft photon emission during the $\alpha\rightarrow\beta$ scattering process can be obtained by adding a single soft photon $\gamma$ of momentum $\vec{q}_{\gamma}$, such that $ |\vec{q}_{\gamma}|<E_d$, and polarization vector $\epsilon_{r\mu}(\vec{q}_{\gamma})$ in the final state:
\begin{equation}\label{eq.2.2.2.1}
    \tilde{S}_{\beta\gamma,\alpha}=
        \tensor[_d]{\braket{\beta\gamma
            |S|
        \alpha}}{_d}
\end{equation}
The case where $|\vec{q}_{\gamma}|>E_d$ does not require further analysis, since the extra photon can be considered to be hard and does not lead to any infrared singularities. 

To calculate this FK amplitude for soft photon emission, we first express the final dressed state in a product form
\begin{equation}\label{eq.2.2.2.2}
    \ket{\beta\gamma}_d=
        \Big(
            \ket{\beta\gamma}-
            f^{*\mu}_{\beta}(\vec{q}_{\gamma})
            \epsilon_{r\mu}(\vec{q}_{\gamma})
            \ket{\beta}
        \Big)\times
        \ket{f_{\beta}}
\end{equation}
acting with the FK operator $e^{R_f}$ on the Fock state $\ket{\beta\gamma}$. It is straightforward to verify that for $\alpha \neq \beta$, the trivial part of the S-matrix vanishes:
\begin{equation}\label{eq.2.2.2.3}
    \tensor[_d]{\braket{\beta\gamma|\alpha}}{_d}=
        \big[
            f_{\alpha}(\vec{q}_{\gamma})-
            f_{\beta}(\vec{q}_{\gamma})
        \big]
                \cdot
            \epsilon^*_r(\vec{q}_{\gamma})
            \braket{f_{\beta}|f_{\alpha}}
            \braket{\beta|\alpha}=
            \big[
                f_{\alpha}(\vec{q}_{\gamma})-
                f_{\alpha}(\vec{q}_{\gamma})
            \big]
                \cdot
            \epsilon^*_r(\vec{q}_{\gamma})
            =0
\end{equation}
since $\braket{\beta|\alpha}=\delta_{\beta\alpha}$. Hence, only the non-trivial part of the S-matrix contributes to this dressed amplitude.

According to \eqref{eq.2.2.2.2}, we split $\tilde{S}_{\beta\gamma,\alpha}$ into two contributions
\begin{equation}\label{eq.2.2.2.4}
    \tilde{S}_{\beta\gamma,\alpha}=
        \tilde{S}_{\beta\gamma,\alpha}
            ^{(1)}+
        \tilde{S}_{\beta\gamma,\alpha}
            ^{(2)}
\end{equation}
where 
\begin{equation}\label{eq.2.2.2.5}
    \tilde{S}_{\beta\gamma,\alpha}
        ^{(1)}=
        -f_{\beta}(\vec{q}_{\gamma})
            \cdot
        \epsilon_r^*(\vec{q}_{\gamma})\
        \tilde{S}_{\beta\alpha}
\end{equation}
and
\begin{equation}\label{eq.2.2.2.6}
    \tilde{S}_{\beta\gamma,\alpha}
        ^{(2)}\!=
        \mathcal{N}_{\beta}\
        \mathcal{N}_{\alpha}
        \!\sum_{m,n=0}^{\infty}
        \frac{1}{m!n!}
        \bra{\beta}
        a_r(\vec{q}_{\gamma})
        \prod_{l=1}^{m}
        \int
        \reallywidetilde{d^3q}_l
            \big[
                f^*_{\beta}(\vec{q}_l)
                    \cdot
                a(\vec{q}_l)
            \big]
        S
        \prod_{s=1}^{n}
        \int
        \reallywidetilde{d^3k}_s
            \big[
                f^*_{\beta}(\vec{k}_s)
                    \cdot
                a^{\dagger}(\vec{k}_s)
            \big]
        \ket{\alpha}
\end{equation}
The second term can be further split into two parts, as the additional soft photon in the final state, created by $a^{\dagger}_r(\vec{q}_{\gamma})$, can be interacting or non-interacting. In the first case, we apply the leading soft photon theorem to get
\begin{equation}\label{eq.2.2.2.7}
    \tilde{S}_1^{(2)}=
        \bigg(
            \frac{E_d}{\lambda}
        \bigg)^{\mathcal{B}_{\beta\alpha}}\
        S_{\beta\alpha}\
        \big[
            f_{\beta}(\vec{q}_{\gamma})-
            f_{\alpha}(\vec{q}_{\gamma})
        \big]
            \cdot
        \epsilon^*_r(\vec{q}_{\gamma})
        +...
\end{equation}
In the second case, we employ combinatorial arguments: when the additional photon is non-interacting, we get overall $l+1$ non-interacting photons, yielding a net contribution
\begin{equation}\label{eq.2.2.2.8}
    (l+1)\ l!\
    f_{\alpha}(\vec{q}_{\gamma})
        \cdot
    \epsilon^*_r(\vec{q}_{\gamma})\
    \bigg(
        \int
            \reallywidetilde{d^3q}
            f_{\alpha}(\vec{q})
                \cdot
            f^*_{\beta}(\vec{q})
    \bigg)^l
\end{equation}
to the FK amplitude. The remaining $n'=n-l-1$ and $m'=m-l$ soft photons are absorbed and emitted by external fermion lines, respectively. 
Therefore, we obtain
\begin{equation}\label{eq.2.2.2.9}
\begin{split}
    \tilde{S}_2^{(2)}=
        \sum_{l,m',n'=0}^{\infty}\!
        \frac{1}{(m'+l)!(n'+l+1)!}&
        \frac{(m'+l)!}{m'!l!}
        \frac{(n'+l+1)!}{n'!(l+1)!}
        \\&\times
        (l+1)\ l!\
        f_{\alpha}(\vec{q}_{\gamma})
            \cdot
        \epsilon^*_r(\vec{q}_{\gamma})
        \bigg(
            \int
                \reallywidetilde{d^3q}\,
                f_{\alpha}(\vec{q})
                    \cdot
                f^*_{\beta}(\vec{q})
        \bigg)^l
        \\&
        \times\bigg(
            \int\reallywidetilde{d^3q}\
            f^*_{\beta}(\vec{q})
                \cdot
            \big[
                f_{\beta}(\vec{q})-
                f_{\alpha}(\vec{q})
            \big]
            \bigg)^{m'}
        \\&
        \times\bigg(-
            \int\reallywidetilde{d^3q}\
            \big[
                f^*_{\beta}(\vec{q})-
                f^*_{\alpha}(\vec{q})
            \big]
                \cdot
            f_{\alpha}(\vec{q})
            \bigg)^{n'}
        S_{\beta\alpha}
\end{split}
\end{equation}
where $(n'+l+1)!/[n'!(l+1)!]$ is the number of ways in which $l+1$ soft photons can be chosen among the $n$ incoming photons, and $(m'+l)!/(m'! l!)$ is the number of ways in which $l$ soft photons can be chosen among the $m$ outgoing photons. Simplifying the combinatorial factors and exponentiating as before, yields
\begin{equation}\label{eq.2.2.2.10}
    \tilde{S}_2^{(2)}=
        \bigg(
            \frac{E_d}{\lambda}
        \bigg)^
            {\mathcal{B}_{\beta\alpha}}\
        S_{\beta\alpha}\
        \big[ 
            f_{\alpha}(\vec{q}_{\gamma})
                \cdot
            \epsilon^*_r(\vec{q}_{\gamma})
        \big]
\end{equation}
Therefore, $\tilde{S}_{\beta\gamma,\alpha}^{(2)}$ is given by
\begin{equation}\label{eq.2.2.2.11}
    \tilde{S}_{\beta\gamma,\alpha}^{(2)}=
        \tilde{S}_2^{(1)}+
        \tilde{S}_2^{(2)}=
        \bigg(
            \frac{E_d}{\lambda}
        \bigg)^
            {\mathcal{B}_{\beta\alpha}}\
        S_{\beta\alpha}\
        \big[ 
            f_{\beta}(\vec{q}_{\gamma})
                \cdot
            \epsilon^*_r(\vec{q}_{\gamma})
        \big]+...
\end{equation}
where the ellipses stand for subleading contributions in the soft momentum expansion $\vec{q}_{\gamma}$.
Adding all contributions, we obtain for the dressed amplitude for single photon emission
\begin{equation}\label{eq.2.2.2.12}
    \tilde{S}_{\beta\gamma,\alpha}=
        F_{\beta\alpha}
            \big(
                \vec{q}_{\gamma},
                \epsilon_r(\vec{q}_{\gamma})
            \big)
\end{equation}
where $F_{\beta\alpha}\big( \vec{q}_{\gamma}, \epsilon_r(\vec{q}_{\gamma} )\big)$ is free of any singular poles as $|\Vec{q}_{\gamma}|\rightarrow0$ \cite{Tomaras_2020}. In fact, this function is smooth as $\lambda,|\vec{q}_{\gamma}|\rightarrow0$, at least to leading order in perturbation theory (tree level). (See \secref{sec:conc}
for discussions concerning higher orders.) 

In \cite{Choi:2019rlz}, it has been argued that by appropriately modifying the dressing function to subleading order in the soft photon momentum (and to leading order in the electron charge), $F_{\beta\alpha}\big( \vec{q}_{\gamma}, \epsilon_r(\vec{q}_{\gamma} )\big)$ becomes of order $E_d$, and therefore it is completely negligible. The subleading corrections to the dressing function are expected to suppress the emission of soft photons with energy $\omega_{\gamma}<E_d$ at the tree-level. In \secref{sec:3}, we will apply the arguments in \cite{Choi:2019rlz} to explicit examples of scattering processes in QED, in order to verify that the dressed amplitudes for single soft photon emission are completely suppressed. It would be interesting to incorporate higher order corrections to the dressing function that suppress soft radiation at higher orders in perturbation theory. It is well known that radiation-emitting amplitudes admit higher order corrections, which are logarithmically divergent as the photon momentum vanishes \cite{Bern:2014oka,He:2014bga,Mao:2017wvx,Sahoo:2018lxl,Laddha:2018myi,Saha:2019tub,Sahoo:2019yod,Sahoo:2020ryf,Delisle:2020uui,Krishna:2023fxg,Bianchi:2014gla}. To cancel these logarithmic divergences occurring at the loop-level in perturbation theory and to keep the amplitude suppressed, we must further correct the dressing functions.

\section{Dressed states beyond leading soft order}
    \label{sec:3}
    In this section, we proceed to study soft photon emission during scattering of charged dressed states in QED. We consider three typical cases, the scattering of an electron by a heavier muon, electron-photon scattering and electron-positron annihilation. We will work at tree level. In the soft momentum limit, we expand the corresponding tree-level undressed invariant amplitudes in powers of the soft photon energy $\omega_k=|\vec{k}|$. As dictated by Low's theorem \cite{Bloch:1937pw,Gell-Mann:1954wra,Low:1954kd,Low:1958sn,Weinberg:1965nx,Burnett:1967km,DelDuca:1990gz,Luo:2014wea,Bern:2014vva,Lysov:2014csa,Strominger:2017zoo,AtulBhatkar:2018kfi,Beneke:2021ilf,Beneke:2021umj,Travaglini:2022uwo}, the expansion takes the form
\begin{equation}\label{eq.3.0.1}
    i\mathcal{M}_{\text{tree}}
        (\omega_k,\hat{k})=
        \omega_k^{-1}\ i\mathcal{M}_L(\hat{k})+
        i\mathcal{M}_{SL}(\hat{k})+
        \mathcal{O}(\omega_k)
\end{equation}
where 
the unit vector $\hat{k}$ specifies the direction of the soft photon momentum. 
From \eqref{eq.3.0.1} we see that the leading term in the expansion diverges as $\omega_k^{-1}$, while the first subleading term 
is constant as $\omega_k\rightarrow0$. The remaining terms in the expansion vanish as $\omega_k\rightarrow0$, and therefore, can be neglected in the following calculations.

Following \cite{Choi:2019rlz}, we construct dressed states by correcting the dressing function to subleading order in the soft momentum expansion. We demonstrate explicitly that the dressed elastic amplitudes for electron-muon, electron-photon, and electron-positron annihilation are free of infrared divergences.
Finally, we demonstrate that the emission of an additional soft photon of energy less than $E_d$, the characteristic energy of photons in the  clouds accompanying the charged particles, is completely suppressed, of order $E_d$, when calculating dressed amplitudes at tree-level. This suppression could not be achieved having kept only the leading singular term in the dressing function as proposed in \cite{Dollard:1964jmp,Chung:1965zza,Kibble:1968lka,Kibble:1968npb,Kibble:1968oug,Kibble:1968sfb,Kulish:1970ut}. The subleading corrections to the dressing function are thus important.

\subsection{Application of subleading soft theorem to QED scattering}
    The undressed elastic invariant amplitudes at tree level
are given by the following expressions:
\begin{itemize}
    \item for $e^-(p_1)+\mu^-(p_2)\rightarrow e^-(q_1)+\mu^-(q_2)$:
    \begin{equation}\label{eq.3.1.1(a)}
        i\mathcal{M}_0^{(\mu)}=
            \bar{u}(\vec{q}_1)
                \big[
                    ie\gamma^{\mu}
                \big]
            u(\vec{p}_1)
            \frac{(-i)g_{\mu\nu}}
                {[(p_1-q_1)^2-i\epsilon]}
            \bar{u}(\vec{q}_2)
                \big[
                    ie\gamma^{\nu}
                \big]
            u(\vec{p}_2)
    \end{equation}
    \item for $e^-(p_1)+\gamma(p_2)\rightarrow\gamma(q_1)+ e^-(q_2)$:
    \begin{equation}\label{eq.3.1.1(b)}
    \begin{split}
        i\mathcal{M}_0^{(\gamma)}=
            \bar{u}(\vec{q}_2)
            \bigg\{&
                \big[
                    ie\slashed{\epsilon}_{\lambda_2}(\Vec{p}_2)
                \big]
                \frac{(-i)(-\slashed{p}_1+\slashed{q}_1+m)}
                    {[(p_1-q_1)^2+m^2-i\epsilon]}
                \big[
                    ie\slashed{\epsilon}_{r_1}^*(\Vec{q}_1)
                \big]
                \\&+
                \big[
                    ie\slashed{\epsilon}_{r_1}^*(\Vec{q}_1)
                \big]
                \frac{(-i)(-\slashed{p}_1-\slashed{p}_2+m)}
                    {[(p_1+p_2)^2+m^2-i\epsilon]}
                \big[
                    ie\slashed{\epsilon}_{\lambda_2}(\Vec{p}_2)
                \big]
            \bigg\}
            u(\vec{p}_1)
    \end{split}
    \end{equation}
    \item for $e^-(p_1)+e^+(p_2)\rightarrow\gamma(q_1)+\gamma(q_2)$:
    \begin{equation}\label{eq.3.1.1(c)}
    \begin{split}
        i\mathcal{M}_0^{(e)}=
            \bar{\upsilon}(\vec{p}_2)
            \bigg\{&
                \big[
                    ie\slashed{\epsilon}_{r_2}^*(\Vec{q}_2)
                \big]
                \frac{(-i)(-\slashed{p}_1+\slashed{q}_1+m)}
                    {[(p_1-q_1)^2+m^2-i\epsilon]}
                \big[
                    ie\slashed{\epsilon}_{r_1}^*(\Vec{q}_1)
                \big]
            \\&+
                \big[
                    ie\slashed{\epsilon}_{r_1}^*(\Vec{q}_1)
                \big]
                \frac{(-i)(-\slashed{p}_1+\slashed{q}_2+m)}
                    {[(p_1-q_2)^2+m^2-i\epsilon]}
                \big[
                    ie\slashed{\epsilon}_{r_2}^*(\Vec{q}_2)
                \big]
            \bigg\}
            u(\vec{p}_1)
    \end{split}
    \end{equation}
\end{itemize}
We have labeled the three amplitudes 
using the superscripts $\mu$, $\gamma$, and $e$, respectively. 
$\vec{p}_1$ and $\vec{p}_2$ denote the momenta of the incoming particles, while $\vec{q}_1$ and $\vec{q}_2$ the momenta of the outgoing particles. In particular, $\vec{p}_1$ denotes the momentum of the incoming electron in all three processes.
$\vec{p}_2$ denotes the momentum of the incoming muon for the first process, of the photon (with polarization index $\lambda_2$) for the second, and of the positron for the last one. For the electron-muon scattering process, $\vec{q}_1$ denotes the momentum of the outgoing electron and $\vec{q}_2$ of the outgoing muon. For the electron-photon scattering process, $\vec{q}_1$ denotes the momentum of the outgoing photon (of polarization index $r_1$) and $\vec{q}_2$ the momentum of the outgoing electron. Finally, for electron-positron annihilation, $\vec{q}_1$ and $\vec{q}_2$ denote the momenta of the two outgoing photons (with polarization indices $r_1$ and $r_2$). All external particles are on-shell with the energy given in terms of the momentum and mass by the usual relativistic formula. The charge of the electron is denoted by $e$.

The spinor wavefunctions are denoted by the standard notation \cite{Peskin:1995ev}, with $u(\vec{p})$ describing an incoming fermionic particle, and $\bar{u}(\vec{p})$ an outgoing one. $\bar{\upsilon}(\vec{p})$ and $\upsilon(\vec{p})$ describe the incoming and outgoing antiparticles, respectively. We have adopted Feynman's slash notation, $\slashed{p} = \gamma^\mu p_\mu$, where $p_\mu$ is the four-momentum carried by the virtual fermionic particle, and $\gamma^\mu$ are the usual Dirac matrices. These satisfy the anticommutation relation $\{ \gamma^\mu, \gamma^\nu \}=-2 g^{\mu\nu}$, with $g^{\mu\nu}$ being the Minkowski metric. For simplicity, spin polarization indices 
have been suppressed. $\epsilon^*_{\lambda}(\vec{k})$ denotes the polarization vector of an outgoing photon with momentum $\vec{k}$ and helicity $\lambda$. The polarization vectors satisfy $\epsilon_{\lambda}^*(\vec{k})\cdot k = 0$, ensuring that the photon is transverse. 
The leading tree-level Feynman diagrams describing the three elastic processes are shown in \figref{fig.2.1.1}.

Using energy and momentum conservation, we can  express the above scattering amplitudes more symmetrically in terms of the momenta of the external particles, as follows:
\begin{itemize}
    \item for $e^-(p_1)+\mu^-(p_2)\rightarrow e^-(q_1)+\mu^-(q_2)$:
    \begin{equation}\label{eq.3.1.2(a)}
    \begin{split}
        i\mathcal{M}_0^{(\mu)}=
            \bar{u}(\vec{q}_1)
            \big[
                ie\gamma^{\mu}
            \big]
            u(\vec{p}_1)
            \frac{(-i)g_{\mu\nu}}
                {[\frac{1}{2}(p_1-q_1)^2
                    +\frac{1}{2}(q_2-p_2)^2
                    -i\epsilon]}
            \bar{u}(\vec{q}_2)
            \big[
                ie\gamma^{\nu}
            \big]
            u(\vec{p}_2)
    \end{split}
    \end{equation}
    \item for $e^-(p_1)+\gamma(p_2)\rightarrow\gamma(q_1)+ e^-(q_2)$:
    \begin{equation}\label{eq.3.1.2(b)}
    \begin{split}
        i\mathcal{M}_0^{(\gamma)}=
            \bar{u}(\vec{q}_2)
        \bigg\{&
            \big[
                ie\slashed{\epsilon}_{\lambda_2}(\Vec{p}_2)
            \big]
            \frac{(-i)(
                    -\frac{1}{2}\slashed{p}_1
                    +\frac{1}{2}\slashed{q}_1
                    -\frac{1}{2}\slashed{q}_2
                    +\frac{1}{2}\slashed{p}_2
                    +m)}
                {[\frac{1}{2}(p_1-q_1)^2
                    +\frac{1}{2}(q_2-p_2)^2
                    +m^2-i\epsilon]}
            \big[
                ie\slashed{\epsilon}_{r_1}^*(\Vec{q}_1)
            \big]
            \\&\hspace{-0.75em}+
            \big[
                ie\slashed{\epsilon}_{r_1}^*(\Vec{q}_1)
            \big]
            \frac{(-i)(
                    -\frac{1}{2}\slashed{p}_1
                    -\frac{1}{2}\slashed{p}_2
                    -\frac{1}{2}\slashed{q}_1
                    -\frac{1}{2}\slashed{q}_2
                    +m)}
                {[\frac{1}{2}(p_1+p_2)^2
                    +\frac{1}{2}(q_1+q_2)^2
                    +m^2-i\epsilon]}
            \big[
                ie\slashed{\epsilon}_{\lambda_2}(\Vec{p}_2)
            \big]            
        \bigg\}
        u(\vec{p}_1)
    \end{split}
    \end{equation}
    \item for $e^-(p_1)+e^+(p_2)\rightarrow\gamma(q_1)+\gamma(q_2)$:
    \begin{equation}\label{eq.3.1.2(c)}
    \begin{split}
        i\mathcal{M}_0^{(e)}&=
            \bar{\upsilon}(\vec{p}_2)
            \bigg\{
            \big[
                ie\slashed{\epsilon}_{r_2}^*(\Vec{q}_2)
            \big]
            \frac{(-i)(-\frac{1}{2}\slashed{p}_1+
                    \frac{1}{2}\slashed{q}_1+
                    \frac{1}{2}\slashed{p}_2-
                    \frac{1}{2}\slashed{q}_2+m)}
                {[\frac{1}{2}(p_1-q_1)^2
                    +\frac{1}{2}(q_2-p_2)^2+m^2-i\epsilon]}
            \big[
                ie\slashed{\epsilon}_{r_1}^*(\Vec{q}_1)
            \big]
            \\&\hspace{4.25em}+
            \big[
                ie\slashed{\epsilon}_{r_2}^*(\Vec{q}_1)
            \big]
            \frac{(-i)(-\frac{1}{2}\slashed{p}_1+
                    \frac{1}{2}\slashed{q}_1+
                    \frac{1}{2}\slashed{p}_2-
                    \frac{1}{2}\slashed{q}_2+m)}
                {[\frac{1}{2}(p_1-q_1)^2
                    +\frac{1}{2}(q_2-p_2)^2+m^2-i\epsilon]}
            \big[
                ie\slashed{\epsilon}_{r_1}^*(\Vec{q}_2)
            \big]
            \bigg\}
        u(\vec{p}_1)
    \end{split}
    \end{equation}
\end{itemize}

\begin{figure}[t]
    \centering
    \begin{subfigure}[t]{0.31\linewidth}
        \centering
        \begin{tikzpicture}[line width=1pt, 
  fermion/.style={thick},  
  boson/.style={decorate, decoration={snake}, draw=black},
  halffermion/.style={thick, ->}, scale=0.75]]

  \draw[fermion] (0.5,2) -- (2,1.25);    
  \draw[fermion] (0.5,-2) -- (2,-1.25);  
  \draw[fermion] (2,1.25) -- (3.5,2);    
  \draw[fermion] (2,-1.25) -- (3.5,-2);  

  \draw[boson] (2,1.25) -- (2,-1.25) node[midway, right] {$\gamma$};

  \filldraw[black] (2,1.25) circle (2pt);
  \filldraw[black] (2,-1.25) circle (2pt);

  \draw[halffermion] (0.5,2) -- (1.25,1.625);    
  \draw[halffermion] (0.5,-2) -- (1.25,-1.625);  
  \draw[halffermion] (2,1.25) -- (2.75,1.625);    
  \draw[halffermion] (2,-1.25) -- (2.75,-1.625);  

  \node at (-0.1,2.2) {$e^-,\ p_1\ $};
  \node at (-0.1,-2.2) {$\mu^-,\ p_2$};
  \node at (4.1,2.2) {$e^-,\ q_1$};
  \node at (4.1,-2.2) {$\mu^-,\ q_2$};

\end{tikzpicture}
        \caption{}
    \end{subfigure}
    \begin{subfigure}[t]{0.31\linewidth}
        \centering
        \begin{tikzpicture}[line width=1pt, 
  fermion/.style={thick},  
  boson/.style={decorate, decoration={snake}, draw=black},
  halffermion/.style={thick, ->}, scale=0.75]]

  \draw[fermion] (0.5,2) -- (2,1.25);        
  \draw[boson] (0.5,-2) -- (2,-1.25);    
  \draw[boson] (2,1.25) -- (3.5,2);       
  \draw[fermion] (2,-1.25) -- (3.5,-2);     

  \draw[fermion] (2,1.25) -- (2,-1.25) node[midway, right] {$e^-$};
  \draw[halffermion] (2,1.25) -- (2,0);

  \filldraw[black] (2,1.25) circle (2pt);
  \filldraw[black] (2,-1.25) circle (2pt);

  \draw[halffermion] (0.5,2) -- (1.25,1.625);    
  \draw[halffermion] (2,-1.25) -- (2.75,-1.625);     

  \node at (-0.1,2.2) {$e^-,\ p_1\ $};
  \node at (-0.1,-2.2) {$\gamma,\ p_2$};
  \node at (4.1,2.2) {$\gamma,\ q_1$};
  \node at (4.1,-2.2) {$e^-,\ q_2$};

\end{tikzpicture}
        \caption{}
    \end{subfigure}
    \begin{subfigure}[t]{0.31\linewidth}
        \centering
        \begin{tikzpicture}[line width=1pt, 
  fermion/.style={thick},  
  boson/.style={decorate, decoration={snake}, draw=black},
  halffermion/.style={thick, ->}, scale=0.75]]

  \draw[fermion] (0.5,2) -- (2,1.25);        
  \draw[boson] (0.5,-2) -- (2,1.25);    
  \draw[boson] (2,-1.25) -- (3.5,2); 
  \draw[fermion] (2,-1.25) -- (3.5,-2);     

  \draw[fermion] (2,1.25) -- (2,-1.25); 
  \node at (2.325,0.45) {$e^-$};
  \draw[halffermion] (2,1.25) -- (2,0);

  \filldraw[black] (2,1.25) circle (2pt);
  \filldraw[black] (2,-1.25) circle (2pt);

  \draw[halffermion] (0.5,2) -- (1.25,1.625); 
  \draw[halffermion] (2,-1.25) -- (2.75,-1.625); 

  \node at (-0.1,2.2) {$e^-,\ p_1\ $};
  \node at (-0.1,-2.2) {$\gamma,\ p_2$};
  \node at (4.1,2.2) {$\gamma,\ q_1$};
  \node at (4.1,-2.2) {$e^-,\ q_2$};

\end{tikzpicture}
        \caption{}
    \end{subfigure}
    \\ \vspace{1em}
    \begin{subfigure}[t]{0.31\linewidth}
        \centering
        \begin{tikzpicture}[line width=1pt, 
  fermion/.style={thick},  
  boson/.style={decorate, decoration={snake}, draw=black},
  halffermion/.style={thick, ->}, scale=0.75]]

  \draw[fermion] (0.5,2) -- (2,1.25);        
  \draw[fermion] (0.5,-2) -- (2,-1.25);    
  \draw[boson] (2,1.25) -- (3.5,2);       
  \draw[boson] (2,-1.25) -- (3.5,-2);     

  \draw[fermion] (2,1.25) -- (2,-1.25) node[midway, right] {$e^-$};
  \draw[halffermion] (2,1.25) -- (2,0);

  \filldraw[black] (2,1.25) circle (2pt);
  \filldraw[black] (2,-1.25) circle (2pt);

  \draw[halffermion] (0.5,2) -- (1.25,1.625);    
  \draw[halffermion] (2,-1.25) -- (1.25,-1.625);     

  \node at (-0.1,2.2) {$e^-,\ p_1\ $};
  \node at (-0.1,-2.2) {$e^+,\ p_2$};
  \node at (4.1,2.2) {$\gamma,\ q_1$};
  \node at (4.1,-2.2) {$\gamma,\ q_2$};

\end{tikzpicture}
        \caption{}
    \end{subfigure}
    \begin{subfigure}[t]{0.31\linewidth}
        \centering
        \begin{tikzpicture}[line width=1pt, 
  fermion/.style={thick},  
  boson/.style={decorate, decoration={snake}, draw=black},
  halffermion/.style={thick, ->}, scale=0.75]

  \draw[fermion] (0.5,2) -- (2,1.25);        
  \draw[fermion] (0.5,-2) -- (2,-1.25);      
  \draw[boson] (2,-1.25) -- (3.5,2);         
  \draw[boson] (2,1.25) -- (2.49,0.23);      
  \draw[boson] (2.65,-0.2) -- (3.5,-2);      

  \draw[fermion] (2,1.25) -- (2,-1.25) node[midway, left] {$e^-$};
  \draw[halffermion] (2,1.25) -- (2,0);

  \filldraw[black] (2,1.25) circle (2pt);
  \filldraw[black] (2,-1.25) circle (2pt);

  \draw[halffermion] (0.5,2) -- (1.25,1.625);    
  \draw[halffermion] (2,-1.25) -- (1.25,-1.625); 

  \node at (-0.1,2.2) {$e^-,\ p_1\ $};
  \node at (-0.1,-2.2) {$e^+,\ p_2$};
  \node at (4.1,2.2) {$\gamma,\ q_2$};      
  \node at (4.1,-2.2) {$\gamma,\ q_1$};

\end{tikzpicture}
        \caption{}
    \end{subfigure}
    \caption{Tree-level Feynman diagrams for the elastic processes (a) \texorpdfstring{$e^- \mu \rightarrow e^- \mu$}{TEXT}, (b) \& (c) \texorpdfstring{$e^- \gamma \rightarrow e^- \gamma$}{TEXT}, and (d) \& (e) \texorpdfstring{$e^- e^+ \rightarrow \gamma \gamma$}{TEXT}}
    \label{fig.2.1.1}
\end{figure}

To describe single soft-photon emission during the processes described above, we include one additional photon in the final state. This photon
carries momentum $\vec{k}$ and polarization $\epsilon_r^*(\vec{k})$.
For physical on-shell transverse photons, $r = 1,\ 2$.
The undressed photon emission invariant amplitudes for each process under study can be obtained as follows:
\begin{itemize}
    \item for $e^-(p_1)+\mu^-(p_2)\rightarrow e^-(q_1)+\mu^-(q_2)$:
        \begin{equation}\label{eq.3.1.3(a)}
            i\mathcal{M}_{\text{tree}}^{(\mu)}
                (\omega_k,\hat{k})=
                i\mathcal{M}_a^{(\mu)}
                    (\omega_k,\hat{k})+
                i\mathcal{M}_b^{(\mu)}
                    (\omega_k,\hat{k})+
                i\mathcal{M}_c^{(\mu)}
                    (\omega_k,\hat{k})+
                i\mathcal{M}_d^{(\mu)}
                    (\omega_k,\hat{k})
        \end{equation}
        where $ i\mathcal{M}_a^{(\mu)}(\omega_k,\hat{k}),\ i\mathcal{M}_b^{(\mu)}(\omega_k,\hat{k}),\ i\mathcal{M}_c^{(\mu)}(\omega_k,\hat{k})$, and $i\mathcal{M}_d^{(\mu)}(\omega_k,\hat{k})$ denote the amplitudes corresponding to photon emission from the incoming and the outgoing electrons, as well as from the incoming and the outgoing muons, respectively. These amplitudes are given by 
        \begin{equation*}
        \begin{split}
            i\mathcal{M}_a^{(\mu)}&
                (\omega_k,\hat{k})=
                \bar{u}(\vec{q}_1)
                \big[
                    ie\gamma^{\mu}
                \big]
                \frac{(-i)(
                        -\slashed{p}_1
                        +\slashed{k}
                        +m)}
                    {(-2p_1\cdot k)}
                \big[
                    ie\slashed{\epsilon}_r^*(\vec{k})
                \big]
                u(\vec{p}_1)
                \\&\hspace{12.5em}
                \frac{(-i)g_{\mu\nu}}
                    {[\frac{1}{2}(p_1-q_1-k)^2
                        +\frac{1}{2}(q_2-p_2)^2
                        -i\epsilon]}
                \bar{u}(\vec{q}_2)
                \big[
                    ie\gamma^{\nu}
                \big]
                u(\vec{p}_2)
            \\
            i\mathcal{M}_b^{(\mu)}&
                (\omega_k,\hat{k})=
                \bar{u}(\vec{q}_1)
                \big[
                    ie\slashed{\epsilon}_r^*(\vec{k})
                \big]
                \frac{(-i)(
                        -\slashed{q}_1
                        -\slashed{k}
                        +m)}
                    {(+2q_1\cdot k)}
                \big[
                    ie\gamma^{\mu}
                \big]
                u(\vec{p}_1)
                \\&\hspace{12.5em}
                \frac{(-i)g_{\mu\nu}}
                    {[\frac{1}{2}(p_1-q_1-k)^2
                        +\frac{1}{2}(q_2-p_2)^2
                        -i\epsilon]}
                \bar{u}(\vec{q}_2)
                \big[
                    ie\gamma^{\nu}
                \big]
                u(\vec{p}_2)
            \\
            i\mathcal{M}_c^{(\mu)}&
                (\omega_k,\hat{k})=
                \bar{u}(\vec{q}_1)
                \big[
                    ie\gamma^{\mu}
                \big]
                u(\vec{p}_1)
                \frac{(-i)g_{\mu\nu}}
                    {[\frac{1}{2}(p_1-q_1)^2
                        +\frac{1}{2}(q_2-p_2+k)^2
                        -i\epsilon]}
                \\&\hspace{16em}
                \bar{u}(\vec{q}_2)
                \big[
                    ie\gamma^{\nu}
                \big]
                \frac{(-i)(
                        -\slashed{p}_2
                        +\slashed{k}
                        +m)}
                    {(-2p_2\cdot k)}
                \big[
                    ie\slashed{\epsilon}_r^*(\vec{k})
                \big]
                u(\vec{p}_2)
            \\
            i\mathcal{M}_d^{(\mu)}&
                (\omega_k,\hat{k})=
                \bar{u}(\vec{q}_1)
                \big[
                    ie\gamma^{\mu}
                \big]
                u(\vec{p}_1)
                \frac{(-i)g_{\mu\nu}}
                    {[\frac{1}{2}(p_1-q_1)^2
                        +\frac{1}{2}(q_2-p_2+k)^2
                        -i\epsilon]}
                \\&\hspace{16em}
                \bar{u}(\vec{q}_2)
                \big[
                    ie\slashed{\epsilon}_r^*(\vec{k})
                \big]
                \frac{(-i)(
                        -\slashed{q}_2
                        -\slashed{k}
                        +m)}
                    {(+2q_2\cdot k)}
                \big[
                    ie\gamma^{\nu}
                \big]
                u(\vec{p}_2)
        \end{split}
        \end{equation*}
        The diagrams representing each contribution to the photon-emitting invariant amplitude are illustrated in \figref{fig.2.1.2(a)}.
        \begin{figure}[t]
            \centering
            \begin{subfigure}[t]{0.31\linewidth}
                \centering
                \begin{tikzpicture}[line width=1pt, 
  fermion/.style={thick},  
  boson/.style={decorate, decoration={snake}, draw=black},
  halffermion/.style={thick, ->},
  softboson/.style={decorate, decoration={snake}, draw=blue},
  halffermion/.style={thick, ->}, scale=0.75]

  \draw[fermion] (0.5,2) -- (2,1.25);    
  \draw[fermion] (0.5,-2) -- (2,-1.25);  
  \draw[fermion] (2,1.25) -- (3.5,2);    
  \draw[fermion] (2,-1.25) -- (3.5,-2);  

  \draw[boson] (2,1.25) -- (2,-1.25) node[midway, right] {$\gamma$};

  \filldraw[black] (2,1.25) circle (2pt);
  \filldraw[black] (2,-1.25) circle (2pt);
  \filldraw[black] (1.25,1.625) circle (2pt);  

  \draw[halffermion] (0.5,2) -- (0.875,1.8125);    
  \draw[halffermion] (1.25,1.625) -- (1.625,1.4375);    
  \draw[halffermion] (0.5,-2) -- (1.25,-1.625);  
  \draw[halffermion] (2,1.25) -- (2.75,1.625);    
  \draw[halffermion] (2,-1.25) -- (2.75,-1.625);  

  \draw[softboson] (1.25,1.625) -- (2.75,2.375);    

  \node at (1.675, 2.2) {$\gamma,\ k\ $};    

  \node at (-0.1,2.2) {$e^-,\ p_1\ $};
  \node at (-0.1,-2.2) {$\mu^-,\ p_2\ $};
  \node at (4.1,2.2) {$\ e^-,\ q_1$};
  \node at (4.1,-2.2) {$\ \mu^-,\ q_2$};

\end{tikzpicture}
                \caption{}
            \end{subfigure}
            \begin{subfigure}[t]{0.31\linewidth}
                \centering
                \begin{tikzpicture}[line width=1pt, 
  fermion/.style={thick},  
  boson/.style={decorate, decoration={snake}, draw=black},
  halffermion/.style={thick, ->},
  softboson/.style={decorate, decoration={snake}, draw=blue}, scale=0.75]

  \draw[fermion] (0.5,2) -- (2,1.25);    
  \draw[fermion] (0.5,-2) -- (2,-1.25);  
  \draw[fermion] (2,1.25) -- (3.5,2);    
  \draw[fermion] (2,-1.25) -- (3.5,-2);  

  \draw[boson] (2,1.25) -- (2,-1.25) node[midway, right] {$\gamma$};

  \filldraw[black] (2,1.25) circle (2pt);  
  \filldraw[black] (2,-1.25) circle (2pt); 
  \filldraw[black] (2.75,1.625) circle (2pt);  

  \draw[halffermion] (0.5,2) -- (1.25,1.625);    
  \draw[halffermion] (0.5,-2) -- (1.25,-1.625);  
  \draw[halffermion] (2,1.25) -- (2.375,1.4375);
  \draw[halffermion] (2.75,1.625) -- (3.125,1.8125);    
  \draw[halffermion] (2,-1.25) -- (2.75,-1.625);  

  \draw[softboson] (2.75,1.625) -- (4,0.875);

  \node at (4.1,1.3) {$\gamma,\ k$};

  \node at (-0.1,2.2) {$e^-,\ p_1\ $};
  \node at (-0.1,-2.2) {$\mu^-,\ p_2\ $};
  \node at (4.1,2.2) {$\ e^-,\ q_1$};
  \node at (4.1,-2.2) {$\ \mu^-,\ q_2$};

\end{tikzpicture}
                \caption{}
            \end{subfigure}
            \begin{subfigure}[t]{0.31\linewidth}
                \centering
                \begin{tikzpicture}[line width=1pt, 
  fermion/.style={thick},  
  boson/.style={decorate, decoration={snake}, draw=black},
  halffermion/.style={thick, ->},
  softboson/.style={decorate, decoration={snake}, draw=blue}, scale=0.75]

  \draw[fermion] (0.5,2) -- (2,1.25);    
  \draw[fermion] (0.5,-2) -- (2,-1.25);  
  \draw[fermion] (2,1.25) -- (3.5,2);    
  \draw[fermion] (2,-1.25) -- (3.5,-2);  

  \draw[boson] (2,1.25) -- (2,-1.25) node[midway, right] {$\gamma$};

  \filldraw[black] (2,1.25) circle (2pt);  
  \filldraw[black] (2,-1.25) circle (2pt); 
  \filldraw[black] (1.25,-1.625) circle (2pt);  

  \draw[halffermion] (0.5,2) -- (1.25,1.625);  
  \draw[halffermion] (0.5,-2) -- (0.875,-1.8125);  
  \draw[halffermion] (1.25,-1.625) -- (1.625,-1.4375);  
  \draw[halffermion] (2,1.25) -- (2.75,1.625);    
  \draw[halffermion] (2,-1.25) -- (2.75,-1.625);  

  \draw[softboson] (1.25,-1.625) -- (2.75,-2.375);

  \node at (1.6,-2.25) {$\gamma,\ k$};

  \node at (-0.1,2.2) {$e^-,\ p_1\ $};
  \node at (-0.1,-2.2) {$\mu^-,\ p_2\ $};
  \node at (4.1,2.2) {$\ e^-,\ q_1$};
  \node at (4.1,-2.2) {$\ \mu^-,\ q_2$};

\end{tikzpicture}

\hspace{1em}
                \caption{}
            \end{subfigure}
            \\ \vspace{1em}
            \begin{subfigure}[t]{0.31\linewidth}
                \centering
                \begin{tikzpicture}[line width=1pt, 
  fermion/.style={thick},  
  boson/.style={decorate, decoration={snake}, draw=black},
  halffermion/.style={thick, ->},
  softboson/.style={decorate, decoration={snake}, draw=blue}, scale=0.75]

  \draw[fermion] (0.5,2) -- (2,1.25);    
  \draw[fermion] (0.5,-2) -- (2,-1.25);  
  \draw[fermion] (2,1.25) -- (3.5,2);    
  \draw[fermion] (2,-1.25) -- (3.5,-2);  

  \draw[boson] (2,1.25) -- (2,-1.25) node[midway, right] {$\gamma$};

  \filldraw[black] (2,1.25) circle (2pt);  
  \filldraw[black] (2,-1.25) circle (2pt); 
  \filldraw[black] (2.75,-1.625) circle (2pt);  

  \draw[halffermion] (0.5,2) -- (1.25,1.625);  
  \draw[halffermion] (0.5,-2) -- (1.25,-1.625);  
  \draw[halffermion] (2,1.25) -- (2.75,1.625);    
  \draw[halffermion] (2,-1.25) -- (2.375,-1.4375);  
  \draw[halffermion] (2.75,-1.625) -- (3.125,-1.8125);  

  \draw[softboson] (2.75,-1.625) -- (4,-0.875);

  \node at (4.1,-1.3) {$\gamma,\ k$};

  \node at (-0.1,2.2) {$e^-,\ p_1\ $};
  \node at (-0.1,-2.2) {$\mu^-,\ p_2\ $};
  \node at (4.1,2.2) {$\ e^-,\ q_1$};
  \node at (4.1,-2.2) {$\ \mu^-,\ q_2$};

\end{tikzpicture}
                \caption{}
            \end{subfigure}
            \caption{Tree-level Feynman diagrams for single-photon emission in the \texorpdfstring{$e^-\mu\rightarrow e^-\mu $}{TEXT} scattering. The four diagrams show all possible soft photon (blue line) emissions during this scattering.}
            \label{fig.2.1.2(a)}
        \end{figure}
    \item for $e^-(p_1)+\gamma(p_2)\rightarrow\gamma(q_1)+ e^-(q_2)$:
        \begin{equation}\label{eq.3.1.3(b)}
        \begin{split}
            i\mathcal{M}_{\text{tree}}^{(\gamma)}
                (\omega_k,\hat{k})&=
                i\mathcal{M}_a^{(\gamma)}
                    (\omega_k,\hat{k})+
                i\mathcal{M}_b^{(\gamma)}
                    (\omega_k,\hat{k})+
                i\mathcal{M}_c^{(\gamma)}
                    (\omega_k,\hat{k})
                \\& \hspace{5em}+
                i\mathcal{M}_d^{(\gamma)}
                    (\omega_k,\hat{k})+
                i\mathcal{M}_e^{(\gamma)}
                    (\omega_k,\hat{k})+
                i\mathcal{M}_f^{(\gamma)}
                    (\omega_k,\hat{k})
        \end{split}
        \end{equation}
        where $i\mathcal{M}_a^{(\gamma)}(\omega_k,\hat{k}),\ i\mathcal{M}_b^{(\gamma)}(\omega_k,\hat{k}),\ $ and $i\mathcal{M}_c^{(\gamma)}(\omega_k,\hat{k})$ represent the contributions from photon emission by the incoming electron, the virtual electron, and the outgoing electron of \figref{fig.2.1.1} (b), respectively 
        (the rest of the amplitudes correspond to contributions from photon emission by the incoming electron, the virtual electron, and the outgoing electron of \figref{fig.2.1.1} (c), in a similar manner). These contributions are given explicitly by:
        \begin{align*}
            i\mathcal{M}_a^{(\gamma)}&
                (\omega_k,\hat{k})=
                \bar{u}(\vec{q}_2)
                \big[
                    ie\slashed{\epsilon}_{\lambda_2}(\vec{p}_2)
                \big]
                \frac{(-i)
                        (-\frac{1}{2}\slashed{p}_1
                        +\frac{1}{2}\slashed{q}_1
                        +\frac{1}{2}\slashed{k}
                        -\frac{1}{2}\slashed{q}_2
                        +\frac{1}{2}\slashed{p}_2
                        +m)}
                    {[\frac{1}{2}(p_1-q_1-k)^2
                    +\frac{1}{2}(q_2-p_2)^2
                    +m^2-i\epsilon]}
                \\&\hspace{15em}
                \big[
                    ie\slashed{\epsilon}_{r_1}^*(\vec{q}_1)
                \big]
                \frac{(-i)
                        (-\slashed{p}_1
                        +\slashed{k}
                        +m)}
                    {(-2p_1\cdot k)}
                \big[
                    ie\slashed{\epsilon}_r^*(\vec{k})
                \big]
                u(\vec{p}_1)
            \\
            i\mathcal{M}_b^{(\gamma)}&
                (\omega_k,\hat{k})=
                \bar{u}(\vec{q}_2)
                \big[
                    ie\slashed{\epsilon}_{\lambda_2}(\vec{p}_2)
                \big]
                \frac{(-i)(
                        -\frac{1}{2}\slashed{p}_1
                        +\frac{1}{2}\slashed{q}_1
                        -\frac{1}{2}\slashed{q}_2
                        +\frac{1}{2}\slashed{p}_2
                        +\frac{1}{2}\slashed{k}
                        +m)}
                    {[\frac{1}{2}(p_1-q_1-k)^2
                    +\frac{1}{2}(q_2-p_2)^2
                    +m^2-i\epsilon]}
                \big[
                    ie\slashed{\epsilon}^*_{r}(\vec{k})
                \big]
                \\&\hspace{10em}
                \frac{(-i)(
                        -\frac{1}{2}\slashed{p}_1
                        +\frac{1}{2}\slashed{q}_1
                        -\frac{1}{2}\slashed{q}_2
                        +\frac{1}{2}\slashed{p}_2
                        -\frac{1}{2}\slashed{k}
                        +m)}
                    {[\frac{1}{2}(p_1-q_1)^2
                    +\frac{1}{2}(q_2-p_2+k)^2
                    +m^2-i\epsilon]}
                \big[
                    ie\slashed{\epsilon}^*_{r_1}(\vec{q}_1)
                \big]
                u(\vec{p}_1)
            \displaybreak[1]
            \\
            i\mathcal{M}_c^{(\gamma)}&
                (\omega_k,\hat{k})=
                \bar{u}(\vec{q}_2)
                \big[
                    ie\slashed{\epsilon}_{r}^*(\Vec{k})
                \big]
                \frac{(-i)(
                        -\slashed{q}_2
                        -\slashed{k}
                        +m)}
                    {(+2q_2\cdot k)}
                \big[
                    ie\slashed{\epsilon}_{\lambda_2}(\Vec{p}_2)
                \big]
                \\&\hspace{10em}
                \frac{(-i)(
                        -\frac{1}{2}\slashed{p}_1
                        +\frac{1}{2}\slashed{q}_1
                        -\frac{1}{2}\slashed{q}_2
                        +\frac{1}{2}\slashed{p}_2
                        -\frac{1}{2}\slashed{k}
                        +m)}
                    {[\frac{1}{2}(p_1-q_1)
                        +\frac{1}{2}(q_2-p_2+k)^2
                        +m^2-i\epsilon]}
                \big[
                    ie\slashed{\epsilon}_{r_1}^*(\Vec{q}_1)
                \big]
            u(\vec{p}_1)
        \end{align*}
        The diagrams in \figref{fig.2.1.2(b)} illustrate three of the contributions to $i\mathcal{M}_{\text{tree}}^{(\gamma)}(\omega_k,\hat{k})$. The rest are obtained by adding an outgoing soft photon line to \figref{fig.2.1.1} (c) in all possible ways.
        \begin{figure}[t]
            \centering
            \begin{subfigure}[t]{0.31\linewidth}
                \centering
                \begin{tikzpicture}[line width=1pt, 
  fermion/.style={thick},  
  boson/.style={decorate, decoration={snake}, draw=black},
  softboson/.style={decorate, decoration={snake}, draw=blue},
  halffermion/.style={thick, ->}, scale=0.75]

  \draw[fermion] (0.5,2) -- (2,1.25);        
  \draw[boson] (0.5,-2) -- (2,-1.25);        
  \draw[boson] (2,1.25) -- (3.5,2);          
  \draw[fermion] (2,-1.25) -- (3.5,-2);      

  \draw[fermion] (2,1.25) -- (2,-1.25) node[midway, right] {$e^-$};
  \draw[halffermion] (2,1.25) -- (2,0);

  \filldraw[black] (2,1.25) circle (2pt);
  \filldraw[black] (2,-1.25) circle (2pt);
  \filldraw[black] (1.25,1.625) circle (2pt); 

  \draw[halffermion] (0.5,2) -- (0.875,1.8125); 
  \draw[halffermion] (1.25,1.625) -- (1.625,1.4375); 
  \draw[halffermion] (2,-1.25) -- (2.75,-1.625); 

  \draw[softboson] (1.25,1.625) -- (2.75,2.375);
  \node at (1.8,2.3) {$\gamma,\ k\ \ $};

  \node at (-0.1,2.2) {$e^-,\ p_1\ $};
  \node at (-0.1,-2.2) {$\gamma,\ p_2\ $};
  \node at (4.1,2.2) {$\ \gamma,\ q_1$};
  \node at (4.1,-2.2) {$\ e^-,\ q_2$};

\end{tikzpicture}
                \caption{}
            \end{subfigure}
            \begin{subfigure}[t]{0.31\linewidth}
                \centering
                \begin{tikzpicture}[line width=1pt, 
  fermion/.style={thick},  
  boson/.style={decorate, decoration={snake}, draw=black},
  softboson/.style={decorate, decoration={snake}, draw=blue},
  halffermion/.style={thick, ->}, scale=0.75]

  \draw[fermion] (0.5,2) -- (2,1.25);        
  \draw[boson] (0.5,-2) -- (2,-1.25);        
  \draw[boson] (2,1.25) -- (3.5,2);          
  \draw[fermion] (2,-1.25) -- (3.5,-2);      

  \draw[fermion] (2,1.25) -- (2,0);          
  \draw[fermion] (2,0) -- (2,-1.25);         
  
  \filldraw[black] (2,1.25) circle (2pt);
  \filldraw[black] (2,-1.25) circle (2pt);
  \filldraw[black] (2,0) circle (2pt); 

  \draw[halffermion] (2,1.25) -- (2,0.625);     
  \draw[halffermion] (2,0) -- (2,-0.625);       

  \draw[halffermion] (0.5,2) -- (1.25,1.625); 
  \draw[halffermion] (2,-1.25) -- (2.75,-1.625); 

  \draw[softboson] (2,0) -- (3.25,0);
  \node at (3.4,0.4) {$\gamma,\ k$};

  \node at (-0.1,2.2) {$e^-,\ p_1\ $};
  \node at (-0.1,-2.2) {$\gamma,\ p_2\ $};
  \node at (4.1,2.2) {$\ \gamma,\ q_1$};
  \node at (4.1,-2.2) {$\ e^-,\ q_2$};

\end{tikzpicture}
                \caption{}
            \end{subfigure}
            \begin{subfigure}[t]{0.31\linewidth}
                \centering
                \begin{tikzpicture}[line width=1pt, 
  fermion/.style={thick},  
  boson/.style={decorate, decoration={snake}, draw=black},
  softboson/.style={decorate, decoration={snake}, draw=blue},
  halffermion/.style={thick, ->}, scale=0.75]

  \draw[fermion] (0.5,2) -- (2,1.25);        
  \draw[boson] (0.5,-2) -- (2,-1.25);        
  \draw[boson] (2,1.25) -- (3.5,2);          
  \draw[fermion] (2,-1.25) -- (3.5,-2);      

  \draw[fermion] (2,1.25) -- (2,-1.25) node[midway, right] {$e^-$};
  \draw[halffermion] (2,1.25) -- (2,0);

  \filldraw[black] (2,1.25) circle (2pt);
  \filldraw[black] (2,-1.25) circle (2pt);
  \filldraw[black] (2.75,-1.625) circle (2pt);  

  \draw[halffermion] (0.5,2) -- (1.25,1.625); 
  \draw[halffermion] (2,-1.25) -- (2.375,-1.4375); 
  \draw[halffermion] (2.75,-1.625) -- (3.125,-1.8125); 

  \draw[softboson] (2.75,-1.625) -- (4.25,-0.875);
  \node at (4.1,-1.4) {$\ \gamma,\ k$};

  \node at (-0.1,2.2) {$e^-,\ p_1\ $};
  \node at (-0.1,-2.2) {$\gamma,\ p_2\ $};
  \node at (4.1,2.2) {$\ \gamma,\ q_1$};
  \node at (4.1,-2.2) {$\ e^-,\ q_2$};

\end{tikzpicture}
                \caption{}
            \end{subfigure}
            \caption{Representative tree-level Feynman diagrams for single-photon emission in the \texorpdfstring{$e^-\gamma\rightarrow e^-\gamma $}{TEXT} scattering. The three diagrams shown correspond to half of the possible soft photon (blue line) emissions.
            }
            \label{fig.2.1.2(b)}
        \end{figure}
    \item for $e^-(p_1)+e^+(p_2)\rightarrow\gamma(q_1)+\gamma(q_2)$:
        \begin{equation}\label{eq.3.1.3(c)}
        \begin{split}
            i\mathcal{M}_{\text{tree}}^{(e)}
                (\omega_k,\hat{k})&=
                i\mathcal{M}_a^{(e)}
                    (\omega_k,\hat{k})+
                i\mathcal{M}_b^{(e)}
                    (\omega_k,\hat{k})+
                i\mathcal{M}_c^{(e)}
                    (\omega_k,\hat{k})
                \\& \hspace{5em}+
                i\mathcal{M}_d^{(e)}
                    (\omega_k,\hat{k})+
                i\mathcal{M}_e^{(e)}
                    (\omega_k,\hat{k})+
                i\mathcal{M}_f^{(e)}
                    (\omega_k,\hat{k})
        \end{split}
        \end{equation}
        where $i\mathcal{M}_a^{(e)}(\omega_k,\hat{k}),\ i\mathcal{M}_b^{(e)}(\omega_k,\hat{k})$, and $i\mathcal{M}_c^{(e)}(\omega_k,\hat{k})$ represent the photon emission contributions from the incoming electron, the virtual electron and the incoming positron, respectively 
        (see diagram (d) in \figref{fig.2.1.1}). The contributions $\smash{i\mathcal{M}_d^{(e)}(\omega_k,\hat{k})}$,  $\smash{i\mathcal{M}_e^{(e)}(\omega_k,\hat{k})}$, and $\smash{i\mathcal{M}_f^{(e)}(\omega_k,\hat{k})}$, can be obtained by exchanging the final-state photon momenta, $q_1\leftrightarrow q_2$, in $\smash{i\mathcal{M}_a^{(e)}(\omega_k,\hat{k}),\ i\mathcal{M}_b^{(e)}(\omega_k,\hat{k})}$, and $\smash{i\mathcal{M}_c^{(e)}(\omega_k,\hat{k})}$, respectively (see diagram (e) in \figref{fig.2.1.1}). Explicitly, the first three contributions are given by
        \begin{equation*}
        \begin{split}
            i\mathcal{M}_a^{(e)}&
                (\omega_k,\hat{k})=
                \bar{\upsilon}(\vec{p}_2)
                \big[
                    ie\slashed{\epsilon}_{r_2}^*(\vec{q}_2)
                \big]
                \frac{(-i)
                        (-\frac{1}{2}\slashed{p}_1
                        +\frac{1}{2}\slashed{q}_1
                        +\frac{1}{2}\slashed{k}
                        -\frac{1}{2}\slashed{q}_2
                        +\frac{1}{2}\slashed{p}_2
                        +m)}
                    {[\frac{1}{2}(p_1-q_1-k)^2
                    +\frac{1}{2}(q_2-p_2)^2
                    +m^2-i\epsilon]}
                \\&\hspace{15em}
                \big[
                    ie\slashed{\epsilon}_{r_1}^*(\vec{q}_1)
                \big]
                \frac{(-i)
                        (-\slashed{p}_1
                        +\slashed{k}
                        +m)}
                    {(-2p_1\cdot k)}
                \big[
                    ie\slashed{\epsilon}_r^*(\vec{k})
                \big]
                u(\vec{p}_1)
            \\
            i\mathcal{M}_b^{(e)}&
                (\omega_k,\hat{k})=
                \bar{\upsilon}(\vec{p}_2)
                \big[
                    ie\slashed{\epsilon}^*_{r_1}(\vec{q}_1)
                \big]
                \frac{(-i)(
                        -\frac{1}{2}\slashed{p}_1
                        +\frac{1}{2}\slashed{q}_1
                        -\frac{1}{2}\slashed{q}_2
                        +\frac{1}{2}\slashed{p}_2
                        -\frac{1}{2}\slashed{k}
                        +m)}
                    {[\frac{1}{2}(p_1-q_1)^2
                    +\frac{1}{2}(q_2-p_2+k)^2
                    +m^2-i\epsilon]}
                \big[
                    ie\slashed{\epsilon}^*_{r}(\vec{k})
                \big]
                \\&\hspace{10em}
                \frac{(-i)(
                        -\frac{1}{2}\slashed{p}_1
                        +\frac{1}{2}\slashed{q}_1
                        -\frac{1}{2}\slashed{q}_2
                        +\frac{1}{2}\slashed{p}_2
                        -\frac{1}{2}\slashed{k}
                        +m)}
                    {[\frac{1}{2}(p_1-q_1)^2
                    +\frac{1}{2}(q_2-p_2+k)^2
                    +m^2-i\epsilon]}
                \big[
                    ie\slashed{\epsilon}^*_{r_2}(\vec{q}_2)
                \big]
                u(\vec{p}_2)
            \\
            i\mathcal{M}_c^{(e)}&
                (\omega_k,\hat{k})=
                \bar{\upsilon}(\vec{p}_2)
                \big[
                    ie\slashed{\epsilon}_{r}^*(\Vec{k})
                \big]
                \frac{(-i)(
                        +\slashed{p}_2
                        -\slashed{k}
                        +m)}
                    {(-2p_2\cdot k)}
                \big[
                    ie\slashed{\epsilon}_{r_2}^*(\Vec{q}_2)
                \big]
                \\&\hspace{10em}
                \frac{(-i)(
                        -\frac{1}{2}\slashed{p}_1
                        +\frac{1}{2}\slashed{q}_1
                        -\frac{1}{2}\slashed{q}_2
                        +\frac{1}{2}\slashed{p}_2
                        -\frac{1}{2}\slashed{k}
                        +m)}
                    {[\frac{1}{2}(p_1-q_1)
                        +\frac{1}{2}(q_2-p_2+k)^2
                        +m^2-i\epsilon]}
                \big[
                    ie\slashed{\epsilon}_{r_1}^*(\Vec{q}_1)
                \big]
            u(\vec{p}_1)
        \end{split}
        \end{equation*}
        In \figref{fig.2.1.2(c)}, we present the diagrams corresponding to $i\mathcal{M}_a^{(e)}(\omega_k,\hat{k})$, $i\mathcal{M}_b^{(e)}(\omega_k,\hat{k})$, and $\smash{i\mathcal{M}_c^{(e)}(\omega_k,\hat{k})}$. The rest are obtained by exchanging $q_1$ and $q_2$.
        \begin{figure}[t]
            \centering
            \begin{subfigure}[t]{0.31\linewidth}
                \centering
                \begin{tikzpicture}[line width=1pt, 
  fermion/.style={thick},  
  boson/.style={decorate, decoration={snake}, draw=black},
  softboson/.style={decorate, decoration={snake}, draw=blue},
  halffermion/.style={thick, ->}, scale=0.75]

  \draw[fermion] (0.5,2) -- (2,1.25);        
  \draw[fermion] (0.5,-2) -- (2,-1.25);            
  \draw[boson] (2,1.25) -- (3.5,2);                
  \draw[boson] (2,-1.25) -- (3.5,-2);              

  \draw[fermion] (2,1.25) -- (2,-1.25) node[midway, right] {$e^-$};

  \filldraw[black] (2,1.25) circle (2pt);
  \filldraw[black] (2,-1.25) circle (2pt);
  \filldraw[black] (1.25,1.625) circle (2pt);   

  \draw[halffermion] (0.5,2) -- (0.875, 1.8125);
  \draw[halffermion] (1.25,1.625) -- (1.625,1.4375);

  \draw[softboson] (1.25,1.625) -- (2.75,2.375);
  \node at (1.8,2.4) {$\gamma,\ k\ $};

  \draw[halffermion] (2,-1.25) -- (1.25,-1.625);     

  \node at (-0.1,2.2) {$e^-,\ p_1\ $};
  \node at (-0.1,-2.2) {$\gamma,\ p_2\ $};
  \node at (4.1,2.2) {$\ \gamma,\ q_1$};
  \node at (4.1,-2.2) {$\ e^-,\ q_2$};

\end{tikzpicture}
                \caption{}
            \end{subfigure}
            \begin{subfigure}[t]{0.31\linewidth}
                \centering
                \begin{tikzpicture}[line width=1pt, 
  fermion/.style={thick},  
  boson/.style={decorate, decoration={snake}, draw=black},
  softboson/.style={decorate, decoration={snake}, draw=blue},
  halffermion/.style={thick, ->}, scale=0.75]

  \draw[fermion] (0.5,2) -- (2,1.25);        
  \draw[fermion] (0.5,-2) -- (2,-1.25);        
  \draw[boson] (2,1.25) -- (3.5,2);          
  \draw[boson] (2,-1.25) -- (3.5,-2);      

  \draw[fermion] (2,1.25) -- (2,0);          
  \draw[fermion] (2,0) -- (2,-1.25);         
  
  \filldraw[black] (2,1.25) circle (2pt);
  \filldraw[black] (2,-1.25) circle (2pt);
  \filldraw[black] (2,0) circle (2pt); 

  \draw[halffermion] (2,1.25) -- (2,0.625);     
  \draw[halffermion] (2,0) -- (2,-0.625);       

  \draw[halffermion] (0.5,2) -- (1.25,1.625); 
  \draw[halffermion] (2,-1.25) -- (1.25,-1.625);     

  \draw[softboson] (2,0) -- (3.25,0);
  \node at (3.4,0.4) {$\gamma,\ k$};

  \node at (-0.1,2.2) {$e^-,\ p_1\ $};
  \node at (-0.1,-2.2) {$\gamma,\ p_2\ $};
  \node at (4.1,2.2) {$\ \gamma,\ q_1$};
  \node at (4.1,-2.2) {$\ e^-,\ q_2$};

\end{tikzpicture}

\hspace{1em}
                \caption{}
            \end{subfigure}
            \begin{subfigure}[t]{0.31\linewidth}
                \centering
                \begin{tikzpicture}[line width=1pt, 
  fermion/.style={thick},  
  boson/.style={decorate, decoration={snake}, draw=black},
  halffermion/.style={thick, ->},
  softboson/.style={decorate, decoration={snake}, draw=blue}, scale=0.75]

  \draw[fermion] (0.5,2) -- (2,1.25);    
  \draw[fermion] (0.5,-2) -- (2,-1.25);  
  \draw[boson] (2,1.25) -- (3.5,2);    
  \draw[boson] (2,-1.25) -- (3.5,-2);  

  \draw[fermion] (2,1.25) -- (2,-1.25) node[midway, right] {$e^-$};

  \filldraw[black] (2,1.25) circle (2pt);
  \filldraw[black] (2,-1.25) circle (2pt); 
  \filldraw[black] (1.25,-1.625) circle (2pt);  

  \draw[halffermion] (0.5,2) -- (1.25,1.625);  
  \draw[halffermion] (2,1.25) -- (2,0);  
  \draw[halffermion] (1.25,-1.625) -- (0.875,-1.8125);  
  \draw[halffermion] (2,-1.25) -- (1.625,-1.4375);  

  \draw[softboson] (1.25,-1.625) -- (2.75,-2.375);

  \node at (1.6,-2.3) {$\gamma,\ k\ $};

  \node at (-0.1,2.2) {$e^-,\ p_1\ $};
  \node at (-0.1,-2.2) {$\gamma,\ p_2\ $};
  \node at (4.1,2.2) {$\ \gamma,\ q_1$};
  \node at (4.1,-2.2) {$\ e^-,\ q_2$};

\end{tikzpicture}
                \caption{}
            \end{subfigure}
            \caption{Representative tree-level Feynman diagrams for single-photon emission in the \texorpdfstring{$e^-e^+\rightarrow\gamma\gamma$}{TEXT} scattering. 
            }
            \label{fig.2.1.2(c)}
        \end{figure}
\end{itemize}

In the limit where the momentum of the additional emitted photon becomes soft, the invariant amplitudes factorize according to the subleading soft photon theorem. See \hyperref[sec:B]{Appendix B} for a review of the derivation. This result facilitates the expansion of the amplitudes in powers of $\omega_k$, producing both the leading singular and the subleading behavior. The relevant soft factors are universal, determined by properties of the external charged particles. The resulting expressions for the amplitudes of the three processes are given below \cite{Gell-Mann:1954wra,Low:1954kd,Low:1958sn,Weinberg:1965nx,Burnett:1967km,DelDuca:1990gz,Luo:2014wea,Bern:2014vva,Lysov:2014csa,AtulBhatkar:2018kfi,Beneke:2021ilf,Beneke:2021umj,Travaglini:2022uwo}.
\begin{itemize}
    \item for $e^-(p_1)+\mu^-(p_2)\rightarrow e^-(q_1)+\mu^-(q_2)$:
    \begin{equation}\label{eq.3.1.4(a)}
    \begin{split}
        i\mathcal{M}_{\text{tree}}^{(\mu)}
            (\omega_k,\hat{k})=
            e\bigg\{
                \Big[&
                    \frac{q_1\cdot
                            \epsilon_r^*(\vec{k})}
                        {q_1\cdot k}+
                    \frac{q_2\cdot
                            \epsilon_r^*(\vec{k})}
                        {q_2\cdot k}-
                    \frac{p_1\cdot
                            \epsilon_r^*(\vec{k})}
                        {p_1\cdot k}-
                    \frac{p_2\cdot
                            \epsilon_r^*(\vec{k})}
                        {p_2\cdot k}
                \Big]\\&\!+
                \frac{2k\cdot(p_1-q_1)}
                    {(p_1-q_1)^2}
                \Big[
                    \frac{q_1\cdot
                            \epsilon_r^*(\vec{k})}
                        {q_1\cdot k}-
                    \frac{q_2\cdot
                            \epsilon_r^*(\vec{k})}
                        {q_2\cdot k}-
                    \frac{p_1\cdot
                            \epsilon_r^*(\vec{k})}
                        {p_1\cdot k}+
                    \frac{p_2\cdot
                            \epsilon_r^*(\vec{k})}
                        {p_2\cdot k}
                \Big]\\&\!+
                i\epsilon^*_{r\mu}(\vec{k})\ 
                k_{\nu}
                \Big[
                    \frac{\bar{S}_{q_1}^{\mu\nu}}
                        {q_1\cdot k}+
                    \frac{\bar{S}_{q_2}^{\mu\nu}}
                        {q_2\cdot k}-
                    \frac{S_{p_1}^{\mu\nu}}
                        {p_1\cdot k}-
                    \frac{S_{p_2}^{\mu\nu}}
                        {p_2\cdot k}
                \Big]
            \bigg\}
            i\mathcal{M}_0^{(\mu)}+
            \mathcal{O}(\omega_k)
    \end{split}
    \end{equation}
    \item for $e^-(p_1)+\gamma(p_2)\rightarrow\gamma(q_1)+ e^-(q_2)$:
    \begin{equation}\label{eq.3.1.4(b)}
    \begin{split}
        i\mathcal{M}_{\text{tree}}^{(\gamma)}
            (\omega_k,\hat{k})\!=\!
            e \bigg\{
                \Big[&
                    \frac{q_2\cdot
                            \epsilon_r^*(\vec{k})}
                        {q_2\cdot k}\!-\!
                    \frac{p_1\cdot
                            \epsilon_r^*(\vec{k})}
                        {p_1\cdot k}
                \Big]\!+\!
                i\epsilon^*_{r\mu}(\vec{k}) 
                k_{\nu}
                \Big[
                    \frac{\bar{J}_{q_2}^{\mu\nu}}
                        {q_2\cdot k}\!-\!
                    \frac{J_{p_1}^{\mu\nu}}
                        {p_1\cdot k}
                \Big]
            \bigg\}
            i\mathcal{M}_0^{(\gamma)}\!+\!
            \mathcal{O}(\omega_k)
    \end{split}
    \end{equation}
    \item for $e^-(p_1)+e^+(p_2)\rightarrow\gamma(q_1)+\gamma(q_2)$:
    \begin{equation}\label{eq.3.1.4(c)}
    \begin{split}
        i\mathcal{M}_{\text{tree}}^{(e)}
            (\omega_k,\hat{k})\!=\!
            e \bigg\{
                \Big[&
                    \frac{p_2\cdot
                            \epsilon_r^*(\vec{k})}
                        {p_2\cdot k}\!-\!
                    \frac{p_1\cdot
                            \epsilon_r^*(\vec{k})}
                        {p_1\cdot k}
                \Big]\!+\!
                i\epsilon^*_{r\mu}(\vec{k}) 
                k_{\nu}
                \Big[
                    \frac{\bar{J}_{\bar{p}_2}^{\mu\nu}}
                        {p_2\cdot k}\!-\!
                    \frac{J_{p_1}^{\mu\nu}}
                        {p_1\cdot k}
                \Big]
            \bigg\}
            i\mathcal{M}_0^{(e)}\!+\!
            \mathcal{O}(\omega_k)
    \end{split}
    \end{equation}
\end{itemize}
In \eqref{eq.3.1.4(a)}, \eqref{eq.3.1.4(b)} and \eqref{eq.3.1.4(c)}, we have introduced the total angular momentum operator, which is given by the sum of the orbital and the spin components. For the incoming and the outgoing particles, this operator is defined, respectively, by
\begin{equation}\label{eq.3.1.5}
\begin{split}
    J_p^{\mu\nu}=
        L_p^{\mu\nu}+
        S_p^{\mu\nu}
    , \hspace{5em}
    \bar{J}_p^{\mu\nu}=
        \bar{L}_p^{\mu\nu}+
        \bar{S}_p^{\mu\nu}
\end{split}
\end{equation}
while for the incoming and the outgoing antiparticles we have
\begin{equation}\label{eq.3.1.6}
\begin{split}
    \bar{J}_{\bar{p}}^{\mu\nu}=
        \bar{L}_{\bar{p}}^{\mu\nu}+
        \bar{S}_{\bar{p}}^{\mu\nu}
    \ \hspace{5em}
    J_{\bar{p}}^{\mu\nu}=
        L_{\bar{p}}^{\mu\nu}+
        S_{\bar{p}}^{\mu\nu}
\end{split}
\end{equation}
The orbital angular momentum operators act on the amplitude, which depends on the momenta of the particles, excluding any momentum dependence arising from the spinor wavefunctions (see also \hyperref[sec:B]{Appendix B}). They are given by
\begin{equation}\label{eq.3.1.7}
\begin{split}
    L_p^{\mu\nu}=
    \bar{L}_{\bar{p}}^{\mu\nu}=+
        i\Big(p^{\mu}\frac{\partial}{\partial p_{\nu}}-
        p^{\nu}\frac{\partial}{\partial p_{\mu}}\Big)
    \\
    \bar{L}_q^{\mu\nu}=
    L_{\bar{q}}^{\mu\nu}=-
        i\Big(q^{\mu}\frac{\partial}{\partial q_{\nu}}-
        q^{\nu}\frac{\partial}{\partial q_{\mu}}\Big)
    \\
\end{split}
\end{equation}
The spin angular momentum operators can be obtained by utilizing the commutator of the Dirac gamma matrices. They act directly on the spinor wavefunctions. For the incoming and the outgoing particles, the explicit expressions are \cite{Bern:2014vva}
\begin{equation}\label{eq.3.1.8}
\begin{split}
    S_p^{\mu\nu}=+
        \frac{i}{4}[\gamma^{\mu},\gamma^{\nu}]\
        u(\vec{p})\
        \circ\
        \frac{\partial}
            {\partial u(\vec{p})}
    \\
    \bar{S}_q^{\mu\nu}=+
        \bar{u}(\vec{q})\
        \frac{i}{4}[\gamma^{\mu},\gamma^{\nu}]\
        \circ\
        \frac{\partial}
            {\partial \bar{u}(\vec{q})}
\end{split}
\end{equation}
For the incoming and the outgoing antiparticles,
\begin{equation}\label{eq.3.1.9}
\begin{split}
    \bar{S}_{\bar{p}}^{\mu\nu}&=-
        \bar{\upsilon}(\vec{p})\
        \frac{i}{4}[\gamma^{\mu},\gamma^{\nu}]\
        \circ\
        \frac{\partial}
            {\partial \bar{\upsilon}(\vec{p})}
    \\
    S_{\bar{q}}^{\mu\nu}&=-
        \frac{i}{4}[\gamma^{\mu},\gamma^{\nu}]\
        \upsilon(\vec{q})
        \circ\
        \frac{\partial}
            {\partial \upsilon(\vec{q})}
\end{split}
\end{equation}
We have suppressed spinor indices 
(see \hyperref[sec:B]{Appendix B} for the expressions for which the spinor indices appear explicitly). 
The symbol $\circ$ indicates that the free spinor index from the commutator is contracted with the spinor index carried by the derivative with respect to the fermionic wavefunction. 

The first term in \eqref{eq.3.1.4(a)}, \eqref{eq.3.1.4(b)} and \eqref{eq.3.1.4(c)} is the leading contribution, $\smash{\omega_k^{-1}i\mathcal{M}_L(\hat{k})}$, in the soft momentum expansion of the photon emitting tree-level invariant amplitude, \eqref{eq.3.0.1}. The remaining terms comprise the subleading constant contribution, $\smash{i\mathcal{M}_{SL}(\hat{k})}$, in the soft photon energy. Specifically, each term in the second and third lines of \eqref{eq.3.1.4(a)} can be obtained by the action of the total angular momentum operator
on the elastic amplitude $i\mathcal{M}_0$ (see \hyperref[sec:B]{Appendix B} for more details on the action of the total angular momentum operator on the elastic amplitude). 

The leading contribution to the soft momentum expansion is exact to all orders in perturbation theory. However, the subleading correction in \eqref{eq.3.1.4(a)} receives corrections at the higher loop level. 
Indeed, the loop-level amplitudes
admit additional contributions that depend logarithmically on the energy of the soft photon. 
Such loop corrections have been thoroughly studied in \cite{Sahoo:2019yod,Krishna:2023fxg}.

\eqref{eq.3.1.4(a)}, \eqref{eq.3.1.4(b)}, and \eqref{eq.3.1.4(c)} comprise relations between single photon-emitting invariant amplitudes and the corresponding elastic ones. 
For the corresponding full S-matrix elements we need to include
the energy and momentum conserving $\delta$ functions. 
Let $\smash{S_0^{(\alpha)}},\ \alpha=\{\mu,\gamma,e\}$ denote the non-trivial part of the S-matrix element corresponding to the three elastic processes above, at tree level. Then, 
\begin{equation}\label{eq.3.1.11}
    S_0^{(\alpha)}=
        (2\pi)^4
        \delta^{(4)}(p_1+p_2-q_1-q_2)\
        i\mathcal{M}_{0}^{(\alpha)}
\end{equation}
Likewise, for the non-trivial part of the S-matrix element corresponding to single photon emission at tree level, we have 
\begin{equation}
S_{\text{tree}}^{(\alpha)}
            (\omega_k,\hat{k})=(2\pi)^4
        \delta^{(4)}(p_1+p_2-q_1-q_2-k)\ i\mathcal{M}_{\text{tree}}^{(\alpha)}
                (\omega_k,\hat{k})
\end{equation}
As we have remarked before, all external particles are on-shell with the energy given in terms of the momentum and mass. The presence of the $\delta$ function ensures that the S-matrix element for single photon emission vanishes if energy and momentum are not conserved. In other words, the presence of the $\delta$ function imposes energy and momentum conservation: $p_1+p_2-q_1-q_2-k=0$.

The soft photon theorem can be applied to the full S-matrix elements, which include the energy-momentum conserving $\delta$ function. For the S-matrix elements associated with single photon emission, 
the soft-photon momentum appears in the argument of the $\delta$ function: $$\delta^{(4)}(p_1+p_2-q_1-q_2-k)$$ As a result, we must take into account the  expansion of the $\delta$ function in terms of the soft-photon momentum $k$, in order to obtain the full subleading contribution to the expansion of the S-matrix element in terms of $k$. In particular, we formally treat the $\delta$ function as a distribution and Taylor-expand it in terms of the soft-photon momentum to order $k$. The 
subleading contribution to the S-matrix element will then include extra terms given 
by the product of the leading contribution to $i\mathcal{M}^{(\alpha)}_{\text{tree}}(\omega_k,\hat{k})$ 
and derivatives of the $\delta$ function with respect to the four components of the soft-photon momentum $k$ (evaluated at $k=0$). The latter can be related to derivatives of the $\delta$ function with respect to the momenta of the hard particles. It can be shown that all the extra terms that involve derivatives of the $\delta$ function can be reproduced by the action of the total angular momentum operators (one for each external hard particle) on the $\delta$ function \cite{Elvang:2013cua,Cachazo:2014fwa,Chakrabarti:2017zmh,Sahoo:2019yod}. In all intermediate steps of the calculation, the full energy and momentum conservation relation $p_1+p_2-q_1-q_2-k=0$ must be applied, and all terms up to order $\mathcal{O}(\omega_k^0)$ in the expansion must be retained. See the end of \hyperref[sec:B]{Appendix B} for a derivation of the extra terms\footnote{
Reference \cite{Burnett:1967km} derives the subleading soft-photon theorem by defining $k$-depended shifts between the hard momenta of the elastic process and that of the radiative process. 
It turns out that this results in a different, but equivalent, approach for expanding radiative amplitudes to subleading order in the soft-momentum expansion, since the additional shift-dependent terms appearing in the subleading soft-photon theorem can be chosen such that they do not contribute to physically measurable quantities.
}.

Taking into account the extra terms from the expansion of the $\delta$ function,
the subleading soft-photon theorem 
in terms of S-matrix elements takes exactly the same form as the one expressed in terms of the invariant amplitudes \cite{Elvang:2013cua,Cachazo:2014fwa,Chakrabarti:2017zmh,Sahoo:2019yod}.
The corresponding expressions 
are given as follows:
\begin{itemize}
    \item for $e^-(p_1)+\mu^-(p_2)\rightarrow e^-(q_1)+\mu^-(q_2)$:
    \begin{equation}\label{eq.3.1.13(a)}
    \begin{split}
        S_{\text{tree}}^{(\mu)}
            (\omega_k,\hat{k})=
            e\ \bigg\{
                \Big[&
                    \frac{q_1\cdot
                            \epsilon_r^*(\vec{k})}
                        {q_1\cdot k}+
                    \frac{q_2\cdot
                            \epsilon_r^*(\vec{k})}
                        {q_2\cdot k}-
                    \frac{p_1\cdot
                            \epsilon_r^*(\vec{k})}
                        {p_1\cdot k}-
                    \frac{p_2\cdot
                            \epsilon_r^*(\vec{k})}
                        {p_2\cdot k}
                \Big]\\&+
                i\epsilon^*_{r\mu}(\vec{k})\ 
                k_{\nu}
                \Big[
                    \frac{\bar{J}_{q_1}^{\mu\nu}}
                        {q_1\cdot k}+
                    \frac{\bar{J}_{q_2}^{\mu\nu}}
                        {q_2\cdot k}-
                    \frac{J_{p_1}^{\mu\nu}}
                        {p_1\cdot k}-
                    \frac{J_{p_2}^{\mu\nu}}
                        {p_2\cdot k}
                \Big]
            \bigg\}\
            S_0^{(\mu)}+
            \mathcal{O}(\omega_k)
    \end{split}
    \end{equation}
    \item for $e^-(p_1)+\gamma(p_2)\rightarrow\gamma(q_1)+ e^-(q_2)$:
    \begin{equation}\label{eq.3.1.13(b)}
    \begin{split}
        S_{\text{tree}}^{(\gamma)}
            (\omega_k,\hat{k})=\!
            e\ \bigg\{
                \Big[&
                    \frac{q_2\cdot
                            \epsilon_r^*(\vec{k})}
                        {q_2\cdot k}-
                    \frac{p_1\cdot
                            \epsilon_r^*(\vec{k})}
                        {p_1\cdot k}
                \Big]\!+
                i\epsilon^*_{r\mu}(\vec{k})\ 
                k_{\nu}
                \Big[
                    \frac{\bar{J}_{q_2}^{\mu\nu}}
                        {q_2\cdot k}-
                    \frac{J_{p_1}^{\mu\nu}}
                        {p_1\cdot k}
                \Big]
            \bigg\}\
            S_0^{(\gamma)}\!+\!
            \mathcal{O}(\omega_k)
    \end{split}
    \end{equation}
    \item for $e^-(p_1)+e^+(p_2)\rightarrow\gamma(q_1)+\gamma(q_2)$:
    \begin{equation}\label{eq.3.1.13(c)}
    \begin{split}
        S_{\text{tree}}^{(e)}
            (\omega_k,\hat{k})=\!
            e\ \bigg\{
                \Big[&
                    \frac{p_2\cdot
                            \epsilon_r^*(\vec{k})}
                        {p_2\cdot k}-
                    \frac{p_1\cdot
                            \epsilon_r^*(\vec{k})}
                        {p_1\cdot k}
                \Big]\!+
                i\epsilon^*_{r\mu}(\vec{k})\ 
                k_{\nu}
                \Big[
                    \frac{\bar{J}_{\bar{p}_2}^{\mu\nu}}
                        {p_2\cdot k}-
                    \frac{J_{p_1}^{\mu\nu}}
                        {p_1\cdot k}
                \Big]
            \bigg\}\
            S_0^{(e)}\!+\!
            \mathcal{O}(\omega_k)
    \end{split}
    \end{equation}
\end{itemize}
We emphasize that in the above expressions, the angular momentum operators act also on the energy-momentum conserving $\delta$ function included in $S_0^{(\alpha)}$, resulting in the appearance of $\delta$ function derivatives with respect to the momenta of the hard particles. The latter can be treated via integration by parts when computing rates and cross-sections, which involve integration in terms of the momenta of the external particles. 

\subsection{Subleading corrections to Faddeev-Kulish dressings}

We define final and initial dressed states, for the three processes we study, as follows \cite{Kulish:1970ut,Choi:2019rlz,Tomaras_2020}
{
\begin{equation}\label{eq.3.2.2}
\begin{split}
    \ket{q_1,q_2}_d&=
        \mathlarger{e^{
            \int\reallywidetilde{d^3k}\
            [(f_q(\vec{k})+g_q(\vec{k}))
                \cdot 
            a^{\dagger}(\vec{k})
            -h.c.]
        }}
        \ket{q_1,q_2}
    \\&=
        \Big\{
            1+
            \int\reallywidetilde{d^3k}\
                [(f_q(\vec{k})+g_q(\vec{k}))
                    \cdot 
                a^{\dagger}(\vec{k})
                -h.c.]\\& \hspace{3em}+
            \frac{1}{2}\Big(
            \int\reallywidetilde{d^3k}\
                [(f_q(\vec{k})+g_q(\vec{k}))
                    \cdot 
                a^{\dagger}(\vec{k})
                -h.c.]\Big)^2+\mathcal{O}(e^3)
        \Big\}
        \ket{q_1,q_2}
    \\ 
    \ket{p_1,p_2}_d&=
        \mathlarger{e^{
            \int\reallywidetilde{d^3k}\
            [(f_p(\vec{k})+g_p(\vec{k}))
                \cdot 
            a^{\dagger}(\vec{k})
            -h.c.]
        }}
        \ket{p_1,p_2}
    \\&=
        \Big\{
            1+
            \int\reallywidetilde{d^3k}\
                [(f_p(\vec{k})+g_p(\vec{k}))
                    \cdot 
                a^{\dagger}(\vec{k})
                -h.c.]\\& \hspace{3em}+
            \frac{1}{2}\Big(
            \int\reallywidetilde{d^3k}\
                [(f_p(\vec{k})+g_p(\vec{k}))
                    \cdot 
                a^{\dagger}(\vec{k})
                -h.c.]\Big)^2+\mathcal{O}(e^3)
        \Big\}
        \ket{p_1,p_2}
\end{split}
\end{equation}}%
where $\ket{q_1,q_2}, \ket{p_1,p_2}$ are two-particle Fock-basis states, which are taken to be hard. That is, their energies are greater than $E_d$, the infrared energy scale characterizing the soft photons in the clouds. Here, $f$ and $g$ denote the leading and subleading dressing functions. As in \eqref{eq.2.1.4}, the dot product between the subleading dressing function and the photon creation operator is given by
\begin{equation}\label{eq.3.2.3}
    g_q(\vec{k})
        \cdot 
    a^{\dagger}(\vec{k})=
        g_q^{\mu}
            (\vec{k})
        \sum_{r}
            \epsilon_{r\mu}^*
                (\vec{k})
            a_r^{\dagger}
                (\vec{k}),
    \hspace{3em}
    g_p(\vec{k})
        \cdot 
    a^{\dagger}(\vec{k})=
        g_p^{\mu}
            (\vec{k})
        \sum_{r}
            \epsilon_{r\mu}^*
                (\vec{k})
            a_r^{\dagger}
                (\vec{k})
\end{equation}
The subscript $q$ in the dressing functions indicates dependence on both the momenta of the hard outgoing particles, while $p$ indicates dependence on both the momenta of the hard incoming particles. As we will see, only the terms linear in the subleading dressing function $g$ (and, therefore, the the electron charge $e$) will be needed for the suppression of single photon-emitting amplitudes, where the photon has energy below $E_d$, at tree level. 

The leading dressing functions for the outgoing and incoming states are explicitly given as follows \cite{Kulish:1970ut, Tomaras_2020}:
\begin{itemize}
    \item for $e^-(p_1)+\mu^-(p_2)\rightarrow e^-(q_1)+\mu^-(q_2)$:
    \begin{equation}\label{eq.3.2.4(a)}
    \begin{split}
        f_q^{\mu}(\vec{k})&=
            e\bigg[
                \mathlarger{e^
                    {-iq_1\cdot kt_0/q_1^0}}
                \Big(
                    \frac{q_1^{\mu}}
                        {q_1\cdot k}
                    -c^{\mu}
                \Big)+
                \mathlarger{e^
                    {-iq_2\cdot kt_0/q_2^0}}
                \Big(
                    \frac{q_2^{\mu}}
                        {q_2\cdot k}
                    -c^{\mu}
                \Big)
            \bigg]
        \\
        f_p^{\mu}(\vec{k})&=
            e\bigg[
                \mathlarger{e^
                    {-ip_1\cdot kt_0/p_1^0}}
                \Big(
                    \frac{p_1^{\mu}}
                        {p_1\cdot k}
                    -c^{\mu}
                \Big)+
                \mathlarger{e^
                    {-ip_2\cdot kt_0/p_2^0}}
                \Big(
                    \frac{p_2^{\mu}}
                        {p_2\cdot k}
                    -c^{\mu}
                \Big)
            \bigg]
    \end{split}
    \end{equation}
    \item for $e^-(p_1)+\gamma(p_2)\rightarrow\gamma(q_1)+ e^-(q_2)$:
    \begin{equation}\label{eq.3.2.4(b)}
    \begin{split}
        f_q^{\mu}(\vec{k})&=
            e\
                \mathlarger{e^
                    {-iq_2\cdot kt_0/q_2^0}}
                \Big(
                    \frac{q_2^{\mu}}
                        {q_2\cdot k}
                    -c^{\mu}
                \Big)
        \\
        f_p^{\mu}(\vec{k})&=
            e\
                \mathlarger{e^
                    {-ip_1\cdot kt_0/p_1^0}}
                \Big(
                    \frac{p_1^{\mu}}
                        {p_1\cdot k}
                    -c^{\mu}
                \Big)
    \end{split}
    \end{equation}
    \item for $e^-(p_1)+e^+(p_2)\rightarrow\gamma(q_1)+\gamma(q_2)$:
    \begin{equation}\label{eq.3.2.4(c)}
    \begin{split}
        f_q^{\mu}(\vec{k})&=
            0
        \\
        f_p^{\mu}(\vec{k})&=
            e\bigg[
                \mathlarger{e^
                    {-ip_1\cdot kt_0/p_1^0}}
                \Big(
                    \frac{p_1^{\mu}}
                        {p_1\cdot k}
                    -c^{\mu}
                \Big)-
                \mathlarger{e^
                    {-ip_2\cdot kt_0/p_2^0}}
                \Big(
                    \frac{p_2^{\mu}}
                        {p_2\cdot k}
                    -c^{\mu}
                \Big)
            \bigg]
    \end{split}
    \end{equation}
\end{itemize}
Since the integrand in \eqref{eq.3.2.2} is dominated by low momenta, we take $E_d < 1/t_0$, approximating, thus, the phases $e^{-iq_1\cdot k t_0/q_1^0}$, etc,
in \eqref{eq.3.2.4(a)}, \eqref{eq.3.2.4(b)}, and \eqref{eq.3.2.4(c)} with unity. To obtain the subleading dressing functions, we follow \cite{Choi:2019rlz}: 
\begin{itemize}
    \item for $e^-(p_1)+\mu^-(p_2)\rightarrow e^-(q_1)+\mu^-(q_2)$:
    \begin{equation}\label{eq.3.2.5(a)}
    \begin{split}
        g_q^{\mu}(\vec{k})&=
            ie k_{\nu}
            \bigg[
                \mathlarger{e^
                    {-iq_1\cdot kt_0/q_1^0}}
                \frac{\bar{J}_{q_1}^{\mu\nu}}
                    {q_1\cdot k}+
                \mathlarger{e^
                    {-iq_2\cdot kt_0/q_2^0}}
                \frac{\bar{J}_{q_2}^{\mu\nu}}
                    {q_2\cdot k}
            \bigg]\\
        g_p^{\mu}(\vec{k})&=
            ie k_{\nu}
            \bigg[
                \mathlarger{e^
                    {-ip_1\cdot kt_0/p_1^0}}
                \frac{J_{p_1}^{\mu\nu}}
                    {p_1\cdot k}+
                \mathlarger{e^
                    {-ip_2\cdot kt_0/p_2^0}}
                \frac{J_{p_2}^{\mu\nu}}
                    {p_2\cdot k}
            \bigg]
    \end{split}
    \end{equation}
    \item for $e^-(p_1)+\gamma(p_2)\rightarrow\gamma(q_1)+ e^-(q_2)$:
    \begin{equation}\label{eq.3.2.5(b)}
    \begin{split}
        g_q^{\mu}(\vec{k})=
            ie k_{\nu}\
                \mathlarger{e^
                    {-iq_2\cdot kt_0/q_2^0}}
                \frac{\bar{J}_{q_2}^{\mu\nu}}
                    {q_2\cdot k}\\
        g_p^{\mu}(\vec{k})=
            ie k_{\nu}\
                \mathlarger{e^
                    {-ip_1\cdot kt_0/p_1^0}}
                \frac{J_{p_1}^{\mu\nu}}
                    {p_1\cdot k}
    \end{split}
    \end{equation}
    \item for $e^-(p_1)+e^+(p_2)\rightarrow\gamma(q_1)+\gamma(q_2)$:
    \begin{equation}\label{eq.3.2.5(c)}
    \begin{split}
        g_q^{\mu}(\vec{k})&=
            0
        \\
        g_p^{\mu}(\vec{k})&=
            ie k_{\nu}
            \bigg[
                \mathlarger{e^
                    {-ip_1\cdot kt_0/p_1^0}}
                \frac{J_{p_1}^{\mu\nu}}
                    {p_1\cdot k}-
                \mathlarger{e^
                    {-ip_2\cdot kt_0/p_2^0}}
                \frac{\bar{J}_{\bar{p}_2}^{\mu\nu}}
                    {p_2\cdot k}
            \bigg]
    \end{split}
    \end{equation}
\end{itemize}
where the phases appearing in \eqref{eq.3.2.5(a)}, \eqref{eq.3.2.5(b)}, and \eqref{eq.3.2.5(c)} can be approximated by unity. The total angular momentum operators $\bar{J}_{q_1},\ \bar{J}_{q_2},\ J_{p_1},\ J_{p_2}$, and $\bar{J}_{\bar{p}_2}$ are defined as in \eqref{eq.3.1.5} and \eqref{eq.3.1.6}. The orbital angular momentum operators, 
satisfy 
\begin{equation}\label{eq.3.2.7}
\begin{split}
    \bra{\Psi}
    L_p^{\mu\nu}
    \ket{p}=&
        +i\Big(
            p^{\mu}
            \frac{\partial}
                {\partial p_{\nu}}-
            p^{\nu}
            \frac{\partial}
                {\partial p_{\mu}}
        \Big)\
        \braket{\Psi|p}
    \\
    \bra{q}
    \bar{L}_q^{\mu\nu}
    \ket{\Psi}=&
        -i\Big(
            q^{\mu}
            \frac{\partial}
                {\partial q_{\nu}}-
            q^{\nu}
            \frac{\partial}
                {\partial q_{\mu}}
        \Big)\
        \braket{q|\Psi}
\end{split}
\end{equation}
and for the antiparticles 
\begin{equation}
\begin{split}
    \bra{\Psi}
    \bar{L}_{\bar{p}}^{\mu\nu}
    \ket{\bar{p}}=&+i\Big(
            p^{\mu}
            \frac{\partial}
                {\partial p_{\nu}}-
            p^{\nu}
            \frac{\partial}
                {\partial p_{\mu}}
        \Big)\
        \braket{\Psi|\bar{p}}
    \\
    \bra{\bar{q}}
    L_{\bar{q}}^{\mu\nu}
    \ket{\Psi}=&
        -i\Big(
            q^{\mu}
            \frac{\partial}
                {\partial q_{\nu}}-
            q^{\nu}
            \frac{\partial}
                {\partial q_{\mu}}
        \Big)\
        \braket{\bar{q}|\Psi}
\end{split}
\end{equation}
The spin angular momentum operators for charged particles, see  \eqref{eq.3.1.8} and \eqref{eq.3.1.9}, 
satisfy
\begin{equation}\label{eq.3.2.8(a)}
\begin{split}
    \bra{\Psi}
    S_p^{\mu\nu}
    \ket{p}=&
    \frac{i}{4}[\gamma^{\mu},\gamma^{\nu}]\
        u(p)\
        \circ\
        \frac{\partial}
            {\partial u(p)}\
        \braket{\Psi|p}
    \\
    \bra{q}
    \bar{S}_q^{\mu\nu}
    \ket{\Psi}=&
    \bar{u}(q)\
        \frac{i}{4}[\gamma^{\mu},\gamma^{\nu}]\
        \circ\
        \frac{\partial}
            {\partial \bar{u}(q)}\
        \braket{q|\Psi}
\end{split}
\end{equation}
and for antiparticles
\begin{equation}\label{eq.3.2.8(b)}
\begin{split}
    \bra{\Psi}
    \bar{S}_{\bar{p}}^{\mu\nu}
    \ket{\bar{p}}=&-
    \bar{\upsilon}(p)\
        \frac{i}{4}[\gamma^{\mu},\gamma^{\nu}]\
        \circ\
        \frac{\partial}
            {\partial \bar{\upsilon}(p)}\
        \braket{\Psi|\bar{p}}
    \\
    \bra{\bar{q}}
    S_{\bar{q}}^{\mu\nu}
    \ket{\Psi}=&-
    \frac{i}{4}[\gamma^{\mu},\gamma^{\nu}]\
        \upsilon(q)\
        \circ\
        \frac{\partial}
            {\partial \upsilon(q)}\
        \braket{\bar{q}|\Psi}
\end{split}
\end{equation}
$\ket{\Psi}$ stands for an arbitrary state, and 
\begin{equation}\label{eq.3.2.9}
\begin{split}
    \ket{q}&=b^{\dagger}(\vec{q})\ket{0}
    ,\ \hspace{5em}
    \ket{p}=b^{\dagger}(\vec{p})\ket{0}
    \\
    \ket{\bar{q}}&=d^{\dagger}(\vec{q})\ket{0}
    ,\ \hspace{5em}
    \ket{\bar{p}}=d^{\dagger}(\vec{p})\ket{0}
\end{split}
\end{equation}
are single particle/antiparticle states. 

\subsection{Equivalence of dressed and traditional amplitudes}
    We apply the arguments in \cite{Choi:2019rlz} to the case of the three elastic scattering processes, in order to validate the equivalence between the associated  dressed amplitudes and the infrared-finite part of the conventional Fock-basis amplitudes, to order $e^4$ in perturbation theory. Recall that at the one-loop level (order $e^4$), the Fock-basis amplitudes suffer from logarithmic infrared divergences due to virtual soft photons in the loop, with energies below or equal to the infrared reference scale $\Lambda$ ($\lambda \leq E_d \le \Lambda$) \cite{Weinberg:1965nx}. 

The dressed S-matrix elements, associated with the three elastic processes, are given by
\begin{equation}\label{eq.3.3.1}
    \Tilde{S}^{(\alpha)}=
    \tensor[_d]{\braket{q_1,q_2|
        S|
    p_1,p_2}}{_d}
\end{equation}
where $\alpha=\mu,\ \gamma$ or $e$, corresponding, respectively, to an electron scattering off a muon, a photon, or an electron annihilating with a positron. Since the non-trivial part of the S-matrix is of order $e^2$ and we are interested to establish the infrared finiteness of the dressed amplitudes to order $e^4$, we expand the incoming and outgoing dressed states to order $e^2$. Taking into account the fact that the soft cloud-photon annihilation operators annihilate the corresponding Fock-basis state, and omitting terms quadratic in the cloud-photon creation operators, since these will produce contributions of order $e^6$ to the dressed S-matrix elements, the relevant terms are given by
\begin{equation}\label{eq.3.3.2}
\begin{split}
    \ket{q_1,q_2}_d&=
    \Big(
        1+\!
        \int\reallywidetilde{d^3k}\,
            \Big\{\big[f_q(\Vec{k})+g_q(\Vec{k})\big]
                \!\cdot\!
            a^{\dagger}(\Vec{k})
        -\frac{1}{2}
            \big[f^*_q(\Vec{k})+g^*_q(\Vec{k})\big]
                \!\cdot\!
            \big[f_q(\Vec{k})+g_q(\Vec{k})\big]
        \\&\hspace{16.25em}
        +\mathcal{O}(e^3)
    \Big\}\Big)
    \ket{q_1,q_2}
    \\
    \ket{p_1,p_2}_d&=
        \Big(
        1+\!
        \int\reallywidetilde{d^3k}\,
            \Big\{\big[f_p(\Vec{k})+g_p(\Vec{k})\big]
                \!\cdot\!
            a^{\dagger}(\Vec{k})
        -\frac{1}{2}
            \big[f^*_p(\Vec{k})+g^*_p(\Vec{k})\big]
                \!\cdot\!
            \big[f_p(\Vec{k})+g_p(\Vec{k})\big]
        \\&\hspace{16.25em}
        +\mathcal{O}(e^3)
    \Big\}\Big)
    \ket{p_1,p_2}
\end{split}
\end{equation}
To order $e^4$, the dressed S-matrix elements can, therefore, be written as follows
\begin{equation}\label{eq.3.3.3}
\begin{split}
    \Tilde{S}^{(\alpha)}=
    \bra{q_1,q_2}
        \Big(1\!+\!\!
            \int\!\reallywidetilde{d^3k}\,
                \Big\{\!
                \big[f_q^*(\vec{k})\!+\!g_q^*(\vec{k})\big]
                    \!\cdot\!
                a(\vec{k})\!-\!
                \frac{1}{2}\big[
                    f_q^*(\vec{k})\!+\!g_q^*(\vec{k})\big]
                    \!\cdot \!
                \big[f_q(\vec{k})\!+\!g_q(\vec{k})\big]
                \!\Big\}
            \Big)&
    \\
    S
        \Big(1\!+\!\!
            \int\!\reallywidetilde{d^3k'}\,
                \Big\{\!
                \big[f_p(\pvec{k}')\!+\!g_p(\pvec{k}')\big]
                    \!\cdot\!
                a^{\dagger}(\pvec{k}')\!-\!
                \frac{1}{2}\big[
                    f_p(\pvec{k}')\!+\!g_p(\pvec{k}')\big]
                    \!\cdot\! 
                \big[f_p^*(\pvec{k}')\!+\!g_p^*(\pvec{k}')\big]
                \!\Big\}
            \Big)&
        \ket{p_1,p_2}
\end{split}
\end{equation}
This, in turn, can be written as a sum of conventional Fock-basis S-matrix elements, some of which involve the emission or absorption of soft photons from the clouds. Each such S-matrix element involves the corresponding energy and momentum conserving delta function. In applying the subleading soft-photon theorem to these contributions, we must take into account the full dependence on the soft photon momentum and expand the $\delta$ function appropriately, as we have discussed in the previous section. Also, in implementing the action of the angular momentum operators appearing in the subleading dressing functions, we must take into account the action on the $\delta$ functions appearing in the S-matrix elements.

Let us compute all contributions to the dressed S-matrix element in \eqref{eq.3.3.3}, to order $e^4$: 
\begin{enumerate}[label=(\alph*)]
    \item The first contribution is the conventional Fock-basis S-matrix element
    \begin{equation*}
        S^{(\alpha)}=\bra{q_1,q_2}S\ket{p_1,p_2}
    \end{equation*}
    to be computed to order $e^4$. The cloud photons do not contribute to this part. 
    Recall that the dressed charged states can be thought of as superpositions of multiparticle states, involving arbitrary numbers of soft cloud photons, including the state with no such photons. 
    See the diagrams (a) and (b) of \figref{fig.2.3.1} for examples of the relevant Feynman diagrams associated with this contribution, for the case of electron-muon scattering.
    \item Next, there are three contributions that describe the emission and the absorption of a single cloud photon by an outgoing or an incoming charged particle, and the propagation of an initial state cloud photon to a final state cloud photon. 
    \begin{itemize}
        \item The contributions describing the emission and absorption of a single soft cloud photon are given, respectively, by
        \begin{gather*}
            \bra{q_1,q_2}
            \int\!\reallywidetilde{d^3k}\,
                \big[f_q^*(\vec{k})+g_q^*(\vec{k})\big]
                    \cdot
                a(\vec{k})\, 
            S
            \ket{p_1,p_2}
            \\
            \bra{q_1,q_2}
            S\,
            \int\!\reallywidetilde{d^3k}\,
                \big[f_p(\pvec{k})+g_p(\pvec{k})\big]
                    \cdot
                a^{\dagger}(\pvec{k})
            \ket{p_1,p_2}
        \end{gather*}
        As we mentioned before, the energy-momentum conserving delta function is present in the S-matix elements, imposing $p_1+p_2-q_1-q_2 \mp k=0$, including the soft cloud photon momentum.
        Using the subleading soft photon theorem, which can be expressed in the following form,
        \begin{equation*}
        \begin{split}
            \braket{q_1,q_2|\
                a_r(\Vec{k})\ S
            |p_1,p_2}=
                \big[f_q(\Vec{k})+g_q(\Vec{k})
                        -f_p(\Vec{k})-g_p(\Vec{k})\big]
                    \cdot
                \epsilon_r^*(\Vec{k})\ 
                S^{(\alpha)}
            \hspace{1em}
            \\
            \braket{q_1,q_2|\
                S\ a_r^{\dagger}(\Vec{k})
            |p_1,p_2}=
                \big[f_p^*(\Vec{k})+g_p^*(\Vec{k})
                    -f_q^*(\Vec{k})-g_q^*(\Vec{k})\big]
                    \cdot
                \epsilon_r(k)\ 
                S^{(\alpha)}
            \hspace{0.75em}
        \end{split}
        \end{equation*}
        we obtain for the two contributions
        \begin{gather*}
            \int\reallywidetilde{d^3k}\
                [f_q^*(\Vec{k})+g_q^*(\Vec{k})]
                    \cdot
                [f_q(\Vec{k})+g_q(\Vec{k})-f_p(\Vec{k})-g_p(\Vec{k})]
            \, S^{(\alpha)}
            \\
            \int\reallywidetilde{d^3k}\
                [f_p(\Vec{k})+g_p(\Vec{k})]
                    \cdot
                [f_p^*(\Vec{k})+g_p^*(\Vec{k})-f_q^*(\Vec{k})-g_q^*(\Vec{k})]
            \, S^{(\alpha)}
        \end{gather*}
        Since the product of the dressing functions is of order $e^2$, the non-trivial part of the conventional Fock basis elastic S-matrix element $S^{(\alpha)}$ must be computed at the tree level, i.e. at order $e^2$. The angular momentum operators present in the subleading dressing functions $g$ act on $S^{(\alpha)}$, including the $\delta$-function factor.
        The diagrams of \figref{fig.2.3.1} (c) and (d) describe the emission of a single cloud photon, while the diagrams (e) and (f) describe the absorption of a single cloud photon, during the dressed electron-muon scattering process.
        \item The third contribution involving the free propagation of a cloud photon arises from the S-matrix element
        \begin{gather*}
            \bra{q_1,q_2}
            \int\!\reallywidetilde{d^3k}\,
                \big[f_q^*(\vec{k})+g_q^*(\vec{k})\big]
                    \cdot
                a(\vec{k})\,
            S
            \int\reallywidetilde{d^3k'}\,
                \big[f_p(\pvec{k}')+g_p(\pvec{k}')\big]
                    \cdot
                a^{\dagger}(\pvec{k}')
            \ket{p_1,p_2}
        \end{gather*}
        The relevant Feynman diagram is a disconnected one, involving a free photon propagator and a tree-level elastic scattering diagram, contributing to the non-trivial part of the S-matrix. The contributions for which the two cloud photons interact with external charged particles result in terms of order $e^6$; hence we ignore them. Using the commutation relation for the soft photon creation and annihilation operators, the  contribution of order $e^4$ can be reduced to
        \begin{gather*}
            \int\reallywidetilde{d^3k}\
            [f_q^*(\Vec{k})+g_q^*(\Vec{k})]
                \cdot
            [f_p(\Vec{k})+g_p(\Vec{k})]
            \ S^{(\alpha)}
        \end{gather*}
        The diagram of \figref{fig.2.3.1} (g) depicts a representative Feynman diagram for this contribution. Here also, the Fock basis elastic S-matrix element $S^{(\alpha)}$ must be computed at the tree level, i.e. at order $e^2$.
    \end{itemize}
    For all three cases, the completeness relation for the polarization vectors was used (see \hyperref[sec:A]{Appendix A}), taking also into account the fact that the dressing functions are transverse. 
    \item Finally, there are two extra contributions that are quadratic in the dressing functions. They describe single photon exchanges between the clouds of the outgoing charged particles and single photon exchanges between the clouds of the incoming charged particles, respectively. They are given by 
        \begin{align*}
            -\frac{1}{2}
                \int\reallywidetilde{d^3k}\
                [f_q^*(\Vec{k})+g_q^*(\vec{k})]
                    \cdot
                [f_q(\Vec{k})+g_q(\Vec{k})]\,
            S^{(\alpha)}
            \\
            -\frac{1}{2}
                \int\reallywidetilde{d^3k}\
                [f_p(\Vec{k})+g_p(\Vec{k})]
                    \cdot
                [f_p^*(\Vec{k})+g_p^*(\Vec{k})]\,
            S^{(\alpha)}
        \end{align*}
        Since both of these contributions are quadratic in the dressing functions, the conventional Fock basis elastic S-matrix element $S^{(\alpha)}$ must be computed to order $e^2$. 
        See representative examples of Feynman diagrams associated with these contributions, for the electron-muon process, in \figref{fig.2.3.1} (h) and (i).
\end{enumerate}
Collecting all contributions up to order $e^4$, the non-trivial part of the dressed S-matrix element can be written as
\begin{equation}\label{eq.3.3.4}
\begin{split}
    \tilde{S}^{(\alpha)} = 
        \bigg(1+\frac{1}{2}
        \int\reallywidetilde{d^3k}\
            \Big\{&
            [f_q^*(\vec{k})\!+\!g_q^*(\vec{k})]
                \cdot
            [f_q(\vec{k})\!+\!g_q(\vec{k})]+
            [f_p^*(\vec{k})\!+\!g_p^*(\vec{k})]
                \cdot
            [f_p(\vec{k})\!+\!g_p(\vec{k})]\\&-
            2\,
            [f_q^*(\vec{k})\!+\!g_q^*(\vec{k})]
                \cdot
            [f_p(\vec{k})\!+\!g_p(\vec{k})]
            \Big\}
        \bigg)
        \,S^{(\alpha)}
\end{split}
\end{equation}
where $S^{(\alpha)}$ is the corresponding undressed S-matrix element to be computed to order $e^4$. The terms that are linear and quadratic in the subleading dressing functions will yield suppressed contributions of order $\mathcal{O}(E_d)$ upon integrating over the soft-photon momentum. Since the infrared energy scale $E_d$ can be taken to be very small compared to all other hard energy scales, we can neglect these contributions.
To see that a term involving a subleading dressing function is actually of order $\mathcal{O}(E_d)$, consider the following representative example  
\begin{equation}
\begin{split}
        \int\reallywidetilde{d^3k}\
            f_q^*(\Vec{k})
                \cdot
            g_q(\Vec{k})
            \, S^{(\alpha)}&\sim
        \int\reallywidetilde{d^3k}\
            \frac{(q_{1\mu})(ik_{\nu}\bar{J}_{q_1}^{\mu\nu})}
                {(q_1\cdot k)(q_1\cdot k)}\,
        S^{(\alpha)}
        \\&\sim
        \frac{1}{2}
        \int
            \frac{d\Omega_k}
                {(2\pi)^3}
            \frac{(q_{1\mu})(i\hat{k}_{\nu}\bar{J}_{q_1}^{\mu\nu})}
                {(q_1^0-\vec{q}_1\cdot\hat{k})
                (q_1^0-\vec{q}_1\cdot\hat{k})}\,
        S^{(\alpha)}
        \int_{\lambda}^{E_d}
            \omega_kd\omega_k
            \frac{1}{\omega_k}
\end{split}
\end{equation}
where $\hat{k}=\vec{k}/\omega_k$. Performing the integral over $\omega_k$, and taking the $\lambda\to0$ limit, reveals that this contribution is of order $\mathcal{O}(E_d)$, and hence it can be neglected. More precisely, the term will be suppressed with ratios of the energy scale $E_d$ with hard energy scales, which are taken to be very small\footnote{Recall that in the limit $\lambda \to 0$, we keep the ratios $E_d/\Lambda$, $E_d/m_e$ and $\Lambda/m_e$ fixed with $E_d/m_e,\, \Lambda/m_e << 1$.}. 
Similarly, one can show that all terms involving products of the leading dressing functions with the subleading ones are of order $\mathcal{O}(E_d)$. Moreover, the terms involving a product of two subleading dressing functions yield even more suppressed contributions in the limit $\lambda\to0$, since they are of order $\mathcal{O}(E_d^2)$. 
Hence, to order $e^4$, we obtain
\begin{equation}
\tilde{S}^{(\alpha)} = S^{(\alpha)}+\frac{1}{2}\int\reallywidetilde{d^3k}\
            [f_q(\vec{k})-f_p(\vec{k})]
                \cdot
            [f_q^*(\vec{k})-f_p^*(\vec{k})] 
            \ S_0^{(\alpha)}\, + \mathcal{O}(E_d)
\end{equation} 
where $S_0^{(\alpha)}$ denotes the non-trivial part of the undressed elastic S-matrix element at tree-level. The first factor $S^{(\alpha)}$ must be computed up to the one-loop level, equivalently, up to order $e^4$.  

\begin{figure}[!tbp]
    \begin{subfigure}[t]{.32\linewidth}
        \centering
        \input{Figures/Fig5/Fig_5a}
    \caption{}
    \end{subfigure}
    \hfill
    \begin{subfigure}[t]{.32\linewidth}
        \centering
        \input{Figures/Fig5/Fig_5b}
    \caption{}
    \end{subfigure}
    \hfill
    \begin{subfigure}[t]{.32\linewidth}
        \centering
        \input{Figures/Fig5/Fig_5d}
    \caption{}
    \end{subfigure}
    \\ \vspace{1em}
    \begin{subfigure}[t]{.32\linewidth}
        \centering
        \input{Figures/Fig5/Fig_5d2}
    \caption{}
    \end{subfigure}
    \hfill
    \begin{subfigure}[t]{.32\linewidth}
        \centering
        \input{Figures/Fig5/Fig_5c}
    \caption{}
    \end{subfigure}
    \hfill
    \begin{subfigure}[t]{.32\linewidth}
        \centering
        \input{Figures/Fig5/Fig_5c2}
    \caption{}
    \end{subfigure}
    \\ \vspace{1em}
    \begin{subfigure}[t]{.32\linewidth}
        \centering
        \input{Figures/Fig5/Fig_5e}
    \caption{}
    \end{subfigure}
    \hfill
    \begin{subfigure}[t]{.32\linewidth}
        \centering
        \input{Figures/Fig5/Fig_5f}
    \caption{}
    \end{subfigure}
    \hfill
    \begin{subfigure}[t]{.32\linewidth}
        \centering
        \input{Figures/Fig5/Fig_5f2}
    \caption{}
    \end{subfigure}
    \caption{Representative diagrams depicting the elastic \texorpdfstring{$e^-\mu^-\rightarrow e^-\mu^-$}{TEXT} process, in which both interacting particles are accompanied by clouds of soft photons. }
    \label{fig.2.3.1}
\end{figure}

The one-loop contributions to the undressed elastic S-matrix element $S^{(\alpha)}$
are infrared divergent. These divergences are due to soft virtual photons running in the loops of the Feynman diagrams associated with the three elastic cases. Taking into account all possible ways a soft virtual photon propagator can be attached to external fermion lines (see \figref{fig.2.3.1} (b)) and utilizing the leading soft photon theorem, 
to order $e^4$, the undressed elastic S-matrix element can 
be written as \cite{Weinberg:1965nx}
\begin{equation}
S^{(\alpha)}=
    S^{(\alpha,\Lambda)} 
    -\frac{1}{4}
            \int\frac{d\Omega_k}{(2\pi)^3}
                \int_{\lambda}^{\Lambda}d\omega_k\,\omega_k\
            [f_q(\vec{k})-f_p(\vec{k})]
                \cdot
            [f_q^*(\vec{k})-f_p^*(\vec{k})]\ 
        S_0^{(\alpha)}
\end{equation}
where $\smash{S^{(\alpha,\Lambda)}}$ is the finite part of the undressed elastic S-matrix element without the contributions from the virtual soft photons. The integral now is over the soft virtual photon momentum. Any contributions from the subleading soft theorem yield suppressed contributions of order $\Lambda$ upon integration over the soft photon momentum and hence they can be ignored.
The second term in the right hand side is infrared divergent, and cancels exactly the infrared divergent contributions from the dressing factors in the limit $\lambda \to 0$ with $\Lambda/E_d$ kept fixed. To see this, note that
\begin{equation}
\begin{split}
    \frac{1}{2}
        \int\reallywidetilde{d^3k}\
            [&f_q(\vec{k})-f_p(\vec{k})]
                \cdot
            [f_q^*(\vec{k})-f_p^*(\vec{k})]
        \\&=
    \frac{1}{4}
        \int
            \frac{d\Omega_k}{(2\pi)^3}
            \int_{\lambda}^{\Lambda}
                d\omega_k\,\omega_k\ 
                    [f_q(\vec{k})-f_p(\vec{k})]
                        \cdot
                    [f_q^*(\vec{k})-f_p^*(\vec{k})]
        \\&\hspace{1em}+
        \frac{1}{4}
        \int
            \frac{d\Omega_k}{(2\pi)^3}
            \int_{\Lambda}^{E_d}
                d\omega_k\,\omega_k\ 
                    [f_q(\vec{k})-f_p(\vec{k})]
                        \cdot
                    [f_q^*(\vec{k})-f_p^*(\vec{k})]
\end{split}
\end{equation}
where
\begin{equation}
    \frac{1}{4}
        \int
            \frac{d\Omega_k}{(2\pi)^3}
            \int_{\Lambda}^{E_d}
                d\omega_k\,\omega_k\ 
                    [f_q(\vec{k})-f_p(\vec{k})]
                        \cdot
                    [f_q^*(\vec{k})-f_p^*(\vec{k})]
    =\mathcal{O}
        \Big[\ln\Big(\frac{E_d}{\Lambda}\Big)\Big]
\end{equation}

Therefore, to order $e^4$, the dressed S-matrix element reduces to the infrared-finite part of the corresponding undressed one, up to suppressed terms of order $E_d$, and terms of order $\ln{(E_d/\Lambda)}$: 
\begin{equation}\label{eq.3.3.5}
    \Tilde{S}^{(\alpha)}=
        S^{(\alpha,\Lambda)}
        +\mathcal{O}(E_d) 
        +\mathcal{O}
        \Big[\ln\Big(\frac{E_d}{\Lambda}\Big)\Big]
\end{equation}
Taking the reference scale $\Lambda$ to be equal to $E_d$, we get
\begin{equation}\label{eq.3.3.6}
    \Tilde{S}^{(\alpha)}=
        S^{(\alpha,E_d)}
        +\mathcal{O}(E_d)        
\end{equation}
Hence, we have established the equivalence (to fourth order in $e$) between the dressed elastic S-matrix elements and the infrared-finite part of the corresponding conventional Fock-basis S-matrix elements.
This equivalence holds up to very suppressed power-law corrections in the infrared energy scale $E_d$.

\subsection{Soft radiation suppression at the tree level}
    We now use the dressings defined in \eqref{eq.3.2.2} to demonstrate that, to order $e^3$ in the electron coupling, the emission of an additional soft photon with energy smaller than $E_d$ during the scattering of dressed charged states is highly suppressed.
We denote the momentum of the extra radiative photon by $\Vec{k}_{\gamma}$, $|\Vec{k}_{\gamma}|=\omega_{\gamma}\leq E_d$, and the polarization vector by $\epsilon^{\mu}_r(\Vec{k}_{\gamma}),\ r=1,2$. The dressed S-matrix elements corresponding to the soft photon emission are given by
\begin{equation}\label{eq.3.4.1}
    \tilde{S}^{(\alpha)}=
        \tensor[_d]{\braket{q_1,q_2;\gamma|
            S|
        p_1,p_2}}{_d}
\end{equation}
We would like to calculate the non-trivial part to leading non-vanishing order in perturbation theory. The calculation involves tree-level Feynman diagrams, which yield contributions of order $e^3$. To obtain these contributions, it is sufficient to expand the initial and final states to linear order in the electron charge $e$.

To order $e$, the final state is given by 
\begin{equation}
\ket{q_1,q_2;\gamma}_d=-\!\Big(f_q^*(\vec{k}_{\gamma}) + g_q^*(\vec{k}_{\gamma})\Big)
            \!\cdot\!
\epsilon_r(\vec{k}_{\gamma})\ket{q_1,q_2}
+\!
    \Big(
        1\!+\!\!\int\reallywidetilde{d^3k}\
            [f_q(\vec{k})+g_q(\vec{k})]
                \!\cdot\!
            a^{\dagger}(\vec{k})
    \Big)\ket{q_1,q_2;\gamma}
\end{equation}
as obtained by letting the dressing operator, including the subleading correction to the dressing function,
act on the Fock basis state $\ket{q_1,q_2;\gamma}=a^{\dagger}_r(\Vec{k}_{\gamma})\ket{q_1,q_2}$. Likewise, to linear order in the electron charge, the initial state is given by
\begin{equation}
\ket{p_1,p_2}_d=\Big(
        1\!+\!\!\int\reallywidetilde{d^3k}\
            [f_p(\vec{k})+g_p(\vec{k})]
                \!\cdot\!
            a^{\dagger}(\vec{k})
    \Big)\ket{p_1,p_2}
\end{equation}
Taking into account that the emitted additional photon is soft, the non-trivial part of the dressed S-matrix element at order $e^3$ splits
into three contributions
\begin{equation}\label{eq.3.4.3}
\begin{split}
    \Tilde{S}_{\text{tree}}^{(\alpha)}=
    \Tilde{S}_{\text{tree}}^{(\alpha,1)}+
    \Tilde{S}_{\text{tree}}^{(\alpha,2)}+
    \Tilde{S}_{\text{tree}}^{(\alpha,3)}
\end{split}
\end{equation}
where 
\begin{gather*}
    \Tilde{S}_{\text{tree}}^{(\alpha,1)}=
    -f_q(\Vec{k}_{\gamma})
        \cdot
    \epsilon^*_r(\Vec{k}_{\gamma})
    \ {S}_0^{(\alpha)}
    \\
    \Tilde{S}_{\text{tree}}^{(\alpha,2)}=
    -g_q(\Vec{k}_{\gamma})
        \cdot
    \epsilon^*_r(\Vec{k}_{\gamma})
    \ S_0^{(\alpha)}
\end{gather*}
These two contributions arise from the following part of the S-matrix element
\begin{equation}
-\!\Big(f_q(\vec{k}_{\gamma}) + g_q(\vec{k}_{\gamma})\Big)
            \!\cdot\!
\epsilon_r^*(\vec{k}_{\gamma})\bra{q_1,q_2}S\Big(
        1\!+\!\!\int\reallywidetilde{d^3k}\
            [f_p(\vec{k})+g_p(\vec{k})]
                \!\cdot\!
            a^{\dagger}(\vec{k})
    \Big)\ket{p_1,p_2}
\end{equation}
The terms quadratic in the dressing functions yield higher order contributions than $e^3$, and can, therefore, be omitted. The third contribution arises from
\begin{equation}
    \begin{split}
        \Tilde{S}_{\text{tree}}^{(\alpha,3)}=
    \bra{q_1,q_2}a_r(\vec{k}_{\gamma})
            \Big(
                1+\int\reallywidetilde{d^3k}\
                    [f_q^*(\vec{k})+g_q^*(\vec{k})]
                        \cdot
                    a(\vec{k})
                \Big)S\
            &
        \\
            \Big(
            1+\int\reallywidetilde{d^3k}\
                   [f_p(\vec{k})+ g_p(\vec{k})]
                        \cdot
                    a^{\dagger}(\vec{k})
                \Big)&
            \ket{p_1,p_2}
    \end{split}
    \end{equation}
In \hyperref[sec:D]{Appendix C}, we show in detail that to order $e^3$, this yields the third contribution
\begin{equation*}
    \Tilde{S}^{(\alpha,3)}_{\text{tree}}=
        [f_q(\Vec{k}_{\gamma})+
        g_q(\Vec{k}_{\gamma})]
            \cdot
        \epsilon^*_r(\Vec{k}_{\gamma})
        \ {S}_0^{(\alpha)} + \mathcal{O}(\omega_\gamma)
\end{equation*}
Therefore, adding the three  contributions yields
\begin{equation}\label{eq.3.4.4}
    \Tilde{S}_{\text{tree}}^{(\alpha)}=
        \mathcal{O}(E_d)
\end{equation}
since $\omega_\gamma \leq E_d$.
In other words, during the scattering of dressed charged states, the emission of a soft photon with energy less than $E_d$ is vanishingly small, modulo power-law corrections in the infrared scale $E_d$, at least at tree-level. 

We conclude that, \textit{by appropriately extending the dressing functions to subleading order in the soft momentum expansion, the emission of soft photons with energy smaller than the characteristic energy of the cloud photons is highly suppressed in QED scattering processes}\footnote{
    This result can be extended to other gauge theories and gravity \cite{Choi:2019rlz,Tropper:2024kxy}. Since the subleading soft theorem is universal, radiation suppression from subleading soft dressings is expected to take place in any theory with long-range interactions mediated by massless gauge bosons and/or gravitons. 
}. 
As a result, the soft, deep infrared sector of QED, consisting of the cloud photons that accompany the hard charged particles, is highly correlated with the hard sector, since the latter fixes the former. The entanglement between any additional soft radiation with energies above $E_d$ with the hard particles is suppressed, as shown in \cite{Tomaras_2020,Irakleous:2021ggq,Toumbas:2023qbo}.

We proceed now to check that the dressed amplitudes 
reproduce the Bloch-Nordsieck inclusive cross sections of the Fock-basis formalism. Recall that these are finite, free of any infrared divergences, order by order in perturbation theory \cite{Weinberg:1965nx}. We will compute the rate for the emission of any number of soft photons of total energy less than $E_T$ during the scattering process $\alpha$ involving dressed states. We take the total energy $E_T > E_d$ to be sufficiently small.
We will assume that the emission of soft photons of energy less than $E_d$ is suppressed. This has been shown to hold at tree level, but here we will assume that this suppression continues to hold at higher orders in perturbation theory, as well. Thus, the emitted soft photons have energy between $E_d$ and $E_T$. The total inclusive rate is given by
\cite{Weinberg:1965nx,Choi:2019rlz} 
\begin{equation}\label{eq.3.4.5}
    \tilde{\Gamma}^{(\alpha)}(\le E_T)=
        \frac{1}{\pi}
        \sum_{N=0}^{\infty}
        \int_{E_d}^{E_T}\!\!d\omega_1
        ...
        \int_{E_d}^{E_T}\!\!d\omega_N
        \int_{-\infty}^{\infty}\!\! d\sigma\ 
            \frac{\sin (E_T\sigma)}{\sigma}
        \exp\bigg\{
            i\sigma\sum_{i=1}^N\omega_i
        \bigg\}\
        \tilde{\Gamma}^{(\alpha)}
            (\omega_1,...,\omega_N)
\end{equation}
where $\tilde{\Gamma}^{(\alpha)}(\omega_1, ..., \omega_N)$ are the rates for the emission of $N$ soft photons with energies $\omega_1$, ..., $\omega_N$, during the scattering process $\alpha$ between dressed states. The rates for the emission of any number of soft photons during scattering of Fock basis states have been computed in \cite{Weinberg:1965nx}. Note that since the energies of the $N$ soft photons $\omega_1$, ..., $\omega_N$ are greater than $E_d$, the corresponding creation/annihilation operators commute with the creation/annihilation operators of the photons in the clouds that accompany the charged particles. Moreover, since these photons are soft, we can apply the soft theorem and the results of \cite{Weinberg:1965nx} apply to the dressed rates $\tilde{\Gamma}^{(\alpha)}(\omega_1, ..., \omega_N)$, as well. As in the previous section, subleading contributions to the soft theorem will yield suppressed contributions, of order $\mathcal{O}(E_T-E_d)$, upon integration over the soft-photon energies $\omega_1$, ..., $\omega_N$. Therefore, we ignore these contributions.

Using the results of \cite{Weinberg:1965nx},
and computing the integrals and the sum, we obtain
\begin{equation}\label{eq.3.4.6}
    \tilde{\Gamma}^{(\alpha)}(\le E_T)=
        \Big(\frac{E_T}{E_d}\Big)
            ^{2\mathcal{B}^{(\alpha)}}\
        b\big(
            2\mathcal{B}^{(\alpha)}
        \big)\
        \tilde{\Gamma}^{(\alpha)}
\end{equation}
where $b(x)=1-\frac{1}{12}\pi^2 x^2+...\ $. The dressing has suppressed the emission of soft photons of energies less than $E_d$. Since the dressed rates are finite, we see that the dressed inclusive rate is finite.  
\begin{itemize}
    \item For the $e^-(p_1)+\mu^-(p_2)\rightarrow e^-(q_1)+\mu^-(q_2)$ scattering process, we have
    \begin{equation*}
        \mathcal{B}^{(\mu)}=
        -\frac{e^2}{8\pi^2}
            \bigg[
                4+
                \frac{4}{\upsilon}
                    \ln\Big(
                        \frac{1+\upsilon}
                            {1-\upsilon}
                    \Big)-
                \frac{2}{u}
                    \ln\Big(
                        \frac{1+u}
                            {1-u}
                    \Big)-
                \frac{1}{\omega}
                    \ln\Big(
                        \frac{1+\omega}
                            {1-\omega}
                    \Big)-
                \frac{1}{w}
                    \ln\Big(
                        \frac{1+w}
                            {1-w}
                    \Big)
            \bigg]
    \end{equation*}
    with the relative velocities given by
    \begin{gather*}
        \upsilon^2=1-\frac{m^2M^2}{(p_1\cdot p_2)^2}
            =1-\frac{m^2M^2}{(q_1\cdot q_2)^2},\\
        u^2=1-\frac{m^2M^2}{(p_1\cdot q_2)^2}
            =1-\frac{m^2M^2}{(q_1\cdot p_2)^2},\\
        \omega^2=1-\frac{m^4}{(p_1\cdot q_1)^2},
        \hspace{2.5em}
        w^2=1-\frac{M^4}{(p_2\cdot q_2)^2}
    \end{gather*}
    \item for the $e^-(p_1)+\gamma(p_2)\rightarrow\gamma(q_1)+ e^-(q_2)$ scattering process,
    \begin{equation*}
        \mathcal{B}^{(\gamma)}=
        -\frac{e^2}{8\pi^2}
            \bigg[
                2-
                \frac{1}{\upsilon}
                    \ln\Big(
                        \frac{1+\upsilon}
                            {1-\upsilon}
                    \Big)
            \bigg]
    \end{equation*}
    with the relative velocity given by
    \begin{gather*}
        \upsilon^2=1-\frac{m^4}{(p_1\cdot q_2)^2}
            =1-\frac{m^4}{(p_2\cdot q_1)^2}
    \end{gather*}
    \item and for the $e^-(p_1)+e^+(p_2)\rightarrow\gamma(q_1)+\gamma(q_2)$ scattering process:
    \begin{equation*}
        \mathcal{B}^{(e)}=
        -\frac{e^2}{8\pi^2}
            \bigg[
                2-
                \frac{1}{\upsilon}
                    \ln\Big(
                        \frac{1+\upsilon}
                            {1-\upsilon}
                    \Big)
            \bigg]
    \end{equation*}
    with the relative velocity given by
    \begin{gather*}
        \upsilon^2=1-\frac{m^4}{(p_1\cdot p_2)^2}
            =1-\frac{m^4}{(q_1\cdot q_2)^2}
    \end{gather*}
\end{itemize}
In the above expressions, we have denoted the electron mass with $m$ and the muon mass with $M$. $\tilde{\Gamma}^{(\alpha)}$ is the dressed rate for the elastic scattering process $\alpha$. As we have seen in the previous section, modulo suppressed contributions of order $E_d$, this is given by the infrared finite part of the undressed rate, without contributions from soft virtual photons carrying energies below $\Lambda$:
\begin{equation}\label{eq.3.4.7}
\tilde{\Gamma}^{(\alpha)}=\Gamma^{(\alpha,\Lambda)}
\end{equation}

Taking $\Lambda = E_d$, the dressed inclusive rates, then, take the form
\begin{equation}\label{eq.3.4.8}
    \tilde{\Gamma}^{(\alpha)}(\le E_T)=
        \Big(\frac{E_T}{E_d}\Big)
            ^{2\mathcal{B}^{(\alpha)}}\
        b\big(2\mathcal{B}^{(\alpha)}\big)\
        \Gamma^{(\alpha,E_d)}
\end{equation}
Thus, choosing $\Lambda$ to be equal to the characteristic cloud energy scale $E_d$, we recover the Bloch-Nordsieck inclusive rate, as obtained in the conventional Fock-state basis \cite{Weinberg:1965nx}. 


It is worth emphasizing that the suppression of soft radiation with energy below $E_d$ during dressed state scattering processes has been established at tree level in perturbation theory. It would be nice to see whether this suppression continues to hold at the loop level. For this to be true, it is likely that the dressing functions need to be suitably corrected. Indeed, the subleading soft photon theorem receives corrections at the loop level that diverge logarithmically as $\omega_k\rightarrow 0$ \cite{Bern:2014oka,He:2014bga,Mao:2017wvx,Sahoo:2018lxl,Laddha:2018myi,Saha:2019tub,Sahoo:2019yod,Sahoo:2020ryf,Delisle:2020uui,Krishna:2023fxg,Bianchi:2014gla}. In \hyperref[sec:conc]{Section 4}, we briefly discuss how to correct the subleading dressing functions at order $e^2$ 
so that the logarithmically infrared-divergent terms are removed from radiative amplitudes. 
To the best of our knowledge, the correction of the subleading soft dressings
at the loop level remains an open problem. 

\section{Conclusions}
    \label{sec:conc}
    In this paper, we analyzed soft-photon emission during explicit QED scattering processes involving dressed charged states. 
Specifically, we focused on three representative examples — electron-muon scattering, Compton scattering (electron-photon scattering), and electron-positron annihilation into two photons. We constructed the dressed states to subleading order in the soft momentum expansion, with the subleading dressing function given to first order in the QED coupling constant. Using the dressed states, we obtained both the elastic and single photon emitting scattering amplitudes. We demonstrated that the elastic amplitudes are equivalent to the infrared-finite parts of the conventional Fock-basis amplitudes. Moreover, we have shown that the radiative amplitudes are vanishingly small when the energy of the emitted photon is smaller than the characteristic energy of the cloud photons that accompany the incoming and outgoing charged particles.
This result strengthens the proposition that the deep infrared and hard sectors of the space of asymptotic scattering states are highly correlated, as revealed also by the large entanglement between the two sectors \cite{Carney:2017oxp,Carney:2017jut,Tomaras_2020,Irakleous:2021ggq,Toumbas:2023qbo}. Our explicit calculations complement the formal analysis of \cite{Choi:2019rlz}, demonstrating this correlation in concrete QED scattering processes. On the other hand, the entanglement of hard states with emitted soft photons that carry energy greater than $E_d$ is small \cite{Tomaras_2020}.
Since emission of soft photons with energies less than $E_d$ is suppressed during scattering of dressed states at tree level, the dressed state formalism, at this order in perturbation theory, produces the same inclusive rates as the Bloch-Nordsieck method. Whether deviations arise at the loop level is yet to be determined. 

As discussed in \secref{sec:3}, conventional photon-emitting amplitudes admit further logarithmic divergences at the loop level as the soft energy vanishes \cite{Bern:2014oka,He:2014bga,Mao:2017wvx,Sahoo:2018lxl,Laddha:2018myi,Saha:2019tub,Sahoo:2019yod,Sahoo:2020ryf,Delisle:2020uui,Krishna:2023fxg,Bianchi:2014gla}. Thus, at the loop level, it is important to investigate whether the S-matrix elements describing photon-emitting processes continue to be well-defined by employing the dressed state formalism. Indeed, this appearance of logarithmic infrared divergences at higher orders in perturbation theory necessitates further corrections of the subleading dressing functions 
to second and higher orders in the electron coupling. Constructing such dressings and computing the associated amplitudes will be an important step toward a well-defined, infrared finite S-matrix in QED, and it is a promising direction for future study.

It is also worth noting that recent work \cite{Campiglia:2019wxe} establishes the loop-corrected subleading soft-photon theorem in QED as a Ward identity of quantum-corrected asymptotic charges \cite{AtulBhatkar:2019vcb}. This naturally raises the question of whether the corresponding loop-corrected dressings yield eigenstates of these charges. In other words, do loop-corrected subleading dressings also define states with definite soft charge, including quantum corrections? Addressing this question would help clarify the role of loop-corrected FK dressings within the dressed-state formalism.

Applications of the dressed state formalism have also been explored in gravitational theories \cite{Choi:2019sjs} and in the context of quantum information \cite{Gabai:2016kuf, Carney:2017jut,Grignani:2016igg}.
In particular, such ideas could be applied to processes involving the formation and evaporation of black holes \cite{Strominger:2017aeh,Toumbas:2023qbo}. Perhaps they could prove useful in resolving Hawking’s information paradox, where a pure state describing the collapsing matter seemingly evolves into a mixed thermal state of Hawking quanta \cite{Strominger:2017aeh}. Recent proposals suggest that the inclusion of soft photons and gravitons — the so-called ‘soft hair’ of black holes — could be relevant for partially restoring the purity of the final state \cite{Hawking:2016msc, Strominger:2017aeh}.


Building on the perspective of studying FK states in gravitational scattering processes, another promising next step is to investigate whether analogous subleading dressings can be systematically constructed in gravitational scattering, thereby providing explicit realizations of the gravitational S-matrix beyond the leading soft approximation. Natural processes to explore in this direction include soft graviton emission during matter scattering, as well as pure $2\rightarrow2$ graviton amplitudes. 
These examples offer concrete tests for the role of soft gravitons in rendering the S-matrix infrared finite, and for exploring how correlations between soft and hard sectors manifest in gravity. Carrying out such explicit calculations may clarify whether gravitational subleading dressings can play a role analogous to their QED counterparts.

\section*{Acknowledgements}
    \label{sec:acknowledgements}
    S.C. wishes to to thank P. Hager and S. Choi for useful discussions on the definition of a factorized form of a 
spin angular momentum operator for spin 1/2 particles. This work was partially supported by the Cyprus Research and Innovation Foundation grant EXCELLENCE/0421/0362. 

\appendix
\section{Notation and conventions}
    \label{sec:A}
    Throughout this work, we use a mostly plus signature metric. Moreover, we adopt the following expansions for the free fields \cite{Srednicki:2007qs}
\begin{equation}\label{eq.A.1}
\begin{split}
    A^{\mu}(x)=\sum_{\lambda=0}^3
        \int\frac{d^3k}{(2\pi)^3(2\omega_k)}
        \Big[
            \epsilon^{\mu}_{\lambda}(\vec{k})
                a_{\lambda}(\vec{k})
                e^{ik\cdot x}+
            \epsilon^{*\mu}_{\lambda}(\vec{k})
                a^{\dagger}_{\lambda}(\vec{k})
                e^{-ik\cdot x}
        \Big]\hspace{-0.675em}
    \\
    \psi(x)=\sum_{s=\pm}
        \int\frac{d^3p}{(2\pi)^3(2\omega_p)}
        \Big[
            u_s(\vec{p})b_s(\vec{p})
                e^{ip\cdot x}+
            \upsilon_s(\vec{p})d^{\dagger}_s(\vec{p})
                e^{-ip\cdot x}
        \Big]
    \\
    \Bar{\psi}(x)=\sum_{s=\pm}
        \int\frac{d^3p}{(2\pi)^3(2\omega_p)}
        \Big[
            \Bar{u}_s(\vec{p})b^{\dagger}_s(\vec{p})
                e^{-ip\cdot x}+
            \Bar{\upsilon}_s(\vec{p})d_s(\vec{p})
                e^{ip\cdot x}
        \Big]
\end{split}
\end{equation}
where $\omega_p=\sqrt{|\Vec{p}|^2+m^2}$ and $\omega_k=|\Vec{k}|$. The four polarization vectors $\epsilon^{\mu}_{\lambda}(k),\ \lambda=0,...,3$, satisfy the following orthonormality and completeness relations
\begin{equation*}
    \epsilon_{\lambda}(\vec{k})
        \cdot\epsilon^*_{\lambda'}(\vec{k})=
            \zeta_\lambda\delta_{\lambda\lambda'}
    \ ,\hspace{2em} 
    \sum_{\lambda=0}^3\epsilon^{\mu}_{\lambda}(\vec{k})
        \epsilon^{*\nu}_{\lambda}(\vec{k})=
            \eta^{\mu\nu}
\end{equation*}
where $\zeta_1=\zeta_2=\zeta_3=-\zeta_0=1$. When restricted to  transversely polarized photons, the completeness relation reduces to
\begin{equation}
    \sum_{\lambda=1,2}\epsilon^{\mu}_{\lambda}(\vec{k})
        \epsilon^{*\nu}_{\lambda}(\vec{k})=
            \eta^{\mu\nu}-
            k^{\mu}c^{\nu}-
            k^{\nu}c^{\mu},
    \hspace{5em}
    c^{\mu}=\frac{1}{2k^0}
        (-1,\hat{k})
\end{equation}
We further choose to work in the Lorenz gauge, $\partial_{\mu}A^{\mu}=0$, with the free electromagnetic gauge field satisfying $\Box A^{\mu}=0$. The photon creation and annihilation operators are compliant with the usual commutation relations
\begin{equation*}
    \big[a_{\lambda}(\vec{k}),a^{\dagger}_{\lambda'}(\Vec{k}')\big]=
        (2\pi)^3\ (2\omega_k)\ 
        \zeta_{\lambda}
        \delta_{\lambda\lambda'}\
        \delta^3(\Vec{k}-\Vec{k}')
\end{equation*}
In the quantum theory, we impose the Gupta-Bleuler condition:
\begin{equation}
    \big[a_0(\Vec{k})-a_3(\Vec{k})\big]\ket{\Psi}=0
\end{equation}
for an arbitrary Fock basis state $\ket{\Psi}$. Finally, the spinors in \eqref{eq.A.1} satisfy
\begin{equation}
\begin{split}
    \Bar{u}_{s'}(\vec{p})u_{s}(\vec{p})=+2m\delta_{s's}
    \\
    \Bar{\upsilon}_{s'}(\vec{p})\upsilon_s(\vec{p})=-2m\delta_{s's}
\end{split}
\end{equation}
with the remaining combinations $\Bar{u}_{s'}(\vec{p})\upsilon_s(\vec{p})$ and $\Bar{\upsilon}_{s'}(\vec{p})u_{s}(\vec{p})$ vanishing, and 
\begin{equation}
\begin{split}
    \sum_{s=\pm}
        u_s(\vec{p})\Bar{u}_s(\vec{p})=
            -\slashed{p}+m
    \\
    \sum_{s=\pm}
        \upsilon_s(\vec{p})\Bar{\upsilon}_s(\vec{p})=
            -\slashed{p}-m
\end{split}
\end{equation}
The electron/positron creation and annihilation operators satisfy the following anticommutation relations
\begin{equation*}
    \big\{b_r(\vec{p}),b_s^{\dagger}(\vec{p}')\big\}=
        (2\pi)^3\ (2\omega_p)\
            \delta_{rs}\
            \delta^3(\Vec{p}-\pvec{p}'),
    \hspace{2em}
    \big\{d_r(\vec{p}),d_s^{\dagger}(\vec{p}')\big\}=
        (2\pi)^3\ (2\omega_p)\
            \delta_{rs}\
            \delta^3(\Vec{p}-\pvec{p}')
\end{equation*}
Similar spinor and anticommutation relations are satisfied by the muon/anti-muon spinors and creation and annihilation operators.
\section{Review of the subleading soft photon theorem}
    \label{sec:B}
    In this appendix, we review the derivation of the subleading soft photon theorem for an arbitrary scattering process in QED. Following \cite{Gell-Mann:1954wra,Low:1954kd,Low:1958sn,Burnett:1967km},
we study single soft-photon emission during an arbitrary QED scattering process, 
expanding the amplitudes in powers of the photon momentum. Our goal is to obtain an explicit expression for the subleading soft-photon theorem.

Let us emphasize some important properties of the subleading soft-photon theorem. While the leading term in the soft expansion depends only on the momentum and the charge of the incoming and outgoing charged particles, the subleading term also depends on the spin. 
This latter property, and hence the expression of the subleading soft factor, is also universal, describing soft-photon emission irrespective of the matter content of the theory. Similar theorems govern the emission of soft gravitons in perturbative quantum gravity and the emission of soft gluons in non-Abelian gauge theories.

Let the tree-level amplitude, describing an elastic interaction of $n$ incoming fermionic particles and $m$ outgoing ones, be denoted by $i\mathcal{M}_0$. It will be useful to define the amplitude $i\bar{\mathcal{T}}(p)$ by removing the wave-function factor of an incoming fermion with momentum $\vec{p}$. Similarly, we define $i\mathcal{T}(q)$ by removing the wave-function factor of an outgoing fermion with momentum $\vec{q}$. Therefore, we may write
\begin{equation}\label{eq.B.0.1}
    i\mathcal{M}_0=
        \Big[i\bar{\mathcal{T}}(p)\Big]
            _{\alpha}\ u^{\alpha}(\vec{p})=
        \bar{u}^{\alpha}(\vec{q}) 
            \Big[i\mathcal{T}(q)\Big]_{\alpha} 
\end{equation}
where 
$(\alpha=1,...,4)$ is a spinor index 
while the spin polarization indices $(s=\pm)$ have been suppressed for notational simplicity. 
A sum over repeated spinor indices is implied unless explicitly stated. We focus on a scattering process with no incoming or outgoing antiparticles or hard photons. The generalization to these cases is straightforward. 

\begin{figure}[!htbp]
    \centering
    \begin{subfigure}[t]{.3\linewidth}
        \centering
        \includegraphics
            [width=.75\linewidth,scale=1]
                {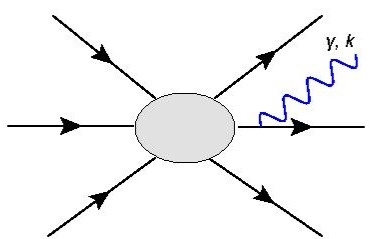}
    \caption{}
    \end{subfigure}
    \begin{subfigure}[t]{.3\linewidth}
        \centering
        \includegraphics
            [width=.75\linewidth,scale=1]
                {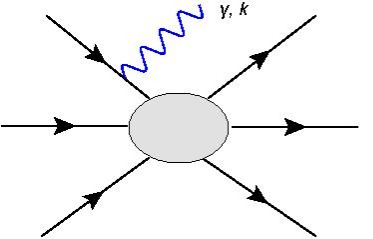}
    \caption{}
    \end{subfigure}
    \begin{subfigure}[t]{.3\linewidth}
        \centering
        \includegraphics
            [width=.75\linewidth,scale=1]
                {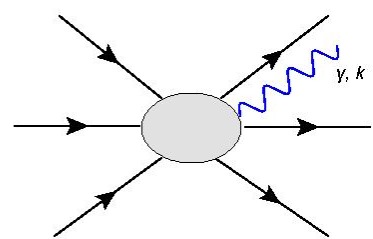}
    \caption{}
    \end{subfigure}
    \caption{Feynman diagrams depicting single photon emission during a generic QED scattering process. 
    The first two diagrams describe soft-photon emission from an outgoing and an incoming fermion, respectively. The last diagram corresponds to photon emission from a virtual fermion.}
    \label{fig.B.1}
\end{figure}

Now suppose that an outgoing soft photon is added to the final state. The amplitude describing the emission 
is given by \cite{Bloch:1937pw,Gell-Mann:1954wra,Low:1954kd,Low:1958sn,Weinberg:1965nx,Burnett:1967km,Weinberg:1995mt,Srednicki:2007qs,Luo:2014wea,Bern:2014vva,Strominger:2017zoo,AtulBhatkar:2018kfi,Beneke:2021ilf,Beneke:2021umj,Travaglini:2022uwo} 
\begin{equation}\label{eq.B.0.2}
\begin{split}
    i\mathcal{M}_{\text{tree}}(\omega_k,\hat{k})=
        &\sum_{i=1}^m
        \bar{u}^{\alpha}(\vec{q}_i)
        \bigg\{
            \slashed{\epsilon}^*_{\lambda}(\vec{k})
            \frac{e_i(-\slashed{q}_i-\slashed{k}+m_i)}
                {(+2q_i\cdot k)}
        \bigg\}_{\alpha\beta}\
        \Big[i\mathcal{T}(q_i+k)\Big]_{\beta}\\&+
        \sum_{i=1}^n
        \Big[i\bar{\mathcal{T}}(p_i-k)\Big]_a\ 
        \bigg\{ 
            \frac{e_i(-\slashed{p}_i+\slashed{k}+m_i)}
                {(-2p_i\cdot k)}
            \slashed{\epsilon}^*_{\lambda}(\vec{k})
        \bigg\}_{\alpha\beta}\
        u^{\beta}(\vec{p}_i)+
        i\mathcal{N}(\omega_k,\hat{k})
\end{split}
\end{equation}
where $\vec{k}$ is the momentum carried by the emitted photon and $\epsilon_{\lambda}^*(\vec{k}),\ \lambda=1,2$ is the polarization vector. 
The first sum includes the contributions to the amplitude when the photon is emitted by an outgoing fermion of momentum $q_i$, mass $m_i$ and charge $e_i,\ i=1,...,m$. In the second sum, the contributions arise due to emission by 
an incoming fermion of momentum $p_i$, mass $m_i$ and charge $e_i,\ i=1,...,n$. The last term 
accounts for the possibility of photon emission by a virtual fermion \cite{Burnett:1967km,Luo:2014wea}. See \figref{fig.B.1}. 

In the soft momentum limit, we expand $i\mathcal{M}_{\text{tree}}(\omega_k,\hat{k})$ in powers of the photon momentum. The leading contribution to the soft factor arises from terms associated with photon emission by external on-shell charged particles. These terms give rise to simple poles at $\omega_k=0$. $i\mathcal{N}(\omega_k,\hat{k})$, on the other hand, remains finite and non-singular as $\omega_k\rightarrow0$, since it describes soft emission by internal, off-shell, virtual fermions. Neglecting all non-singular terms, we derive the leading soft photon theorem,
\begin{equation}\label{eq.B.0.3}
    i\mathcal{M}_{\text{tree}}(\omega_k,\hat{k})=
        \bigg[
        \sum_{i=1}^m
        e_i\frac{q_i\cdot
                \epsilon^*_{\lambda}(\vec{k})}
            {q_i\cdot k}-
        \sum_{i=1}^n
        e_i\frac{p_i\cdot
                \epsilon^*_{\lambda}(\vec{k})}
            {p_i\cdot k}
        \bigg]\
        i\mathcal{M}_0+\mathcal{O}(\omega_k^0)
\end{equation}
Notice that, to leading order in $k$, $i\bar{\mathcal{T}}(p_i-k)\ u(\vec{p}_i)\approx \bar{u}(\vec{q}_i)\ i\mathcal{T}(q_i-k)\approx i\mathcal{M}_0$. 

To derive the subleading soft-photon theorem, we must also keep the next-to-leading terms, which are non-singular. Therefore, we write
\begin{equation}\label{eq.B.0.4}
\begin{split}
    i\mathcal{M}_{\text{tree}}(\omega_k,\hat{k})=
        \bigg[
        &\sum_{i=1}^m
        e_i\frac{q_i\cdot
                \epsilon^*_{\lambda}(\vec{k})}
            {q_i\cdot k}-
        \sum_{i=1}^n
        e_i\frac{p_i\cdot
                \epsilon^*_{\lambda}(\vec{k})}
            {p_i\cdot k}
        \bigg]\
        i\mathcal{M}_0\\&+
        \sum_{i=1}^m
        e_i\frac{q_i\cdot
            \epsilon^*_{\lambda}(\vec{k})}
            {q_i\cdot k}\
        \bar{u}^{\alpha}(\vec{q}_i)
        k\cdot \frac{\partial}
            {\partial q_i}
        \Big[i\mathcal{T}(q_i)\Big]
            _{\alpha}
        \\&\hspace{5em}+
        i\sum_{i=1}^m
        e_i\frac{\epsilon^*_{\lambda\mu}(\vec{k})\
                k_{\nu}}
            {q_i\cdot k}\
        \bar{u}^{\alpha}(\vec{q}_i)\
            \big(\Sigma^{\mu\nu}\big)
                _{\alpha\beta}
        \Big[i\mathcal{T}(q_i)\Big]
            _{\beta}\\&+
        \sum_{i=1}^n
        e_i\frac{p_i\cdot
                \epsilon^*_{\lambda}(\vec{k})}
            {p_i\cdot k}\
        k\cdot \frac{\partial}
            {\partial p_i}
        \Big[i\bar{\mathcal{T}}(p_i)\Big]
            _{\alpha}
        u^{\alpha}(\vec{p}_i)
        \\&\hspace{5em}-
        i\sum_{i=1}^n
        e_i\frac{\epsilon^*_{\lambda\mu}(\vec{k})\
                k_{\nu}}
            {p_i\cdot k}\
        \Big[i\bar{\mathcal{T}}(p_i)\Big]
            _{\alpha}\
            \big(\Sigma^{\mu\nu}\big)
                _{\alpha\beta}
        u^{\beta}(\vec{p}_i)\\&
        +i\mathcal{N}(\omega_k,\hat{k})
        +\mathcal{O}(\omega_k)
\end{split}
\end{equation}
where $(\Sigma^{\mu\nu}) _{\alpha\beta} =\frac{i}{4}[\gamma^{\mu},\gamma^{\nu}] _{\alpha\beta}$ is the spin operator in a four-dimensional spinorial representation of the Lorentz algebra 
\cite{AtulBhatkar:2018kfi}. Note that the derivatives with respect to the hard momenta in \eqref{eq.B.0.4} act only on the stripped amplitudes, defined in \eqref{eq.B.0.1}, and not on the fermion wave-functions. The spin operator $(\Sigma^{\mu\nu})_{\alpha\beta}$ acts only on the spinor wavefunctions associated with the outgoing and the incoming fermions. Furthermore, at this order in the soft momentum expansion, we must consider diagrams describing photon emission by virtual charged particles. Their contribution $i\mathcal{N}(\omega_k,\hat{k})$ can be written as
\begin{equation}\label{eq.B.0.5}
    i\mathcal{N}(\omega_k,\hat{k})=
        \epsilon^*_{\lambda}(\vec{k})
            \cdot
        \mathcal{J}(\omega_k,\hat{k})
\end{equation}
where $\mathcal{J}^{\mu}(\omega_k,\hat{k})$ is a current-like quantity.

We now impose invariance under gauge transformations, in order to fix the form of $i\mathcal{N}(\omega_k,\hat{k})$. At leading order, gauge invariance implies charge conservation,
\begin{equation}\label{eq.B.0.6}
    \sum_{i=1}^me_i-\sum_{i=1}^ne_i=0
\end{equation}
without leading to any further restrictions for the leading soft theorem. At subleading order, however, it implies that \cite{Burnett:1967km,Luo:2014wea,Travaglini:2022uwo}
\begin{equation}\label{eq.B.0.7}
    k\cdot \mathcal{J}(\omega_k,\hat{k})=
        -\sum_{i=1}^m 
            \bar{u}^{\alpha}(\vec{q}_i)\
            k\cdot\frac{\partial}
                {\partial q_i}
            \Big[i\mathcal{T}(q_i)\Big]
                _{\alpha}
        -\sum_{i=1}^n 
            k\cdot\frac{\partial}
                {\partial p_i}\
            \Big[i\bar{\mathcal{T}}(p_i)\Big]
                _{\alpha}
            u^{\alpha}(\vec{p}_i)
\end{equation}
where we have used the antisymmetry of the spin operator, $k_{\mu}k_{\nu}(\Sigma^{\mu\nu})_{\alpha\beta}=0$. Gauge invariance determines $\mathcal{J}^{\mu}(\omega_k,\hat{k})$, up to transverse terms $\upsilon^{\mu}$ that satisfy $k\cdot\upsilon=0$ \cite{Burnett:1967km,Luo:2014wea,Travaglini:2022uwo}. However, no such terms, which are also local in $k$, are possible. Stripping $k_{\mu}$ from $k\cdot \mathcal{J}(\omega_k,\hat{k})$ yields the exact form of the contribution from diagrams for which the soft photon is emitted by an internal fermion line
\begin{equation}\label{eq.B.0.8}
    i\mathcal{N}(\omega_k,\hat{k})=
        -\sum_{i=1}^m 
            \bar{u}^{\alpha}(\vec{q}_i)\
            \epsilon_{\lambda}^*(\vec{k})
            \cdot\frac{\partial}
                {\partial q_i}
            \Big[i\mathcal{T}(q_i)\Big]
                _{\alpha}
        -\sum_{i=1}^n 
            \epsilon_{\lambda}^*(\vec{k})
            \cdot\frac{\partial}
                {\partial p_i}\
            \Big[i\bar{\mathcal{T}}(p_i)\Big]
                _{\alpha}
            u^{\alpha}(\vec{p}_i)
\end{equation}
Substituting into \eqref{eq.B.0.4} yields
\begin{equation}\label{eq.B.0.9}
\begin{split}
    i\mathcal{M}_{\text{tree}}(\omega_k,\hat{k})=
        \bigg[
        &\sum_{i=1}^m
        e_i\frac{q_i\cdot
                \epsilon^*_{\lambda}(\vec{k})}
            {q_i\cdot k}-
        \sum_{i=1}^n
        e_i\frac{p_i\cdot
                \epsilon^*_{\lambda}(\vec{k})}
            {p_i\cdot k}
        \bigg]\
        i\mathcal{M}_0\\&+
        i\sum_{i=1}^m
        e_i\frac{\epsilon^*_{\lambda\mu}(\vec{k})\
                k_{\nu}}
            {q_i\cdot k}\
        \bar{u}^{\alpha}(\vec{q}_i)
        \bar{L}^{\mu\nu}_i
        \Big[i\mathcal{T}(q_i)\Big]
            _{\alpha}
        \\&\hspace{5em}+
        i\sum_{i=1}^m
        e_i\frac{\epsilon^*_{\lambda\mu}(\vec{k})\
                k_{\nu}}
            {q_i\cdot k}\
        \bar{u}^{\alpha}(\vec{q}_i)\
            \big(\Sigma^{\mu\nu}\big)
                _{\alpha\beta}
        \Big[i\mathcal{T}(q_i)\Big]
            _{\beta}\\&-
        i\sum_{i=1}^n
        e_i\frac{\epsilon^*_{\lambda\mu}(\vec{k})\
                k_{\nu}}
            {p_i\cdot k}\
        L^{\mu\nu}_i
        \Big[i\bar{\mathcal{T}}(p_i)\Big]
            _{\alpha}
        u^{\alpha}(\vec{p}_i)
        \\&\hspace{5em}-
        i\sum_{i=1}^n
        e_i\frac{\epsilon^*_{\lambda\mu}(\vec{k})\
                k_{\nu}}
            {p_i\cdot k}\
        \Big[i\bar{\mathcal{T}}(p_i)\Big]
            _{\alpha}\
            \big(\Sigma^{\mu\nu}\big)
                _{\alpha\beta}
        u^{\beta}(\vec{p}_i)\\&
        +\mathcal{O}(\omega_k)
\end{split}
\end{equation}
where the orbital angular momentum operators for outgoing and incoming particles, respectively, are given by
\begin{equation}\label{eq.B.0.10}
\begin{split}
    \bar{L}^{\mu\nu}_i=
        -i\Big(
            q_i^{\mu}
            \frac{\partial}
                {\partial q_{i\nu}}-
            q_i^{\nu}
            \frac{\partial}
                {\partial q_{i\mu}}
        \Big)
    \\
    L^{\mu\nu}_i=
        +i\Big(
            p_i^{\mu}
            \frac{\partial}
                {\partial p_{i\nu}}-
            p_i^{\nu}
            \frac{\partial}
                {\partial p_{i\mu}}
        \Big)
\end{split}
\end{equation}
In analogy with the definition of the gluonic spin operator employed in \cite{Bern:2014vva}, we define the factorizable differential operators, 
\begin{equation}\label{eq.B.0.11}
\begin{split}
    \Bar{S}_i^{\mu\nu}=
        \bar{u}^{\alpha}(\vec{q}_i)\
        \frac{i}{4}[\gamma^{\mu},\gamma^{\nu}]
            _{\alpha\beta}\
        \frac{\partial}
            {\partial \bar{u}^{\beta}(\vec{q}_i)}
    \\
    S_i^{\mu\nu}=
        \frac{i}{4}[\gamma^{\mu},\gamma^{\nu}]
            _{\alpha\beta}\
        u^{\beta}(\vec{p}_i)\
        \frac{\partial}
            {\partial u^{\alpha}(\vec{p}_i)}
\end{split}
\end{equation}
for outgoing and for incoming fermions, where
\begin{equation}\label{eq.B.0.12}
    \frac{\partial}
        {\partial \bar{u}^{\alpha}(\vec{q}_i)}
    \bar{u}^{\beta}(\vec{q}_i)=
    \delta^{\beta}_{\alpha}
    \hspace{5em}
    \frac{\partial}
        {\partial u^{\alpha}(\vec{p}_i)}
    u^{\beta}(\vec{p}_i)=
    \delta^{\beta}_{\alpha}
\end{equation}
with $\delta$ being the usual Kronecker symbol. 
Derivatives with respect to wave-functions give zero when they act on wave-functions of other fermionic particles. 

Using the above results, the subleading soft photon theorem can be expressed in the following convenient factorized form
\begin{equation}\label{eq.B.0.13}
\begin{split}
    i\mathcal{M}_{\text{tree}}(\omega_k,\hat{k})=
        \bigg[
        \sum_{i=1}^m
        e_i\frac{\epsilon^*_{\lambda\mu}(\vec{k})}
            {q_i\cdot k}\
        \big(
            q^{\mu}_i\!+\!
            ik_{\nu}\bar{J}^{\mu\nu}_i
        \big)-
        \sum_{i=1}^n
        e_i\frac{\epsilon^*_{\lambda\mu}(\vec{k})}
            {p_i\cdot k}\
        \big(
            p^{\mu}_i\!+\!
            ik_{\nu}J^{\mu\nu}_i
        \big)
        \bigg]\,
        i\mathcal{M}_0
        +\mathcal{O}(\omega_k)
\end{split}
\end{equation}
where
\begin{equation}\label{eq.B.0.14}
    \bar{J}_i^{\mu\nu}=
        \bar{L}_i^{\mu\nu}+
        \bar{S}_i^{\mu\nu}
    \hspace{7em}
    J_i^{\mu\nu}=
        L_i^{\mu\nu}+
        S_i^{\mu\nu}
\end{equation}
are the total angular momentum operators for outgoing and incoming fermions. The factorization of the amplitudes in the soft momentum limit, including the first subleading order in the soft momentum expansion, is often referred to in the literature as the LBK theorem \cite{Burnett:1967km,Travaglini:2022uwo}. In expressing the subleading soft-photon theorem in its factorized form, we pulled the derivatives with respect to the hard particles' momenta past the fermion wave-functions, with the understanding that the orbital angular momentum operators do not act on the latter \cite{Luo:2014wea}. Namely,
\begin{equation}\label{eq.B.0.15}
    \bar{L}_i^{\mu\nu}\bar{u}(\vec{q}_i)=0
        \hspace{9em}
    L_i^{\mu\nu}u(\vec{p}_i)=0
\end{equation}
This is in line with the fact that the orbital angular momentum of a particle should not depend on its spin degrees of freedom.

As we discussed in \secref{sec:3}, \eqref{eq.B.0.13} provides a relation between  single photon-emitting invariant amplitudes and the corresponding elastic ones. For the full S-matrix element describing soft-photon emission, we need to include the energy and momentum conserving $\delta$ function. Restricting our attention to $2\to2$ processes involving two incoming and two outgoing charged particles of charge $e$ for simplicity, 
like processes described in \secref{sec:3}, the non-trivial part of the S-matrix element for single photon emission at tree level is given by
\begin{equation}
    S_{\rm tree}(\omega_k,\hat{k})=
        (2\pi)^4
        \delta^{(4)}
            (p_1+p_2-q_1-q_2-k)
        \,i\mathcal{M}_{\rm tree}(\omega_k,\hat{k})
\end{equation}

The subleading soft photon theorem applies also to the full S-matrix element. To see this, we expand the full S-matrix element, including the $\delta$ function in powers of the components of the soft-photon momentum $k$. For the subleading contributions, we need to expand the $\delta$ function to linear order in $k$. Taking into account the expansion of the invariant amplitude $i\mathcal{M}_{\rm tree}(\omega_k,\hat{k})$ to subleading order in $k$, 
\begin{equation}
    i\mathcal{M}_{\rm tree}(\omega_k,\hat{k})=
        \omega_k^{-1}\,i\mathcal{M}_{\rm L}(\hat{k})+
        i\mathcal{M}_{\rm SL}(\hat{k})+
        \mathcal{O}(\omega_k)
\end{equation}
we obtain
\begin{equation}\label{eq:S_matrix_el_expansion}
\begin{split}
    S_{\rm tree}(\omega_k,\hat{k})&=
        (2\pi)^4
        \,\delta^{(4)}
            (p_1+p_2-q_1-q_2)
        \,\omega_k^{-1}
        \,i\mathcal{M}_{\rm L}(\hat{k})
    \\&\hspace{1em}+
        (2\pi)^4
        \,\delta^{(4)}
            (p_1+p_2-q_1-q_2)
        \,i\mathcal{M}_{\rm SL}(\hat{k})
    \\&\hspace{1em}+
        (2\pi)^4
        \,k_\nu
        \Big[\frac{\partial}{\partial k_\nu}
        \delta^{(4)}
                (p_1+p_2-q_1-q_2-k)
            \Big|_{k=0}
        \Big]
        \,\omega_k^{-1}
        \,i\mathcal{M}_{\rm L}(\hat{k})
    \\&\hspace{1em}+\mathcal{O}(\omega_k)
\end{split}
\end{equation}
Using relations between the partial derivatives of the $\delta$ function with respect to $k$ and the partial derivatives of the $\delta$ function with respect to the hard particle momenta, e.g.
\begin{equation}
\begin{split}
    \frac{\partial}{\partial k_{\nu}}
        \delta^{(4)}
            (p_1+p_2-q_1-q_2-k)\Big|_{k=0}&=-
    \frac{\partial}{\partial p_{1\nu}}
        \delta^{(4)}
            (p_1+p_2-q_1-q_2)\\&=+
    \frac{\partial}{\partial q_{1\nu}}
        \delta^{(4)}
            (p_1+p_2-q_1-q_2)\\&=\dots
\end{split}
\end{equation}
we can write the last term in \eqref{eq:S_matrix_el_expansion} as follows:
\begin{equation}\label{eq:derivatives_on_delta}
\begin{split}
    ie \mathcal{M}_0\, \epsilon_{r\mu}^*(\vec{k}) \,k_{\nu}\hspace{31.5em}\\
    \times\bigg[
        (-i)
        \frac{q_1^{\mu}}
            {q_1\cdot k}
        \frac{\partial}
            {\partial q_{1\nu}}\!+\!
        (-i)
        \frac{q_2^{\mu}}
            {q_2\cdot k}
        \frac{\partial}
            {\partial q_{2\nu}}\!-\!
        i
        \frac{p_1^{\mu}}
            {p_1\cdot k}
        \frac{\partial}
            {\partial p_{1\nu}}\!-\!
        i
        \frac{p_2^{\mu}}
            {p_2\cdot k}
        \frac{\partial}
            {\partial p_{2\nu}}
    \bigg]
    (2\pi)^4
    \delta^{(4)}(p_1\!+\!p_2\!-\!q_1\!-\!q_2)
    \\=
    ie \mathcal{M}_0\, \epsilon_{r\mu}^*(\vec{k}) \,k_{\nu}
    \bigg[
        \frac{\bar{L}_{q_1}^{\mu\nu}}
            {q_1\cdot k}\!+\!
        \frac{\bar{L}_{q_2}^{\mu\nu}}
            {q_2\cdot k}\!-\!
        \frac{L_{p_1}^{\mu\nu}}
            {p_1\cdot k}\!-\!
        \frac{L_{p_2}^{\mu\nu}}
            {p_2\cdot k}
    \bigg]
    (2\pi)^4
    \delta^{(4)}(p_1\!+\!p_2\!-\!q_1\!-\!q_2)
\end{split}
\end{equation}
where $i\mathcal{M}_0$ is the elastic invariant amplitude. The orbital angular momentum operators associated with the hard particles act on the delta function producing derivatives of it.
Since the action of the spin angular momentum operator on the $\delta$ function is trivial, the orbital angular momentum operators in \eqref{eq:derivatives_on_delta} can be replaced with total angular momentum operators. Combining this result with the second term of \eqref{eq:S_matrix_el_expansion}, we obtain the subleading soft-photon theorem for the full S-matrix elements at tree level:
\begin{equation}
\begin{split}
    S_{\rm tree}(\omega_k,\hat{k})=
    e\ \bigg\{
                \Big[&
                    \frac{q_1\cdot
                            \epsilon_r^*(\vec{k})}
                        {q_1\cdot k}+
                    \frac{q_2\cdot
                            \epsilon_r^*(\vec{k})}
                        {q_2\cdot k}-
                    \frac{p_1\cdot
                            \epsilon_r^*(\vec{k})}
                        {p_1\cdot k}-
                    \frac{p_2\cdot
                            \epsilon_r^*(\vec{k})}
                        {p_2\cdot k}
                \Big]\\&+
                i\epsilon^*_{r\mu}(\vec{k})\ 
                k_{\nu}
                \Big[
                    \frac{\bar{J}_{q_1}^{\mu\nu}}
                        {q_1\cdot k}+
                    \frac{\bar{J}_{q_2}^{\mu\nu}}
                        {q_2\cdot k}-
                    \frac{J_{p_1}^{\mu\nu}}
                        {p_1\cdot k}-
                    \frac{J_{p_2}^{\mu\nu}}
                        {p_2\cdot k}
                \Big]
            \bigg\}\
            S_0+
            \mathcal{O}(\omega_k)
\end{split}
\end{equation}
where $S_0$ is the S-matrix element for the elastic process. The angular momentum operators act both on the invariant amplitude $i\mathcal{M}_0$ and the delta function $(2\pi)^4\delta^{(4)}(p_1+p_2-q_1-q_2)$ in $S_0$ producing the second and third terms of \eqref{eq:S_matrix_el_expansion}. 

\section{Calculation of \texorpdfstring{$\Tilde{S}^{(\alpha,3)}_{\text{tree}}$}{TEXT}}
    \label{sec:D}
    In this Appendix, we calculate the order $e^3$ contribution  $\Tilde{S}^{(\alpha,3)}_{\text{tree}}$, which arises from the S-matrix element
\begin{equation}
    \begin{split}
    \bra{q_1,q_2}a_r(\vec{k}_{\gamma})
            \Big(
                1+\int\reallywidetilde{d^3k}\
                    [f_q^*(\vec{k})+g_q^*(\vec{k})]
                        \cdot
                    a(\vec{k})
                \Big)S\
            &
        \\
            \Big(
            1+\int\reallywidetilde{d^3k}\
                   [f_p(\vec{k})+ g_p(\vec{k})]
                        \cdot
                    a^{\dagger}(\vec{k})
                \Big)&
            \ket{p_1,p_2}
    \end{split}
    \end{equation}
It is easy to check that the terms involving two annihilation operators yield contributions that are higher order than $e^3$. Omitting those, we get
  \begin{equation}
    \bra{q_1,q_2}a_r(\vec{k}_{\gamma})\,S\,\Big(
            1+\int\reallywidetilde{d^3k}\
                   [f_p(\vec{k})+ g_p(\vec{k})]
                        \cdot
                    a^{\dagger}(\vec{k})\Big)\ket{p_1,p_2}
 \end{equation}  
    To order $e^3$, this splits into two contributions, depending on whether the additional photon of momentum $\Vec{k}_{\gamma}$ and polarization index $r$, interacts with an external charge particle. Hence, we write
    \begin{equation}
    \Tilde{S}^{(\alpha,3)}_{\text{tree}}=
        \Tilde{S}^{(\alpha,3)}_1+
        \Tilde{S}^{(\alpha,3)}_2
    \end{equation}
    where 
    the first term corresponds to the case for which the additional soft photon is interacting, and the 
    the second term arises when the additional soft photon is non-interacting. The first term is given by the non-trivial part of the undressed S-matrix element for the emission of a soft photon during the scattering process $\alpha$, at tree level. 
    We can obtain it by applying the subleading soft theorem since the photon is soft. 
    \begin{itemize}
        \item For the $e^-(p_1)+\mu^-(p_2)\rightarrow e^-(q_1)+\mu^-(q_2)$ scattering process, we obtain:
        \begin{equation}
        \begin{split}
            \tilde{S}_1^{(\mu,3)}=
                e\ \bigg\{
                \Big[&
                    \frac{q_1\cdot
                            \epsilon_r^*(\vec{k}_{\gamma})}
                        {q_1\cdot k_{\gamma}}+
                    \frac{q_2\cdot
                            \epsilon_r^*(\vec{k}_{\gamma})}
                        {q_2\cdot k_{\gamma}}-
                    \frac{p_1\cdot
                            \epsilon_r^*(\vec{k}_{\gamma})}
                        {p_1\cdot k_{\gamma}}-
                    \frac{p_2\cdot
                            \epsilon_r^*(\vec{k}_{\gamma})}
                        {p_2\cdot k_{\gamma}}
                \Big]\\&+
                i\epsilon^*_{r\mu}(\vec{k}_{\gamma})\ 
                k_{\gamma\nu}
                \Big[
                    \frac{\bar{J}_{q_1}^{\mu\nu}}
                        {q_1\cdot k_{\gamma}}+
                    \frac{\bar{J}_{q_2}^{\mu\nu}}
                        {q_2\cdot k_{\gamma}}-
                    \frac{J_{p_1}^{\mu\nu}}
                        {p_1\cdot k_{\gamma}}-
                    \frac{J_{p_2}^{\mu\nu}}
                        {p_2\cdot k_{\gamma}}
                \Big]
            \bigg\}\
            {S}_0^{(\mu)}+
            \mathcal{O}(\omega_{\gamma})
        \end{split}
        \end{equation}
        \item for the $e^-(p_1)+\gamma(p_2)\rightarrow\gamma(q_1)+ e^-(q_2)$ scattering process:
    \begin{equation}
    \begin{split}
        \tilde{S}_1^{(\gamma,3)}=\!
            e \bigg\{
                \Big[&
                    \frac{q_2\cdot
                            \epsilon_r^*(\vec{k}_{\gamma})}
                        {q_2\cdot k_\gamma}-
                    \frac{p_1\cdot
                            \epsilon_r^*(\vec{k}_{\gamma})}
                        {p_1\cdot k_\gamma}
                \Big]\!+\!
                i\epsilon^*_{r\mu}(\vec{k}_{\gamma})\ 
                k_{\gamma\nu}
                \Big[
                    \frac{\bar{J}_{q_2}^{\mu\nu}}
                        {q_2\cdot k_\gamma}-
                    \frac{J_{p_1}^{\mu\nu}}
                        {p_1\cdot k_\gamma}
                \Big]
            \bigg\}
            {S}_0^{(\gamma)}\!+\!
            \mathcal{O}(\omega_{\gamma})
    \end{split}
    \end{equation}
    \item for the $e^-(p_1)+e^+(p_2)\rightarrow\gamma(q_1)+\gamma(q_2)$ scattering process:
    \begin{equation}
    \begin{split}
        \tilde{S}_1^{(e,3)}=\!
            e \bigg\{
                \Big[&
                    \frac{p_2\cdot
                            \epsilon_r^*(\vec{k}_{\gamma})}
                        {p_2\cdot k_\gamma}-
                    \frac{p_1\cdot
                            \epsilon_r^*(\vec{k}_{\gamma})}
                        {p_1\cdot k_\gamma}
                \Big]\!+\!
                i\epsilon^*_{r\mu}(\vec{k}_{\gamma})\ 
                k_{\gamma\nu}
                \Big[
                    \frac{\bar{J}_{\bar{p}_2}^{\mu\nu}}
                        {p_2\cdot k_\gamma}-
                    \frac{J_{p_1}^{\mu\nu}}
                        {p_1\cdot k_\gamma}
                \Big]
            \bigg\}
            {S}_0^{(e)}\!+\!
            \mathcal{O}(\omega_\gamma)
    \end{split}
    \end{equation}
    \end{itemize}
    In all three cases, $\tilde{S}_1^{(\alpha,3)}$ can be re-expressed in terms of the dressing functions, as follows
    \begin{equation*}
        \Tilde{S}^{(\alpha,3)}_1=
            [f_q(\vec{k}_{\gamma})+
                g_q(\vec{k}_{\gamma})
            -f_p(\vec{k}_{\gamma})-
                g_p(\vec{k}_{\gamma})]
                \cdot
            \epsilon_r^*(\vec{k}_{\gamma})
            \ {S}_0^{(\alpha)}+
            \mathcal{O}(\omega_{\gamma})
    \end{equation*}
For $\alpha=\mu$, we used the following property
    \begin{equation*}
        \epsilon_r^*(\vec{k}_{\gamma})
            \cdot
        \Big(
            \frac{\partial}
                {\partial p_1}+
            \frac{\partial}
                {\partial q_1}+
            \frac{\partial}
                {\partial p_2}+
            \frac{\partial}
                {\partial q_2}
        \Big)\
        \frac{1}{
            [
            \frac{1}{2}(p_1-q_1)^2
            +\frac{1}{2}(p_2-q_2)^2
            -i\epsilon
            ]}=0
    \end{equation*}
    
    The contribution for the case for which the additional soft photon is not interacting can be obtained by letting the annihilation operator $a_r(\Vec{k}_{\gamma})$ commute with the cloud creation operator acting on the initial state. 
    Taking into account the commutator, we get 
    \begin{equation*}
        \Tilde{S}^{(\alpha,3)}_2=
            f_p(\vec{k}_{\gamma})
                \cdot
            \epsilon^*_r(\vec{k}_{\gamma})
            \ {S}_0^{(\alpha)}
            +g_p(\vec{k}_{\gamma})
                \cdot
            \epsilon^*_r(\vec{k}_{\gamma})
            \ S_0^{(\alpha)}
    \end{equation*}
    Adding the two contributions, 
    we obtain
    \begin{equation}
        \Tilde{S}^{(\alpha,3)}_{\text{tree}}=
            [f_q(\vec{k}_{\gamma})+
                    g_q(\vec{k}_{\gamma})]
                \cdot
            \epsilon^*_r(\vec{k}_{\gamma})
            \ {S}_0^{(\alpha)}\!+\!
            \mathcal{O}(\omega_\gamma)
    \end{equation}


\bibliography{References/References_1,References/References_2,References/References_3,References/References_4}
    \bibliographystyle{jhep}

\end{document}